\documentclass[letterpaper,12pt]{amsart}   

\usepackage{amsmath}
\usepackage{amsthm,amsbsy}
\usepackage{mathcomp}
\usepackage{textcomp}

\usepackage{amssymb}
\usepackage{graphicx}
\usepackage{graphicx}
\usepackage{dcolumn}
\usepackage{bm}
\usepackage{amsfonts}
\usepackage{latexsym}
\usepackage{pdfsync}
    \usepackage{fullpage} 
\usepackage{colordvi}
\usepackage{color}
\usepackage{booktabs}
\usepackage{graphicx}
\usepackage{subfigure}
\usepackage{stmaryrd}
\usepackage{cite}
\input xy
\xyoption{all}

\newcommand{\commentout}[1]{}
\newcommand{\nwc}{\newcommand}
\nwc{\nn}{\nonumber}

\nwc{\nwt}{\newtheorem}
\nwt{cor}{Corollary}
\nwt{proposition}{Proposition}
\nwt{corollary}{Corollary}
\nwt{theorem}{Theorem}
\nwt{summary}{Summary}
\nwt{lemma}{Lemma}
\nwt{definition}{Defintion}
\nwt{remark}{Remark}

\nwc{\FF}{\mathcal{F}}
\nwc{\xx}{\mathbf{x}}
\nwc{\CC}{\mathbb{C}}
\nwc{\ZZ}{\mathbb{Z}}
\nwc{\RR}{\mathbb{R}}
\nwc{\bk}{\mathbf{k}}
\nwc{\bz}{\mathbf{z}}
\nwc{\bom}{\boldsymbol\omega}
\nwc{\bn}{\mathbf{n}}
\nwc{\bN}{\mathbf{N}}
\nwc{\PP}{\mathcal{P}}
\nwc{\PO}{\mathcal{P}_{\rm o}}
\nwc{\POk}{\mathcal{P}_{{\rm o},k}}
\nwc{\QM}{\mathcal{Q}_{\rm m}}
\nwc{\QMb}{\mathcal{\hat Q}_{\rm m}}
\nwc{\QF}{\mathcal{Q}_{\rm f}}
\nwc{\QMk}{\mathcal{Q}_{{\rm m},k}}
\nwc{\QFk}{\mathcal{Q}_{{\rm f},k}}
\nwc{\PF}{\mathcal{P}_{\rm f}}
\nwc{\PFk}{\mathcal{P}_{{\rm f},k}}
\nwc{\RO}{\mathcal{R}_{\rm o}}
\nwc{\RF}{\mathcal{R}_{\rm f}}
\nwc{\ROk}{\mathcal{R}_{{\rm o},k}}
\nwc{\RFk}{\mathcal{R}_{{\rm f},k}}
\nwc{\QQ}{\mathcal{Q}}
\nwc{\PT}{\mathcal{T}}
\nwc{\real}{\text{re}}
\nwc{\imag}{\text{im}}
\nwc{\ep}{\epsilon}
\nwc{\lam}{\lambda}
\nwc{\tlam}{{\lambda}_0}
\nwc{\hlam}{\hat{\lambda}}
\nwc{\tphi}{{{\phi}_0}}
\nwc{\CN}{\mathcal{C}(\cN )}

\nwc{\mf}{\mathbf}
\nwc{\mb}{\mathbf}
\nwc{\ml}{\mathcal}
\nwc{\bj}{{\mb j}}
\nwc{\bA}{{\mb \Phi}}
\nwc{\IA}{\mathbb{A}} 
\nwc{\bi}{\mathbf i}
\nwc{\bo}{\mathbf o}
\nwc{\IS}{\mathbb{S}}
\nwc{\IC}{\mathbb{C}} 
\nwc{\ID}{\mathbb{D}} 
\nwc{\IM}{\mathbb{M}} 
\nwc{\IP}{\mathbb{P}} 
\nwc{\bI}{\mathbf{I}} 
\nwc{\IE}{\mathbb{E}} 
\nwc{\IF}{\mathbb{F}} 
\nwc{\IG}{\mathbb{G}} 
\nwc{\IN}{\mathbb{N}} 
\nwc{\IQ}{\mathbb{Q}} 
\nwc{\IR}{\mathbb{R}} 
\nwc{\IT}{\mathbb{T}} 
\nwc{\IZ}{\mathbb{Z}} 
\nwc{\IV}{\mathbb{V}}
\nwc{\IX}{\mathbb{X}}
\nwc{\IY}{\mathbb{Y}}

\nwc{\cE}{{\ml E}}
\nwc{\cP}{{\ml P}}
\nwc{\cQ}{{\ml Q}}
\nwc{\cL}{{\ml L}}
\nwc{\cX}{{\ml X}}
\nwc{\cW}{{\ml W}}
\nwc{\cZ}{{\ml Z}}
\nwc{\cR}{{\ml R}}
\nwc{\cV}{{\ml V}}
\nwc{\cT}{{\ml T}}
\nwc{\crV}{{\ml L}_{(\delta,\rho)}}
\nwc{\cC}{{\ml C}}
\nwc{\cA}{{\ml A}}
\nwc{\cK}{{\ml K}}
\nwc{\cB}{{\ml B}}
\nwc{\cD}{{\ml D}}
\nwc{\cF}{{\ml F}}
\nwc{\cS}{{\ml S}}
\nwc{\cM}{{\ml M}}
\nwc{\cG}{{\ml G}}
\nwc{\cH}{{\ml H}}

\nwc{\bT}{{\mb T}}
\nwc{\bM}{{\mb M}}
\nwc{\cbz}{\overline{\cB}_z}
\nwc{\supp}{{\hbox{\rm supp}}}
\nwc{\fR}{\mathfrak{R}}
\nwc{\bY}{\mathbf Y}

\nwc{\pft}{\cF^{-1}_2}
\nwc{\bU}{{\mb U}}
\nwc{\bPhi}{{\mb \Phi}}
\nwc{\bPsi}{{\mb \Psi}}
\nwc{\im}{{\rm i}}

\nwc{\bw}{{\mathbf w}}
\nwc{\mbm}{{\mathbf m}}
\nwc{\lbr}{\textlbrackdbl}
\nwc{\rbr}{\textrbrackdbl}
\nwc{\vzero}{{\mathbf 0}}
\nwc{\cN}{{\mathcal N}}
\nwc{\rbra}{\textrbrackdbl}
\nwc{\lbra}{\textlbrackdbl}
\nwc{\conv}{\hbox{conv}}
\nwc{\rank}{\hbox{rank}}

\nwc{\beq}{\begin{eqnarray}}
\nwc{\beqn}{\begin{eqnarray*}}
\nwc{\eeqn}{\end{eqnarray*}}
\nwc{\eeq}{\end{eqnarray}}
\nwc{\cle}{\preccurlyeq}
\nwc{\modpi}{{{\rm mod}\,2\pi}}
\nwc{\lb}{\llbracket}
\nwc{\rb}{\rrbracket}

\begin{document}
\title{Fourier Phasing with Phase-Uncertain Mask$^\dagger$
\footnote{$^\dagger$Inverse Problems {\bf 29} (2013) 125001 
}
}

\author{Albert Fannjiang and Wenjing Liao}
\address{Department of Mathematics, University of California, Davis, CA 95616.}
\email{fannjiang@math.ucdavis.edu}

\date{}
\maketitle 



\begin{abstract} Fourier phasing is the problem of retrieving Fourier phase information from
Fourier intensity data. The standard Fourier phase retrieval (without a mask) is known to have many solutions
which cause the standard phasing algorithms to stagnate and produce wrong or inaccurate solutions. 

In this paper Fourier phase retrieval is carried out with the introduction of a randomly fabricated mask in measurement and reconstruction.
Highly probable uniqueness of solution, up to a global phase, was previously proved with exact knowledge of the mask. Here the 
uniqueness result is extended  to the case where only rough information about the mask's phases is assumed. The exponential probability bound for uniqueness is given in terms of
the uncertainty-to-diversity ratio (UDR) of the unknown mask.  New phasing algorithms alternating between the object update and the mask
update are  systematically tested and demonstrated to have the capability
of recovering both the object and the mask (within the object support) simultaneously, consistent with the uniqueness result.
 Phasing with a phase-uncertain mask is shown to be  robust with respect to the correlation in the mask
 as well as  the Gaussian and Poisson noises.   
\end{abstract}

\section{Introduction}

Fourier phasing is the problem of reconstructing
an unknown object from  its Fourier intensity data and
is fundamental in many applications. Recent breakthroughs  center around
diffractive imaging of non-periodic objects, combining the penetration power of hard X-ray
and the high sensitivity of lensless imaging \cite{Miao99,ptycho08,ptycho10}. 
Since the interaction of X-rays with
matter is weak compared to that of electrons, multiple
scattering can be neglected and the singly  scattered far field is
essentially the Fourier transform of the transmission function of the image
via  proper choice of variables. 
 
 Despite tremendous progresses, 
 many questions, fundamental as well as algorithmic,  remain to be solved. 
The
standard phasing algorithms, based on alternating projections \cite{GS72, Fie82},  are plagued by stagnation and spurious errors partly due to intrinsic non-uniqueness of the standard phasing problem.  The competition among the true and the ambiguous solutions accounts for their slow convergence and possible stagnation  \cite{Fie2,FW}.

We believe that the two problems, non-uniqueness and non-convergence, can be solved
in one stroke with high probability by introducing a random mask. 
We have previously shown that  phasing with a randomly fabricated, but otherwise {\em exactly known},  mask  yields  a unique solution, { up to a global phase factor},  with {\em high probability} as well as
 superior numerical performances, including rapid convergence, much reduced data and noise stability. 
 In particular, the random mask method is robust  to  various types of noise, including Gaussian, Poisson and
 mask noises, with a noise amplification factor about 2  \cite{UniqueRI,FL}. 
 Although uniqueness of solution holds only with high probability (in the mask selection), instead of probability one, it suffices for all practical purposes. 
 
Similar in spirit is the wavefront curvature approach \cite{Nut03, Nut05,Wil06}
which derives uniqueness, up to a global phase, by using cylindrical, in addition to planar,  
incident waves. The cylindrical wave approach, however, requires
$d+1$ Fourier measurements ($d=$ the dimension of the object)
as well as the Neumann boundary condition
of Fourier phase. In contrast, our previous results of  highly probable uniqueness  \cite{UniqueRI} require
just {\em one} Fourier measurement for complex-valued objects whose phases are
limited to any proper interval $[a,b]\subsetneq [0,2\pi)$
and  {\em two}  Fourier measurements for unconstrained complex objects
in any dimension. This is an example of randomized measurement leading to optimal
information retrieval.  Previously, the effect of a random (binary) mask on Fourier phasing has been observed in
\cite{Rod10}. 

Comparison can also be made with ptychography \cite{ptycho10, ptycho2, ptycho08,Thi09}
which  is a coherent diffractive imaging method that uses multiple diffraction patterns obtained through the scan of a localized illumination on the specimen (see also Remark \ref{rmk:pty}).
In ptychography, the adjacent illuminations have to overlap around 60 - 70 \% in every dimension. 
This corresponds to at least 3 illuminations for every point of the object and
roughly more than 3 Fourier measurements in two dimensions.  
In fact, randomly phased masks have been recently  deployed in the ptychographic approach to X-ray microscopy to enhance its
performance with the extra benefit of reduced dynamic range of the recorded diffraction patterns \cite{ptycho-rpi}. 

A critique that can be leveled  against 
the random mask approach is the assumption of exact knowledge of the mask
which is not always available. In the present work, we address the phasing problem with
a random mask whose phases are not exactly known.
We will show that {\em nearly perfect} recovery of both the object and the mask can be achieved with  high probability.  

Our approach is based on  two new highly probable uniqueness results  for the setting with random { phase-uncertain mask} (PUM) whose phases are only {\em roughly known} and satisfy a crude uncertainty constraint.  Instead of running phasing algorithms with a fixed erroneous  mask, we design algorithms to recover the object and the mask simultaneously.  At each iteration, the object and the mask are updated alternatively, aiming at fitting the object constraint, the mask constraint as well as the Fourier intensity data. As shown below  our numerical schemes can accurately recover the object 
with close to 50\% uncertainty in mask phases. 

The paper is organized as follows. We state the uniqueness theorems for phasing with
a random PUM in Section \ref{sec:unique} and give the proofs in Appendices A and B.
We discuss the basic algorithm of Alternating-Error-Reduction (AER) and prove the residual reduction 
property in Section \ref{sec:aer} and Appendix C. We discuss the Douglas-Rachford-Error-Reduction
(DRER) algorithm in Section \ref{sec:drer} and the algorithms with two sets of Fourier intensity data
in Section \ref{sec:2}. We present numerical results in Section \ref{sec:num} and conclude in
Section \ref{sec:last}. A preliminary version of the results
is given in \cite{RKM}.

\section{Random-mask-aided phasing}

Let us consider discrete Fourier phasing first without a mask (the standard setting) and then with a mask. 

 Let $\bn = (n_1,\ldots,n_d) \in \ZZ^d$ and $\bz = (z_1,\ldots,z_d) \in \CC^d$ where $d\geq 2$ is the ambient dimension. Define the multi-index notation $\bz^\bn = z_1^{n_1}z_2^{n_2}\hdots z_d^{n_d}$. Let $\mathcal{C}(\cN )$ denote the set of finite complex-valued functions on $\ZZ^d$ vanishing outside
\[
 \mathcal{N} = \{ \mathbf{0} \le \bn \le \bN\}, \bN = (N_1,N_2,\ldots,N_d). 
\]
Here  $\mathbf{m} \le \bn$ if $m_j \le n_j, \forall j$. Set $|\cN|  =\displaystyle \prod_{j=1}^d (N_j+1)$. 
 
The $\bz$-transform 
$
F(\bz) =  \sum_{\bn} f(\bn) \bz^{-\bn}
$ of  $f\in \mathcal{C}(\cN )$ 
 is an analytic continuation  of 
the Fourier transform with $\bz$ in the $d$-dimensional unit torus $\{(\exp{(2\pi i \omega_1)},\dots,\exp{(2\pi i \omega_d)}),  \omega_j \in [0,1]\}$. The standard Fourier phasing is to determine $F(\bz)$ (and hence $\{f(\bn)\}$) from the data $\{|F(\bz)|\}$ over the $d$-dimensional unit torus. This is a nonlinear inversion problem. Worse still, the problem is
non-convex due to the {\em non-convexity} of
the set of functions satisfying the Fourier intensity data. 
  
But non-uniqueness of phasing solutions may be even more problematic than the non-convexity of
the phasing problem. 
Let us digress to make a simplifying  observation.  Considering   the calculation 
 \beq
  |F(e^{i2\pi\bom})|^2&=& \sum_{\bn =-\bN}^{\bN}\sum_{\mbm\in \cN} f(\mbm+\bn)\overline{f(\mbm)}
   e^{-\im 2\pi \bn\cdot \bom}\nn
   \eeq
where the over-bar notation means
complex conjugacy, 
   we see that the Fourier intensity measurement
   is equivalent to the discrete Fourier measurement of
   the correlation function 
          \beqn
	 \label{aut}
	  \cC_{ f}(\bn)=\sum_{\mbm\in \cN} f(\mbm+\bn)\overline{f(\mbm)}
	  \label{autof}
	  \eeqn
if sampled at the lattice 
 \beqn
\mathcal{L} = \Big\{\bom=(\omega_1,...,\omega_d)\ | \ \omega_j = 0,\frac{1}{2 N_j + 1},\frac{2}{2N_j + 1},...,\frac{2N_j}{2N_j + 1}\Big\}
\label{latice}
\eeqn
which is approximately $2^d$ times of the number of degrees of freedom in $f$. 
Hence sampling on $\cL$ corresponds to  the oversampling ratio 
$
 {{|{\mathcal L}|}}/{|\cN|} \approx 2^d.
$
By the sampling theorem for band-limited signals, the Fourier intensity data over $\cL$ contain
the complete information of the Fourier intensity over the $d$-dimensional unit torus.
So
the standard phasing problem  can be recast as recovering $f$ from its Fourier intensity data  $|F(e^{i 2\pi \bom})|^2,\forall\bom \in \cL$. 

However,  the autocorrelation function $\cC_f$ does not uniquely determine 
 the object $f$. 

First, there are three types of {\em global} ambiguities/associates:
\begin{itemize}
\item[(a)]
 Constant global phase: $
 f(\cdot)\longrightarrow \exp{(i\theta)}f(\cdot),\ \hbox{for some}\,\,\theta\in [0,2\pi),
 $
\item[(b)] Spatial shift:  $
   f(\cdot)\longrightarrow f(\cdot +\mbm),\ \hbox{for some}\,\,\mbm\in \IZ^d,
 $
\item[(c)] Conjugate inversion:
$
    f(\cdot)\longrightarrow \overline{f(\bN-\cdot)}.
$\\
 Conjugate inversion produces the so-called twin image. 
 \end{itemize}
 These trivial,  global associates all share the same global geometric information as the original object and can be viewed as belonging to the same equivalence class of objects.

 The classical result \cite{Hayes82, HM82} says that for {\em generic} objects in dimension two or higher the global ambiguities 
are the only ambiguities in Fourier phasing. Since  the global associates are all 
simple transformations of the original object, one is tempted to believe that
the phasing problem is well-posed relative to the equivalence classes of objects.  There are, however,  two caveats with this result. One,
generic objects  almost surely have a full support ($\cN$) and hence do not include 
objects with zero voxels in $\cN$. This is an unrealistic  restriction. Second, without the exact knowledge of the outer boundary of the object support (i.e. tight
support constraint) the standard phasing algorithms do not perform well even with
noiseless data \cite{Fie82,FW}, indicating ill-posedness relative to the equivalence classes. 

\begin{figure}[th]
\centering
\subfigure{
         \includegraphics[width = 8cm]{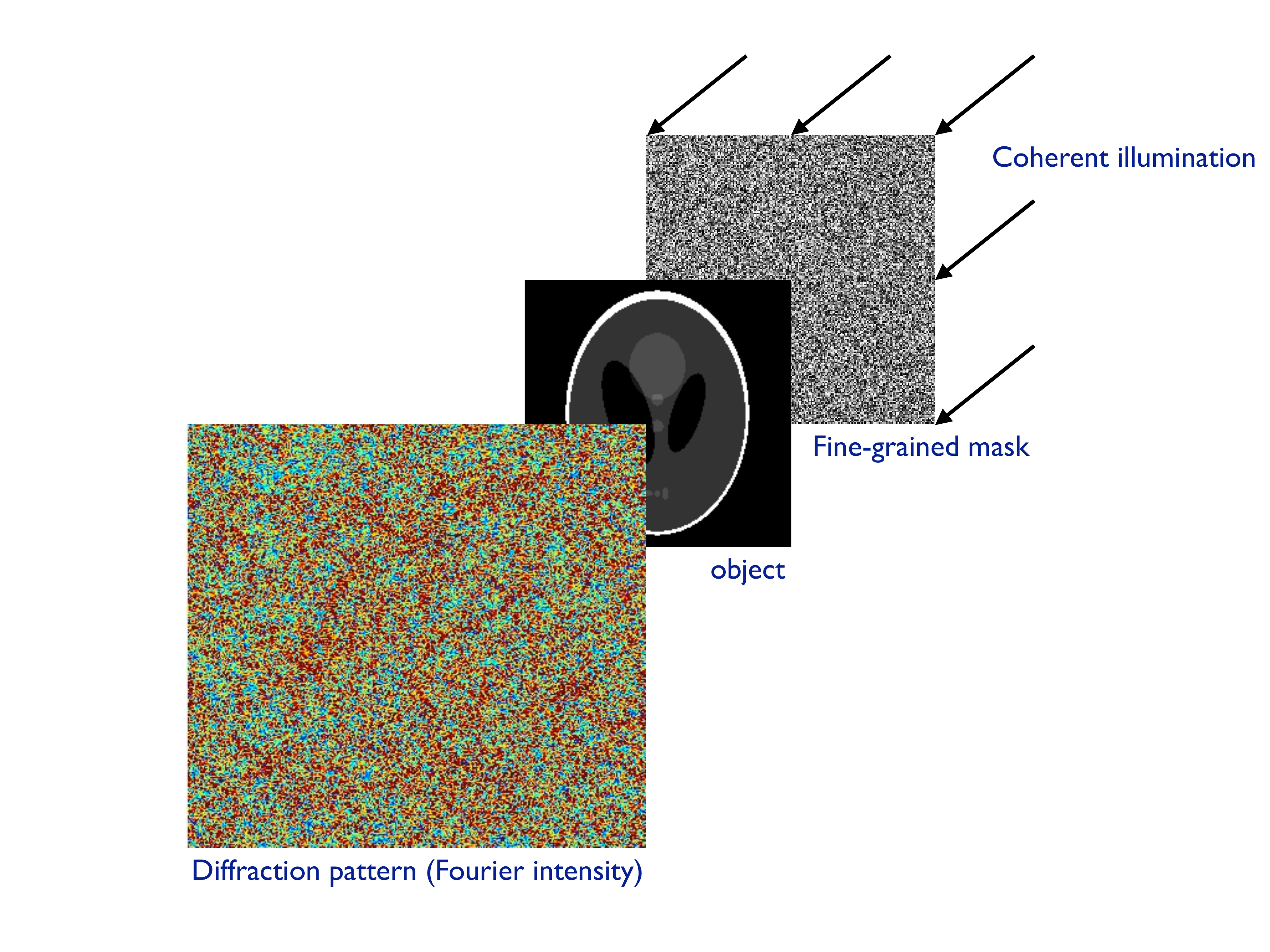}} 
         \subfigure{
         \includegraphics[width = 8cm]{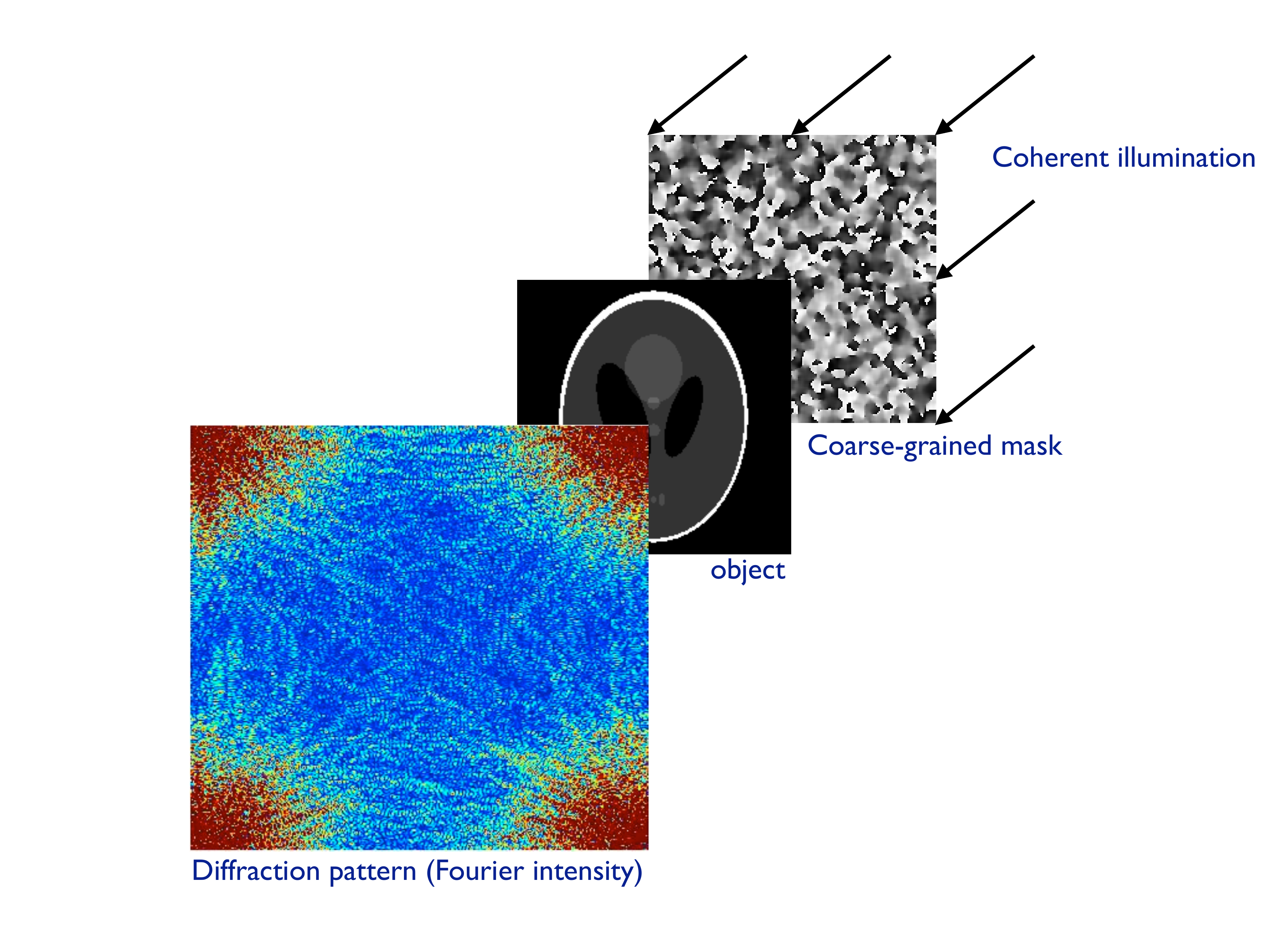}} 
  \caption{Imaging geometry with a fine-grained (high-resolution) mask (left) and coarse-grained (low-resolution) mask (right). The mask can be placed in front of or behind the object. The construction of fine- and coarse-grained masks is given in Section \ref{sec:mask}.} 
 \label{fig:geom}
\end{figure}

The random-mask-aided phasing method \cite{UniqueRI,FL} introduces a random mask into
the Fourier intensity measurement (see Fig. \ref{fig:geom}). 
The effect of a mask  amounts to changing  the original object $f$ to
the masked object
\beq
\label{mo}
g(\bn) = \mu(\bn)f(\bn),\quad\bn\in\IZ^d
\eeq
where $\mu$ is an array representing the mask. The standard phasing set-up is equivalent to
$\mu\equiv 1$, i.e. 
the uniform (hence deterministic) mask (UM). In this paper we assume that  the mask $\mu$ is randomly
fabricated and only
roughly known. We will  focus on  the case of  phase masks  
\beq
\mu(\bn) = \exp{(i\phi(\bn))},\quad \phi(\bn)\in [0,2\pi), \,\, \bn\in \IZ^d
\label{randomphase}
\eeq
whose  true phases $\phi(\bn)$, in radian,  are only known to  lie within
$\delta\pi $ from a {\em known},  initial estimates $\phi_0(\bn)$ for all $\bn$. That is, the random  mask phases $\phi(\bn)$ satisfy the uncertainty constraint
\beq
\phi(\bn)\in \lb \phi_0(\bn)-\delta\pi, \phi_0(\bn)+\delta\pi\rb\equiv  \lb\phi_0(\bn)\pm \delta\pi\rb,\quad \forall\bn,
\label{MaskError}
\eeq
(see Section \ref{sec:ext} for extension to general masks).
Here and below we adopt the following notation:
$\theta\in \lb a, b \rb$ means  
\begin{equation*}
\left\{
\begin{array}{ll}
a (\modpi) \le\theta (\modpi)\le b (\modpi) & \text{ if } a (\modpi)\le b (\modpi)\\
a (\modpi)\le \theta (\modpi) < 2\pi \text{ or } 0 \le \theta (\modpi)\le b (\modpi)& \text{ else. } \\
\end{array}\right.
\end{equation*}

Some words for clarifying the use of  ``random" and ``uncertain": In this paper, ``random" means ``non-deterministic" and  a random mask
is a mask generated by a probabilistic mechanism. For example, each pixel/voxel of a random mask may be independently selected according to a probabilistic distribution which is not a Dirac delta-function. Once a mask (random or not) is generated, it may or may not be exactly known to
the user. In the latter case, we speak of a uncertain mask or a roughly known mask.
In other words,  uncertainty refers to
the calibration while randomness refers to the fabrication of the mask. 

In the original random-mask-aided approach \cite{UniqueRI, FL}, the random mask is exactly known, namely the uncertainty $\delta$ is zero. With an additional random mask in the Fourier intensity measurement, we obtained  not only uniqueness of solution but  also rapid convergence of the phasing algorithms to the true object (up to a global phase), indicating that the use of a random mask renders the phasing problem well-posed.  In the present paper, we extend the results 
to the case of phase-uncertain masks ($\delta>0$).

 \section{Uniqueness}\label{sec:unique}

First we recall the uniqueness results for $\delta=0$ \cite{UniqueRI}.

The {\em rank} of an object 
 is the dimension of the support's convex hull in $\IR^d$. An object is said $\rank \ge 2$ if the convex hull of its support has a dimension $\ge 2$.
The support of a  rank one object  is a subset of a line. The rank 2 property is a key
assumption for our uniqueness results.

The first uniqueness pertains to the real-valued objects.
\begin{proposition}Let $\{\phi(\bn)\}$ be independent, {\bf continuous} random variables on $[0,2\pi]$.  Let $f$ be a real-valued object 
of rank $\geq$   2.  Then, with probability one, $f$ is determined absolutely uniquely up to $\pm$ sign by the Fourier intensity measurement on 
$\cL$. 
\label{UniqueReal}
\end{proposition}
 
A more general constraint is to restrict the object values within a certain sector of the complex plane. For instance, for coherent X-ray diffractive imaging, the electron density is complex with
the real part representing the effective number of electrons that diffract the X-rays in phase and is usually positive and the imaginary part representing  the absorption of the X-rays by the specimen and thus is always positive  \cite{MSC}.

We have the following uniqueness for the so-called sector-constrained objects.
\begin{proposition}Let $\{\phi(\bn)\}$ be independent, {\bf uniform} random variables on $[0,2\pi]$. 
Let  $f$ be a complex-valued object of rank $\geq$ 2 such that $\measuredangle f(\bn)  \in [\alpha,\beta], \forall \bn$. Let $S$ denote the sparsity of the image and let $\lfloor S/2\rfloor$ be the greatest integer at most  half the image sparsity $S$ which is the number of nonzero pixels.   

Then with probability no less than $1-|\cN|  (\beta-\alpha)^{\lfloor S/2\rfloor}(2\pi)^{-\lfloor S/2\rfloor}$, the object $f$ is uniquely determined, up to a global phase, by the Fourier intensity measurement on $\mathcal{L}$. 

\label{UniqueComplexPositive}
\end{proposition}

For general complex-valued images without any sector constraint, measurements with two independent masks are needed to ensure uniqueness.

\begin{proposition}
 Let $\{\phi^{(1)}(\bn)\}$ and $\{\phi^{(2)}(\bn)\}$ be two independent arrays of {\bf continuous} random variables on $[0,2\pi]$. Let $f $ be any complex-valued object of rank $\geq 2$. Then almost surely $f$ is uniquely determined, up to a constant phase factor, by two Fourier intensity measurements on $\mathcal{L}$ with two masks $\mu^{(1)}(\bn)=\exp{[i\phi^{(1)}(\bn)]}$ and $\mu^{(2)}(\bn)=\exp{[i\phi^{(2)}(\bn)]}$. 
 \label{UniqueComplex}
\end{proposition}

Notice that the above uniqueness results deal with any {\em given, deterministic} object of  rank $\geq 2$. 
Moreover, there is substantial flexibility in the mask ensemble in Propositions \ref{UniqueReal}
and \ref{UniqueComplex} since only the {\bf existence} of probability density for the mask phases
is assumed. The uniformity condition in Proposition \ref{UniqueComplexPositive} can also
be relaxed but then the resulting probability bound would be more complicated.

Next we state our main theoretical results that
for proper $\delta>0$ {\em both} the ambiguities for the object and the phase-uncertain mask
can be resolved up to a global phase with overwhelming probability. 

The first result is analogous to Proposition \ref{UniqueReal}.
\begin{theorem}
Let $f$ be a real-valued object of rank $\geq 2$. Suppose  the exact mask phases $\{\phi(\bn)\}$ are independently and {\bf uniformly} distributed on $[-\gamma \pi,\gamma\pi)$. Suppose the  uncertainty of
the mask estimate $\mu_0=\{\exp{(i \phi_0(\bn))}\}$  in (\ref{MaskError}) is $\delta< \gamma/2$. 

Suppose that another real-valued image $\tilde f$ and mask estimate $\tilde \mu=\{\exp{(i\tilde \phi(\bn))} \}$ satisfying the same uncertainty constraint as (\ref{MaskError}), i.e.
\beq
\label{mask2}
\tilde\phi(\bn)\in \lb\phi_0(\bn)\pm \delta\pi\rb,
  \eeq
produce  
 the same Fourier intensity data on $\cL$ as do $f$ and $\mu$.
  Then,  with probability no less than $1-|\cN|(2\delta/\gamma)^{\lfloor S/2\rfloor}$,  $\tilde f(\bn) =\pm f(\bn) \ \forall \bn$ and furthermore $\tilde\phi(\bn) =  \theta+\phi(\bn) $ for a constant $\theta\in [0,2\pi)$ wherever $f(\bn) \neq 0$. 
 \label{TheoremReal}
\end{theorem}
\begin{remark}
If the object is known to be {\bf non-negative}, then $\delta$ can be any number in $[0,\gamma)$ and
 uniqueness holds with probability no less than $1-|\cN|(\delta/\gamma)^{\lfloor S/2\rfloor}$.
\end{remark}

{
Because the phases of the object and the mask are mixed, any uncertainty in the mask phases is automatically transferred to the object. Hence there is no direct extension of  Proposition \ref{UniqueComplexPositive} to the case of uncertain mask. To resolve the phase uncertainty and prove uniqueness we use  two independent sets of Fourier intensity data
 for complex-valued (sector-constrained or not) objects}. 

\begin{theorem}
\label{TheoremComplex2}

Let $f$ be a complex-valued object of  {\em rank} $\ge 2$.
Let the first mask $\mu^{(1)}=\mu$ be as in Theorem \ref{TheoremReal} with the initial mask estimate
$\mu_0$ satisfying (\ref{MaskError}).

Suppose the second mask  $\mu^{(2)}$ is {\bf exactly known} and
the $\bz$-transform of $\mu^{(2)}f$ is irreducible up to a power of $\bz$. Moreover, 
assume the non-degeneracy condition  that there is no $\mbm \neq \mathbf{0} $ such that $\mu^{(2)}(\bn+\mbm)f(\bn+\mbm) = \exp{(i\xi)}\exp{(i \eta(\bn))} \mu^{(2)}(\bn)f(\bn), \forall \bn,$ and no $\mbm$ such that  $ \overline{\mu^{(2)}(\mbm-\bn)f(\mbm-\bn) } = \exp{(i\xi)}\exp{(i \eta(\bn))} \mu^{(2)}(\bn)f(\bn), \forall\bn, $
for  some  $\xi \in [0,2\pi), |\eta(\bn) | \le \pi \delta$. 

Suppose that for  a phase mask  $\tilde\mu$ with (\ref{mask2})
and an object $\tilde f$
the two pairs of masked objects 
$\mu f$ and $\tilde\mu \tilde f $, $\mu^{(2)} f$ and $\mu^{(2)} \tilde f$, respectively, 
produce the same Fourier intensitis on $\mathcal{L}$. Then, with probability no less than $1-|\mathcal{N}|(\delta/\gamma)^{\lfloor S/2\rfloor }$,  $\tilde f(\bn)=\exp{(i\alpha_1)} f(\bn),\forall \bn$,  and
$\tilde\mu(\bn) = \exp{(i\alpha_2)}\mu(\bn)\  \text{if} \ f(\bn) \neq 0$,  where  $\alpha_1,\alpha_2 $ are two real numbers.

\end{theorem}
\begin{remark}
Clearly, most objects and masks obey the non-degeneracy condition. 
\end{remark}
\begin{remark}
\label{rmk:prob}
The probability bounds, in terms of the mask's uncertainty-to-diversity ratio (UDR) $\delta/\gamma$,    in Theorems \ref{TheoremReal} and \ref{TheoremComplex2} are probably not far from optimal. In particular,
the probability bound predicts the threshold UDR $\approx 1$ for reconstruction of nonnegative and complex images which is confirmed by our numerical results (see Fig. \ref{fig:gamma}). 
\end{remark}
The proofs of Theorems \ref{TheoremReal} and \ref{TheoremComplex2} are given in Appendices
A and B, respectively. 

Both theorems assert that not only the uniqueness of the object but also the uniqueness of
the mask, up to a constant phase, {\em inside} the object support. Outside the support,
the mask phase can be arbitrarily assigned without affecting the Fourier intensity data.

The surprising  lesson from Theorems \ref{TheoremReal} and \ref{TheoremComplex2} is that 
a crude constraint on a mask that is sufficiently random  is enough
to enforce uniqueness of solution (up to a global phase) as well as the mask itself (inside the object support). And this mask constraint can be
numerically implemented straightforwardly within the phasing algorithms for $\delta=0$
\cite{FL}. The resulting algorithms turn out to be capable of
nearly perfect recovery of object and mask even in the presence of relatively high uncertainty in mask. 

\section{Extension to general masks}\label{sec:ext}
The preceding discussion is limited  to the case of phase masks. 
It is easy to extend the above results to general masks, if the mask
intensities are strictly positive and certain (i.e. exactly known), as follows.

Let the mask be rewritten as $\mu(\bn)=|\mu|(\bn) \exp{(i\phi(\bn))}$, with $ |\mu(\bn)|>0, \forall \bn\in \cN$, where $|\mu|$ is certain and $\phi$ is uncertain as before.  Define the auxiliary object $\tilde f(\bn)=f(\bn)|\mu|(\bn)$. 
The Fourier phasing problem for the object  $f$ and the mask $\mu$ is  equivalent to
that for the auxiliary  object $\tilde f$ and the phase mask $\exp{(i\phi)}$ which
can be solved as above. The original object can then be recovered by
dividing the recovered auxiliary object by the known, nonzero mask intensities $|\mu|$. 

In this extension, any uncertainty of the mask intensities is converted into
that of the object. So in  case that $|\mu|$ is unknown or highly uncertain 
our approach needs substantial modification  unless the object
intensities are known {\em a priori}. For example, if the object is a phase object ($|f|=1$)
then we can proceed as if the object  were $\tilde f=f|\mu|$
and the mask were  $\exp{(i\phi)}$. After the auxiliary object is recovered, the phase
object can be recovered by normalization. 

\section{Alternating Error Reduction (AER)}\label{sec:aer}

\nwc{\cO}{\mathcal O}

 Let $\Lambda$ be the diagonal matrix with diagonal elements $\{\mu(\bn)\}$ and let $\bPhi$ represent the $d$-dimensional discrete Fourier transform.
Denote the  Fourier magnitude data vector by 
$Y= |\bPhi \Lambda f|$
where $\Lambda f (\bn) = \mu(\bn) f(\bn)$ and the absolute value is taken component-wise. 

A standard way to utilize the oversampled data (over $\cL$)  is to enlarge
the original image by adding corresponding number of zero pixels (i.e. zero padding) which is then enforced as an additional object constraint. 
This procedure  is called the oversampling method \cite{Miao00} and implemented in all our simulations  with the oversampling ratio $|\cL|/|\cN|\approx 4$ (for $d=2$). There are many ways to zero-pad the object. For example, we can extend the definition of the original object $f$ and the mask $\mu$  from
$\cN$ to the larger domain 
\[
 \{ -\bN \le \bn \le \bN\},\quad  \bN = (N_1,N_2,\ldots,N_d)
\]
with the additional object constraint $f=\phi=0$ outside $\cN$. In the framework of the oversampling method,  the $d$-dimensional discrete Fourier transform $\bPhi$ is a $\prod_j (2N_j+1)\times \prod_j (2N_j+1)$ matrix.

\subsection{Object Update}

Given the object estimate $f_k$ and  mask estimate $\mu_k $ at the $k$-th iteration, we use standard phasing algorithms to obtain $f_{k+1}$.

Let $\cO$ denote the ensemble of objects $\tilde f$ satisfying various object constraints (real-valued, sector etc).
Let $\PO$ be the orthogonal projection onto $\mathcal{O}$ (cf. \cite{FL} for details about numerical implementation of  $\PO$) and
$
\PFk= \Lambda_k^{-1} \bPhi^{-1} \PT \bPhi \Lambda_k,
$
where $\PT$ is the intensity fitting operator
\begin{equation}
 \PT G(\bom) = \left \{ \begin{array}{ll} Y(\bom)\exp{(i \measuredangle G(\bom))} & \text{ if } |G(\bom)| > 0 \\
Y(\bom) & \text{ if } |G(\bom)| = 0
\end{array}\right. .
\end{equation}
Here and below $\measuredangle{z} \in [0,2\pi)$ denotes the 
wrapped phase angle  of $z$. When $z=0$, $\measuredangle{z}$ is taken to be $0$ unless specified otherwise.

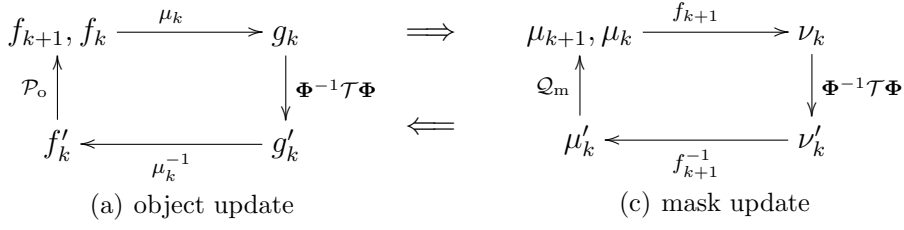
\begin{figure}
\centering
\subfigure[object update]{
$$\xymatrix{
f_{k+1}, f_k\ar[rr]^{\mu_k} && g_k \ar[d]^{\bPhi^{-1}\cT\bPhi} \\
f'_k \ar[u]^{\cP_{\rm o}} && g'_k\ar[ll]^{\mu_k^{-1}} 
}$$
}\subfigure{$$\xymatrix{\Longrightarrow \\
\Longleftarrow}$$}
 \quad 
\subfigure[mask update]{
$$\xymatrix{
\mu_{k+1}, \mu_k\ar[rr]^{f_{k+1}} && \nu_k \ar[d]^{\bPhi^{-1}\cT\bPhi} \\
\mu'_k \ar[u]^{\QM} && \nu'_k\ar[ll]^{f_{k+1}^{-1}}
}$$
}
\caption{Alternating Error Reduction (AER) between object and mask. The object update (a) is
the Error Reduction with a mask.}
\label{fig:aer2}
\end{figure}

Error Reduction (ER) takes the form
$
f_{k+1} = \PO \PFk f_k
$
which is conveniently represented by the
diagram in Figure \ref{fig:aer2}(a).

Let
$r(\tilde{f},\tilde{\mu}) = \| \ |\bPhi\tilde{\Lambda} \tilde{f}| - Y \ \|
$ denote the residual. Here and below $\|\cdot\|$ stands for
the Euclidean norm. 
 With a phase mask, ER enjoys the residual reduction property \cite{Fie82,FL}:

\beq
\label{er1}
&&r(f_{k+1},\mu_k) \le r(f_k,\mu_k)
\label{lemmaer1}
\eeq
and $r( f_{k+1}, \mu_k ) = r( f_{k}, \mu_k )$ if and only if $f_{k+1} = f_k$. 
\subsection{Mask Update}

Based on the newly updated object estimate $f_{k+1}$, the Error Reduction algorithm can 
be similarly applied to update 
the mask. Let  $\QFk$ be defined as
\begin{equation}
\mu_k'=\QFk\mu_k(\bn)  =  \left \{ \begin{array}{ll} 
\Big[{\bPhi^{-1}\PT\bPhi\Lambda_k f_{k+1}\Big](\bn)}/{f_{k+1}(\bn)}
& \text{ if } f_{k+1}(\bn) \neq 0 \\
\\
{\mu}_k(\bn) & \text{ else.} 
\end{array}.\right.
\label{QF}
\end{equation}

 Let $\cM$ be
the ensemble of phase masks  satisfying the phase uncertainty constraint (\ref{mask2}):
\beq
\mathcal{M}& =& \{ \tilde\mu \,\,  |\,\, \forall \bn, |\tilde\mu(\bn)| =1 \text{ and }  \measuredangle\tilde\mu(\bn)\in \lb\phi_0(\bn)\pm\delta\pi\rb \}.\label{man}
\eeq
Let $\QM$ be the orthogonal projection onto $\mathcal{M}$. 

The projector  $\QM$ can be computed pixel by pixel as follows. 
Let $a=(\phi_0(\bn)-\delta\pi) (\hbox{\rm mod\,\,} 2\pi)$, 
$b=(\phi_0(\bn)+\delta\pi ) (\hbox{\rm mod\,\,} 2\pi)$ and
 \[
 c=\left\{\begin{matrix}
  \pi+(a+b)/2\, (\hbox{\rm mod} \,2\pi),& \hbox{\rm if \,\,$a\le b$}\\
   (a+b)/2\, (\hbox{\rm mod}\, 2 \pi), &\hbox{\rm  else.}
   \end{matrix}\right.
   \]
Then $\QM$ can be expressed as
\beq
\QM\mu_k'(\bn) = \left\{
\begin{array}{ll}
\exp{(i\measuredangle\mu_k'(\bn))} & \text{ if }  \measuredangle \mu_k'(\bn) \in \lb a, b\rb \\
\exp{(i b)} & \text{ if } \measuredangle \mu_k'(\bn) \in \lb b,c\rb\\
\exp{(i a)} & \text{ if } \measuredangle \mu_k'(\bn) \in \lb c, a \rb.\\
\end{array}\right.
\label{1.9}
\eeq

Since the the object and the mask have interchangeable roles, we set 
$
\mu_{k+1} = \QM \QFk \mu_k
$
in the spirit of ER (see Fig. \ref{fig:aer2}(b)). 
\begin{remark} \label{rmk:pty}
Note the differences between the mask update rule here and that of the extended ptychographical engine (ePIE) ((4) in \cite{ptycho2}): First, (\ref{QF}) uses the newly updated object $f_{k+1}$ while ePIE uses the previous one. Second, more importantly, the crude prior
information of the mask is enforced by $\QM$ here while ePIE does not consider this aspect.
\end{remark}

We have  the following residual reduction property.
\begin{lemma} With $\QM$ we have
$$r(f_{k+1},\mu_{k+1}) \le r(f_{k+1},\mu_k).$$ 
\label{lemmaer2}
\end{lemma}
The proof of Lemma \ref{lemmaer2} is given in Appendix C. 
Unlike (\ref{er1}) we can not
ascertain that the equality in Lemma \ref{lemmaer2} holds only if $\mu_{k+1}=\mu_k$.

Define  the Alternating Error Reduction (AER) as
\beq
\label{aer}
(f_{k+1},\mu_{k+1})&=&
(\PO\PFk f_k, \QM\QFk \mu_k).
\eeq
In words, AER alternates between updating the object
and the mask estimates.

Lemma \ref{lemmaer2} and  (\ref{lemmaer1}) together  yield the following  residual reduction property 
for AER.
\begin{theorem}
 AER (\ref{aer}) satisfies the residual reduction property:
$r(f_{k+1},\mu_{k+1}) \le r(f_k,\mu_k).$
\label{Thm:er}
\end{theorem}

In our numerical experiments, we find that while $\QM$ works well for  real-valued objects,
but for complex-valued objects
 the following alternative rule is better 
\beq
\QMb\mu_k'(\bn) = \left\{
\begin{array}{ll}
 \exp{(i\measuredangle\mu_k'(\bn))}& \text{ if } \measuredangle \mu_k'(\bn) \in \lb a , b\rb \\
\mu_0(\bn),&\hbox{else}
\end{array}\right.
\label{1.10}
\eeq
where $\mu_0$ is the initial mask estimate. In other words, when the phase of $\mu'_k(\bn)$ falls outside the uncertainty constraint, we keep the initial mask phase instead of updating it.
With $\QMb$  we have the alternative  version of AER 
\beq
\label{aerb}
(f_{k+1},\mu_{k+1})&=&
(\PO\PFk f_k, \QMb\QFk \mu_k).
\eeq

\section{Alternating Douglas-Rachford and Error-Reduction (DRER)}\label{sec:drer}

In practice AER (either version) by itself converges slowly, typically taking up to several thousands steps
for accurate recovery in our numerical tests. 
To speed up convergence we consider  the Douglas-Rachford (DR) algorithm \cite{DR,LM79}, 
also called the averaged alternating reflections \cite{BCL04,BCL02}, 
\begin{equation}
f_{k+1} =\frac{I+\RO\RFk}{2} f_k,\quad\hbox{with}\,\, \RO =  2\PO - I, \quad \RFk =  2\PFk-I
\label{ASR}
\end{equation}
 which  coincides with the hybrid input-output (HIO) algorithm for the parameter $\beta=1$ 
 in the absence of any object value constraint.

Define the DRER iteration as 
\beq
(f_{k+1},\mu_{k+1})&=&\left( \frac{1}{2}(I + \RO\RFk)f_k, \QM  \QFk \mu_k\right)
\label{alterasr}
\label{drer}
\eeq
and the alternative version as
\beq
(f_{k+1},\mu_{k+1})&=&\left( \frac{1}{2}(I + \RO\RFk)f_k, \QMb  \QFk \mu_k\right). 
\label{drerbar}
\eeq

To strictly enforce the mask constraint, we use ER instead of DR for mask update.

\section{AER/DRER with two sets of data}\label{sec:2}
Let $\mu^{(1)}=\mu$ and $\mu^{(2)}$ be two masks with which two sets of Fourier magnitude data $Y= |\bPhi \Lambda f|$ and $Y^{(2)} = |\bPhi \Lambda^{(2)} f|$ are measured on $\cL$. Let $\PT$ and $\PT^{(2)}$ be the intensity fitting operators corresponding to $Y$ and $Y^{(2)}$, respectively.

Suppose $f_k$ and $\mu_k$ are the image and the mask recovered at the end of the $k$-th iteration. At the $(k+1)$-st iteration, the image is first updated  from $f_k$ to $f_{k+1}$ based on $\mu_k$ and $\mu^{(2)}$. Then the first  mask is updated based on $f_{k+1}$ as before. 

For simplicity of presentation we assume  the second mask (random or deterministic) is exactly known and independent from the first mask which is randomly fabricated and only roughly known.   In this case, there is  no need for updating the second mask.

Let 
$\PP_k = \Lambda_k^{-1} \bPhi^{-1} \PT \bPhi \Lambda_k$
and 
$\PP^{(2)} = (\Lambda^{(2)})^{-1} \bPhi^{-1} \PT^{(2)} \bPhi \Lambda^{(2)}.$

The   AER and DRER algorithms with two masks  are  defined respectively as 
\beq
(f_{k+1}, \mu_{k+1}) &=& \left(\PO \PP^{(2)} \PP_k f_k, \QQ_{m} \QFk \mu_k\right),\quad k=0,1,\cdots\label{aer2}\\
(f_{k+1}, \mu_{k+1}) &=&\left( \frac{1}{2}(I + \RO(2\PP^{(2)} \PP_k - I)) f_k,\QQ_{m} \QFk \mu_k\right).\label{drer2}
\eeq
As commented above replacing $\QQ_m$ with $\QMb$ improves the reconstruction of
 complex-valued objects:
\beq
(f_{k+1}, \mu_{k+1}) &=& \left(\PO \PP^{(2)} \PP_k f_k, \QMb\QFk\mu_k\right),\quad k=0,1,\cdots\label{aer2b}\\
(f_{k+1}, \mu_{k+1}) &=&\left( \frac{1}{2}(I + \RO(2\PP^{(2)} \PP_k - I)) f_k,\QMb\QFk \mu_k\right). \label{drer2b}
\eeq

\section{Numerical Simulations}\label{sec:num}
\begin{figure}[th]
\centering
         \subfigure[$282\times 282$ cameraman]{
         \includegraphics[width = 5.4cm]{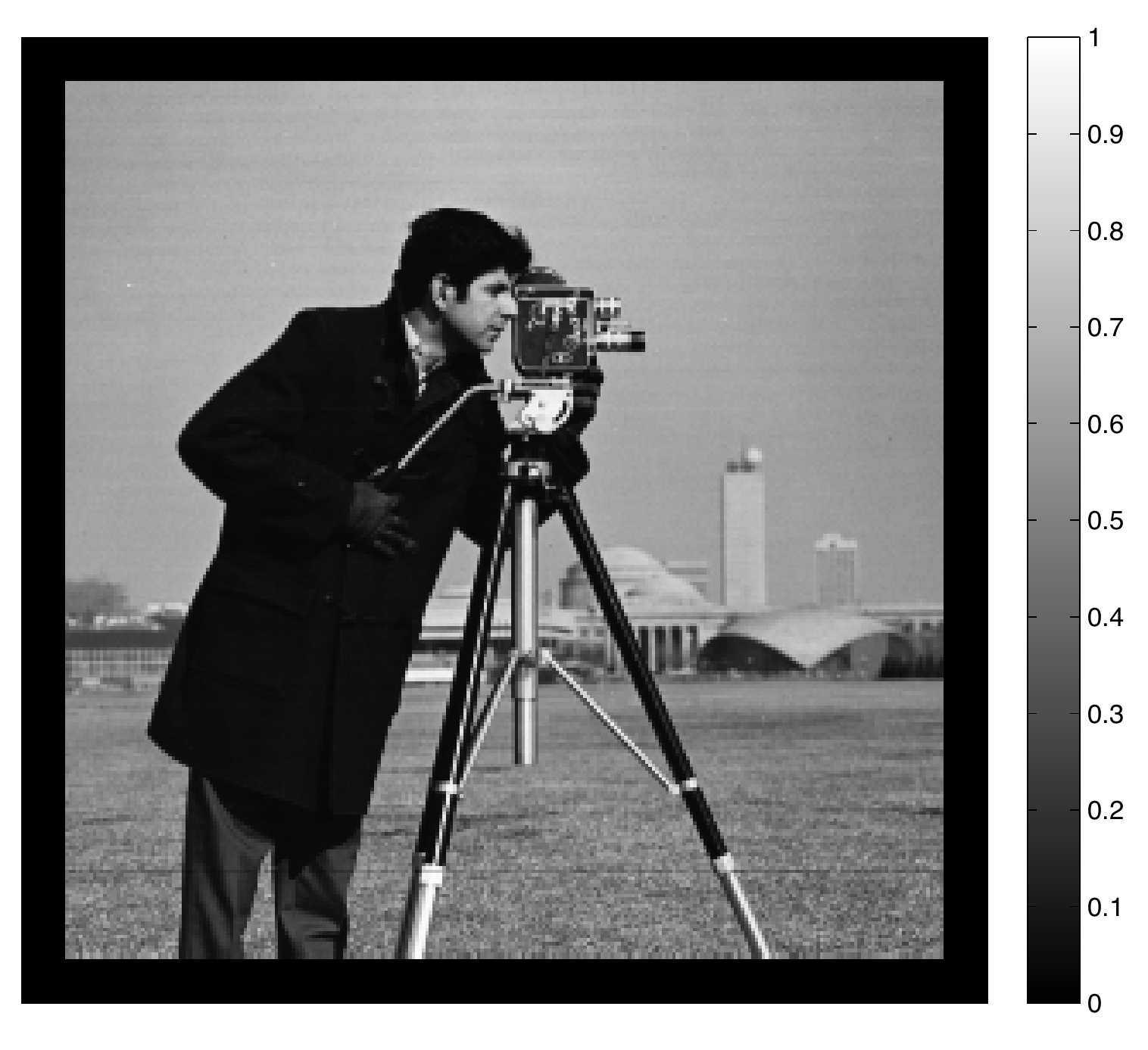}} \hspace{-.5cm}
                  \subfigure[$512\times 512$ mandrill]{
         \includegraphics[width = 5.5cm]{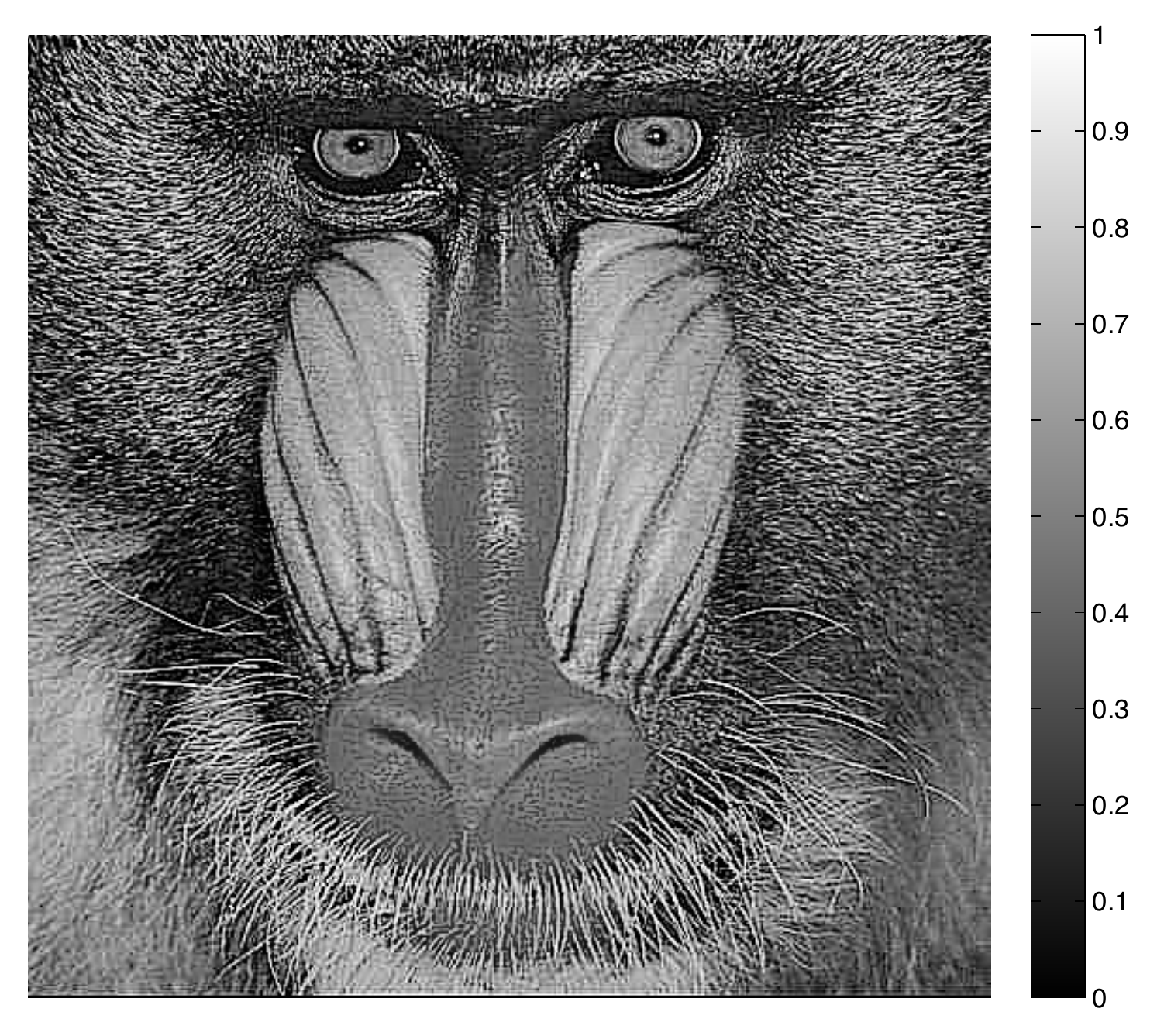}}\hspace{-0.4cm}
                 \subfigure[$200\times 200$ phantom ]{
         \includegraphics[width = 5.6cm]{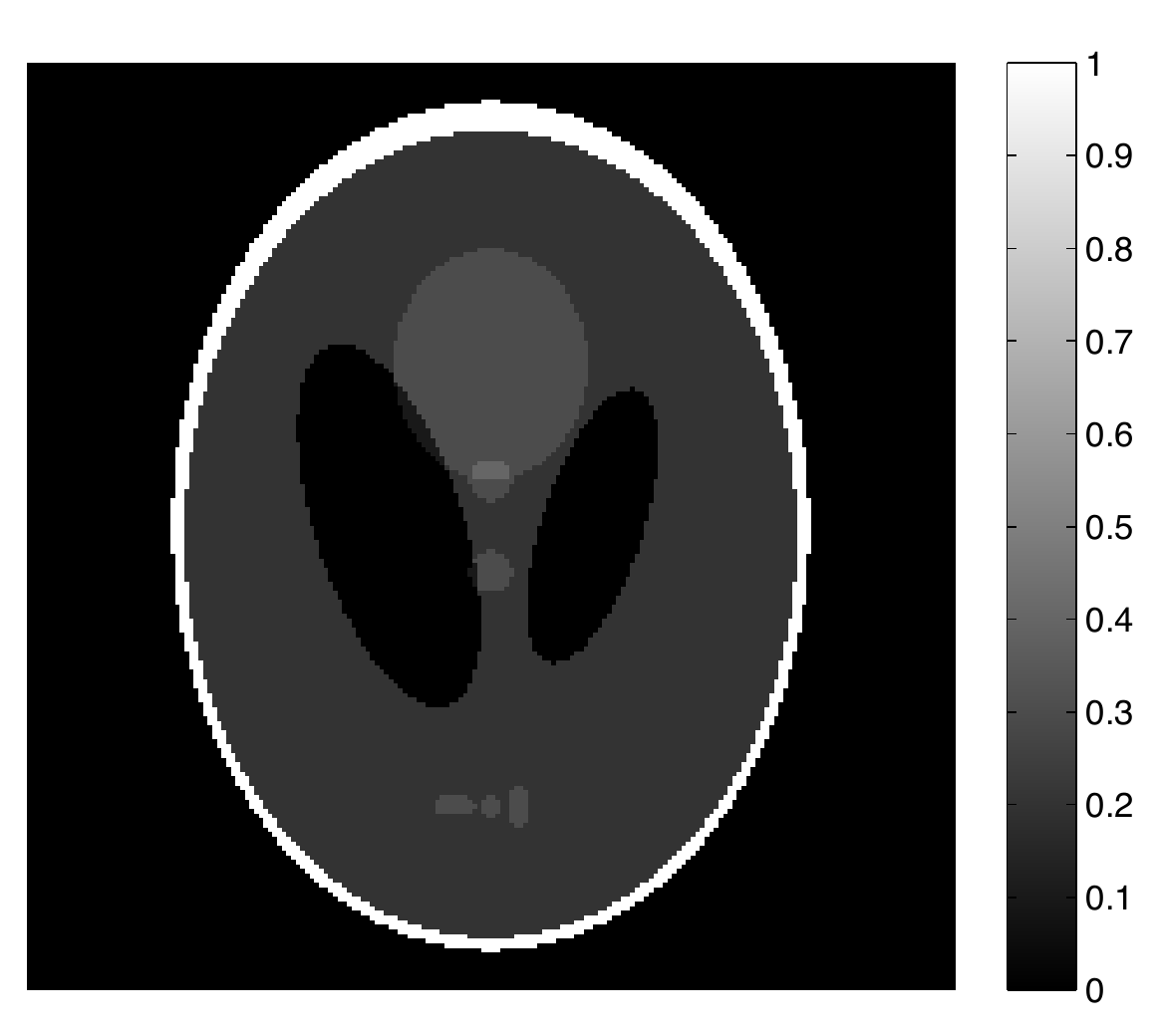}}
 \caption{Test images of loose support (a)(c) and  tight support (b)} 
 \label{TestImage}
\end{figure}

\begin{figure}[h]
\centering
         \subfigure[ HRM]{
         \includegraphics[width = 7cm]{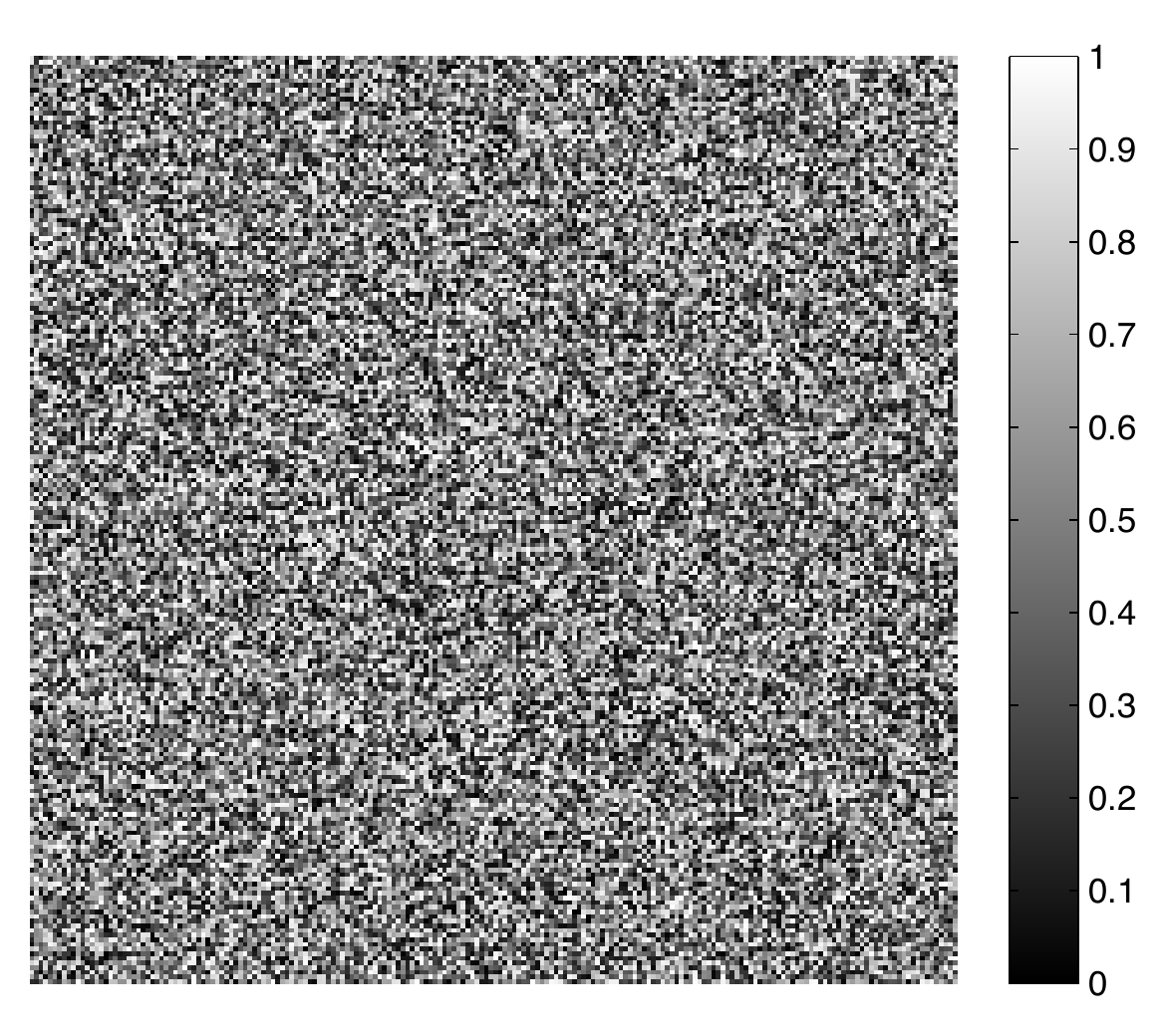}}
       \hspace{1cm}
                 \subfigure[ LRM]{
         \includegraphics[width = 7.2cm]{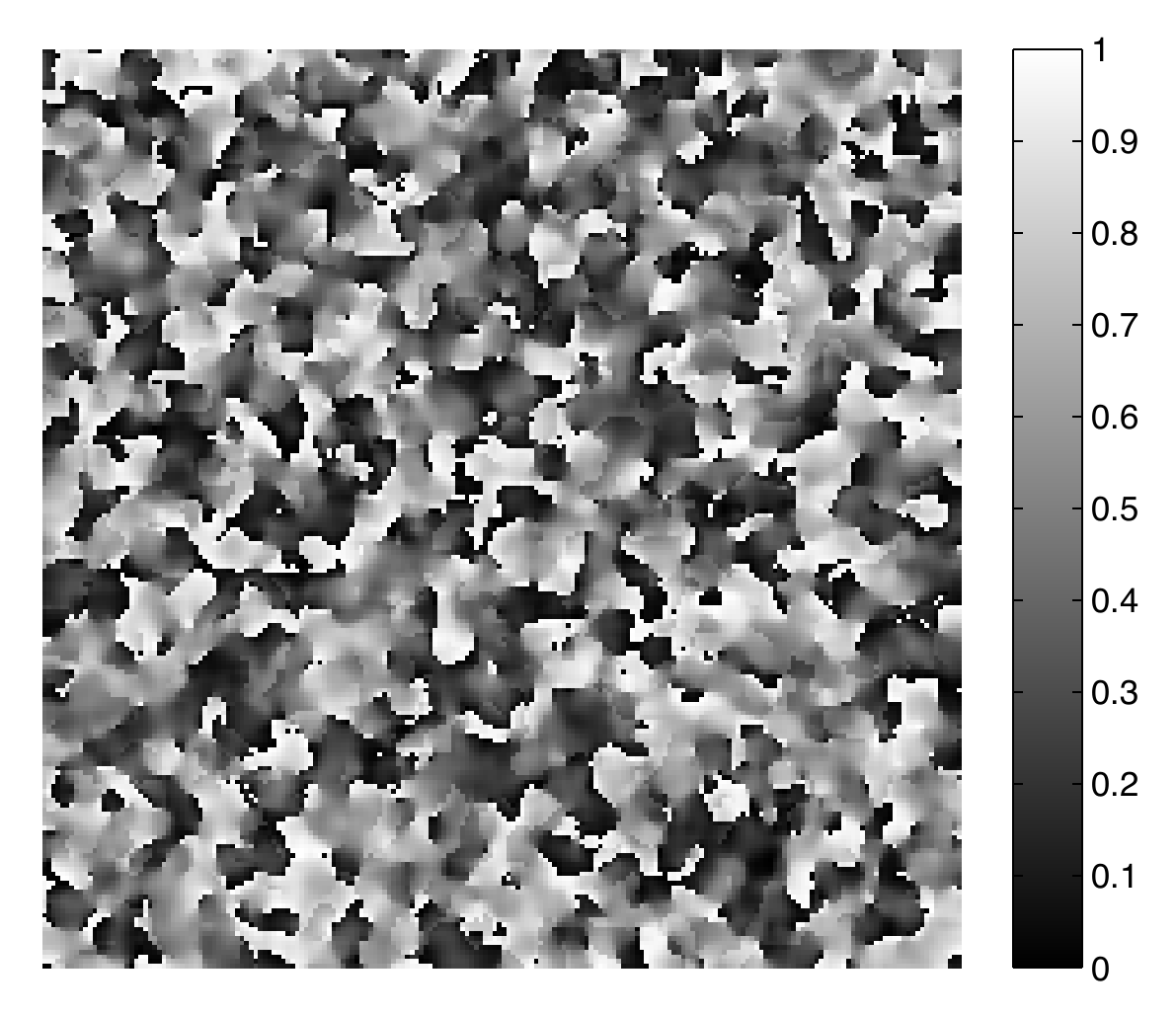}}
 \caption{(a) HRM and (b) LRM.  The gray scale
 represents the phase range $[0,1]$ in the unit of $2\pi$.} 
 \label{fig:mask}
\end{figure}

\begin{figure}[h]
\centering
      \subfigure[UM]{
         \includegraphics[width = 5.7cm]{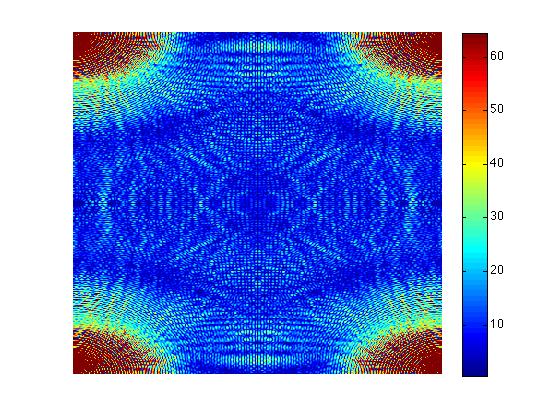}}\hspace{-.7cm}
                         \subfigure[ LRM]{
         \includegraphics[width = 5.7cm]{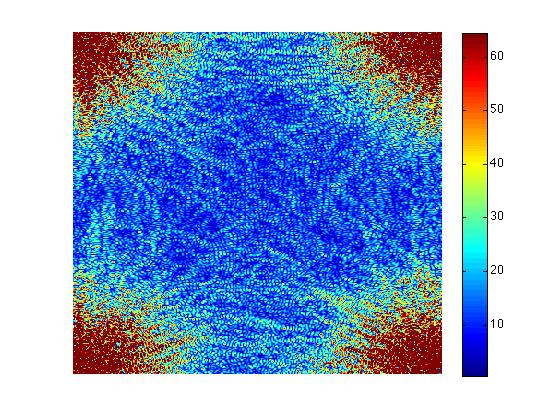}}\hspace{-0.7cm}
           \subfigure[ HRM]{
         \includegraphics[width = 5.7cm]{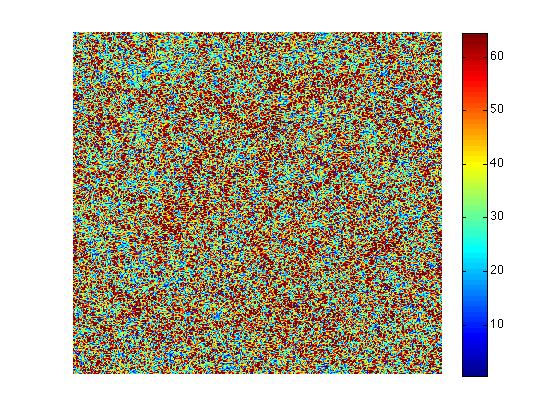}}
 \caption{The diffraction pattern (Fourier intensity) of the non-negative phantom with (a) UM (b) LRM (c) HRM.  } 
 \label{fig:pattern}
\end{figure}

In this section, we present the numerical results of phasing with a PUM .  

 The original  images are the $256\times 256$ cameraman, the $138 \times 184$ phantom
 and the $512\times 512$ mandrll (Fig.\ref{TestImage}(c)). 
We surround the first two images by dark (i.e. zero-valued) borders  to create  the
 $282\times 282$ cameraman and the $200 \times 200$ phantom 
  of {\em loose} supports (Fig. \ref{TestImage}(a)\&(c)). Objects 
of loose support are usually harder to recover than the same objects of {\em tight} support (Fig. \ref{TestImage}(b)). 

\subsection{High and low resolution masks}\label{sec:mask}
First we consider the case of the full mask range $\gamma=1$. 
Let $\{\tphi(\bn)\}$  and $\{\psi(\bn)\}$ be two independent arrays of
 independent uniform random variables over $[-\pi,\pi)$. 
Define the true mask phases $\phi(\bn)=\tphi(\bn)+\delta \psi(\bn),\delta>0$. We refer to the corresponding
mask $\mu=\exp{[i\phi}]$ as a full-ranged,  fine-grained or  {\em high resolution}  mask (HRM), Fig. \ref{fig:mask}(a).  

To demonstrate that the random mask approach is stable with respect to the correlation length of  the mask, we define a full-ranged, coarse-grained or  {\em low resolution}   mask  (LRM) 
as follows. 

Let $\{\tilde\phi_0(\bn)\}$ and $\{\tilde\psi(\bn)\}$ be two other 
independent arrays of independent uniform random variables over $[-\pi,\pi)$.
Convolving $\exp{(i\tilde\phi_0)}$ with the kernel function
\[
g_c({\mb x}) =\left\{ \begin{matrix}
\exp{[-c^2 / (c^2 - |{\mb x}|^2)]}, & |{\mb x}|\leq c\\
  0,&\hbox{else} 
         \end{matrix}\right.
         \]
 with $c=5$ and  normalizing the outcome to have unit modulus we obtain the LRM estimate, still denoted by  $
 \mu_0=\exp{[i\phi_0]}$. Repeating  the same procedure with $\exp{[i\tilde\psi]}$ we obtain $\exp{[i\psi]}$. We then set  the
LRM $\mu=\exp{(i\phi)}$ with phase  $\phi = \tphi+\delta \psi$ (Fig. \ref{fig:mask}(b)). The resulting LRM phases 
and their estimates are uniform random variables over $[-\pi,\pi)$ with a correlation length of about 10 pixel sizes and hence have much lower ($100$ times less) degrees of diversity than HRM. 
Consequently HRM tends to yield a better perform in recovery
 than LRM (cf. Fig. \ref{FigureErrVersusNoise}). 
 
When a second set of Fourier data is used (for complex-valued objects), the data are synthesized  with a UM (i.e. $\mu^{(2)}=1$). 
 
 The diffraction patterns of the non-negative phantom with UM, LRM and HRM are shown
 in Fig. \ref{fig:pattern}. Clearly, the diffraction pattern sensitively depends on how random
 the mask is. 

\commentout{
A standard way to utilize the oversampled data  is to enlarge
the original image by adding corresponding number of zero pixels which is then enforced as an additional object constraint. 
This procedure  is called the oversampling method \cite{Miao00} and implemented in all our simulations
with the oversampling ratio $|\cL|/|\cN|=4$ (for $d=2$). 
}


\subsection{Error and Residual}
We use the relative error and residual as figures of merit. 
Let $\hat{f}$ and $\hat{\mu}$ be the recovered image and mask respectively. The relative error of object reconstruction is defined as  $$e(\hat{f}) = 
\left\{ 
\begin{array}{ll}
{\|f-\hat{f}\|}/{\|f\|} & \text{ if absolute uniqueness holds} \\
{\displaystyle \min_{\alpha\in [0,2\pi)}\|f-\exp{(i\alpha)}\hat{f}\|}/{\|f\|} & \text{ if uniqueness holds only up to a global phase}
\end{array}. \right.
$$

Let $\hat{\Lambda}$ be the diagonal matrix whose diagonal elements are $\hat{\mu}(\bn)$. The relative residual is defined as
\[
\rho(\hat{f},\hat{\mu}) = \frac{\| \ Y - |\bPhi \hat{\Lambda} \PO\hat{f}| \ \| }{\| Y\|}
\]
where $\PO$ is introduced to enforce the object constraints 
 in the case of DRER.

\subsection{Non-negative images}
\begin{figure}[hthp]
\centering
         \subfigure[$e(\hat f) \approx 36.56\%$]{
         \includegraphics[width = 4cm]{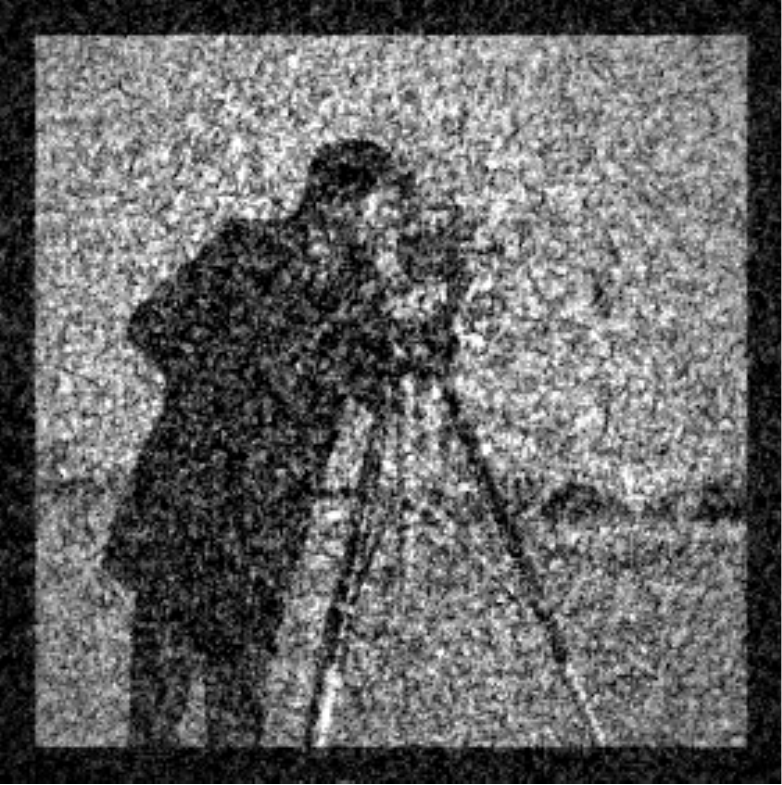}}    
        \subfigure[]{
         \includegraphics[width = 4cm]{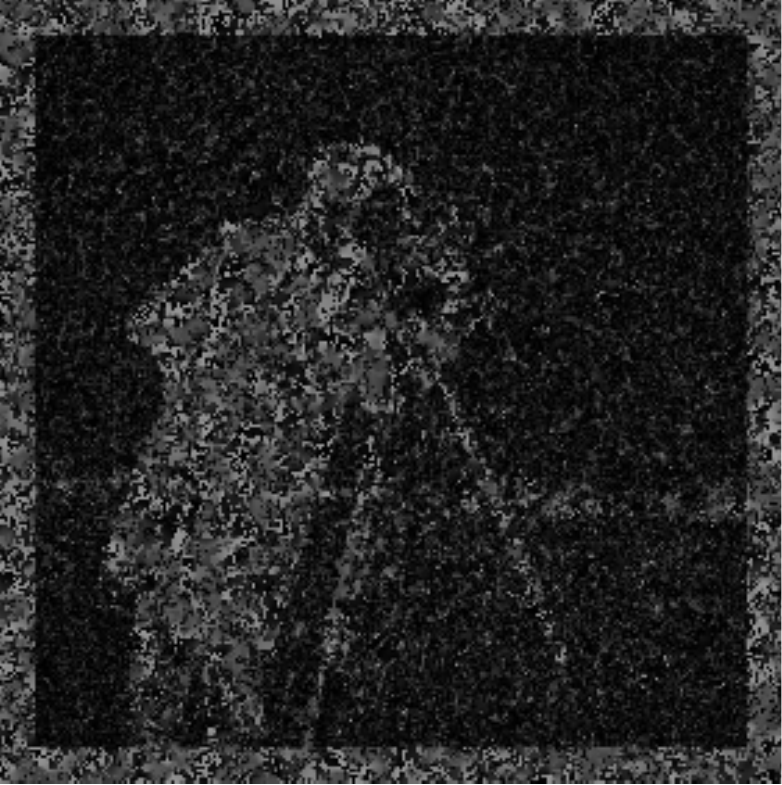}}
          \subfigure[$r(\hat f,\hat \mu) \approx 6.84\%$]{
         \includegraphics[width = 5.3cm]{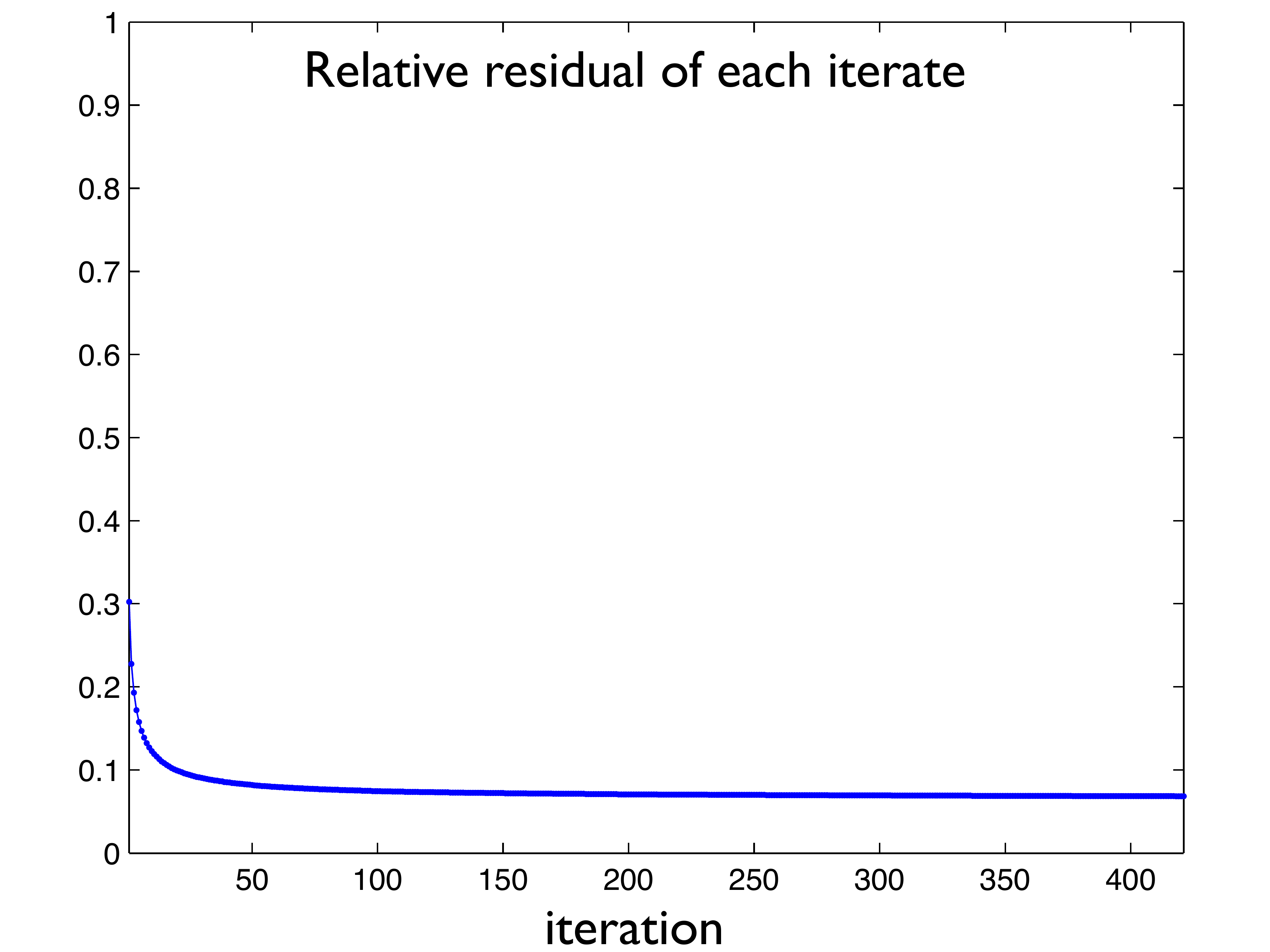}}
                   \subfigure[$e(\hat f) \approx 44.28\%$]{
         \includegraphics[width = 4cm]{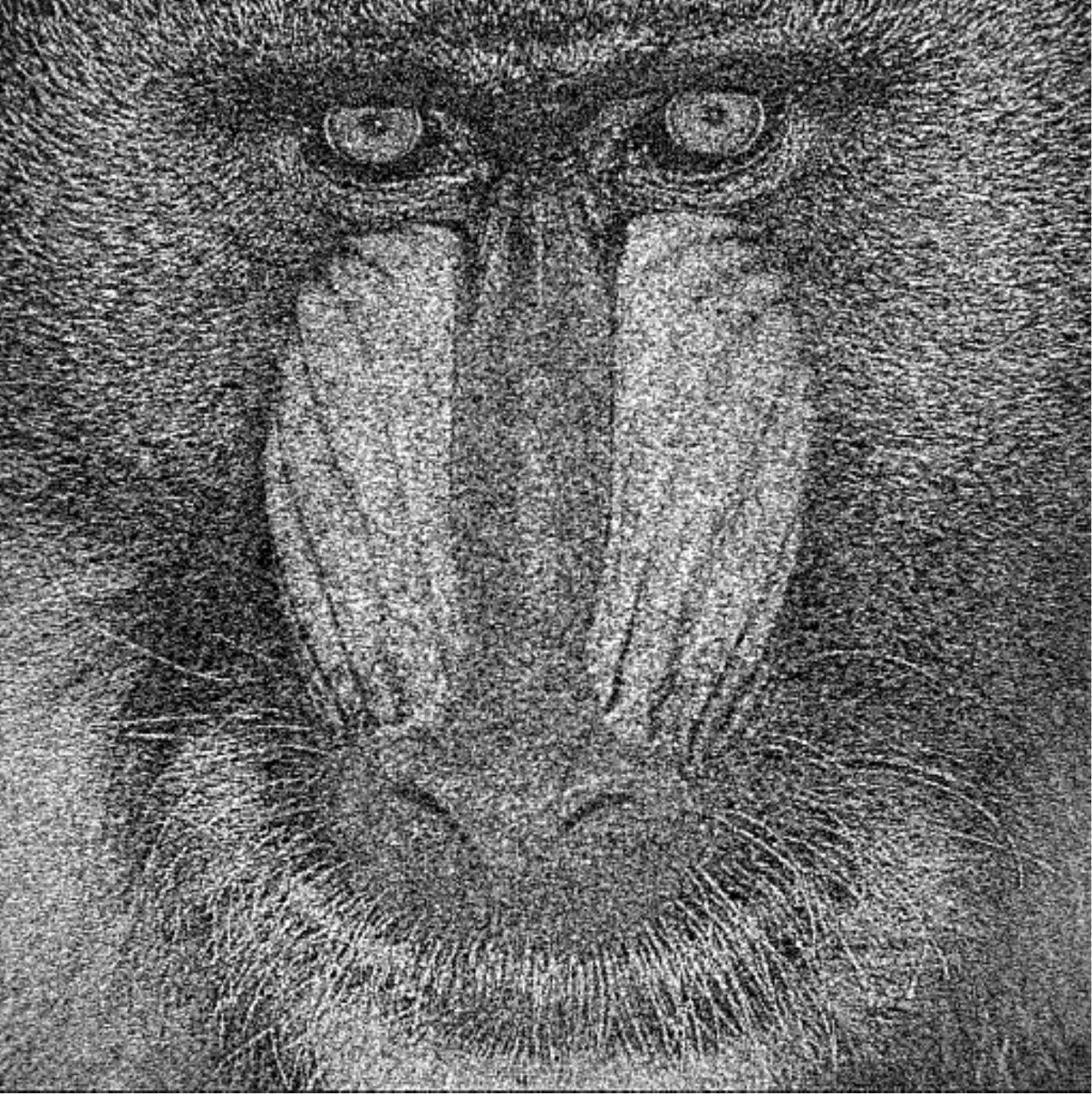}}        
         \subfigure[]{
        \includegraphics[width = 4cm]{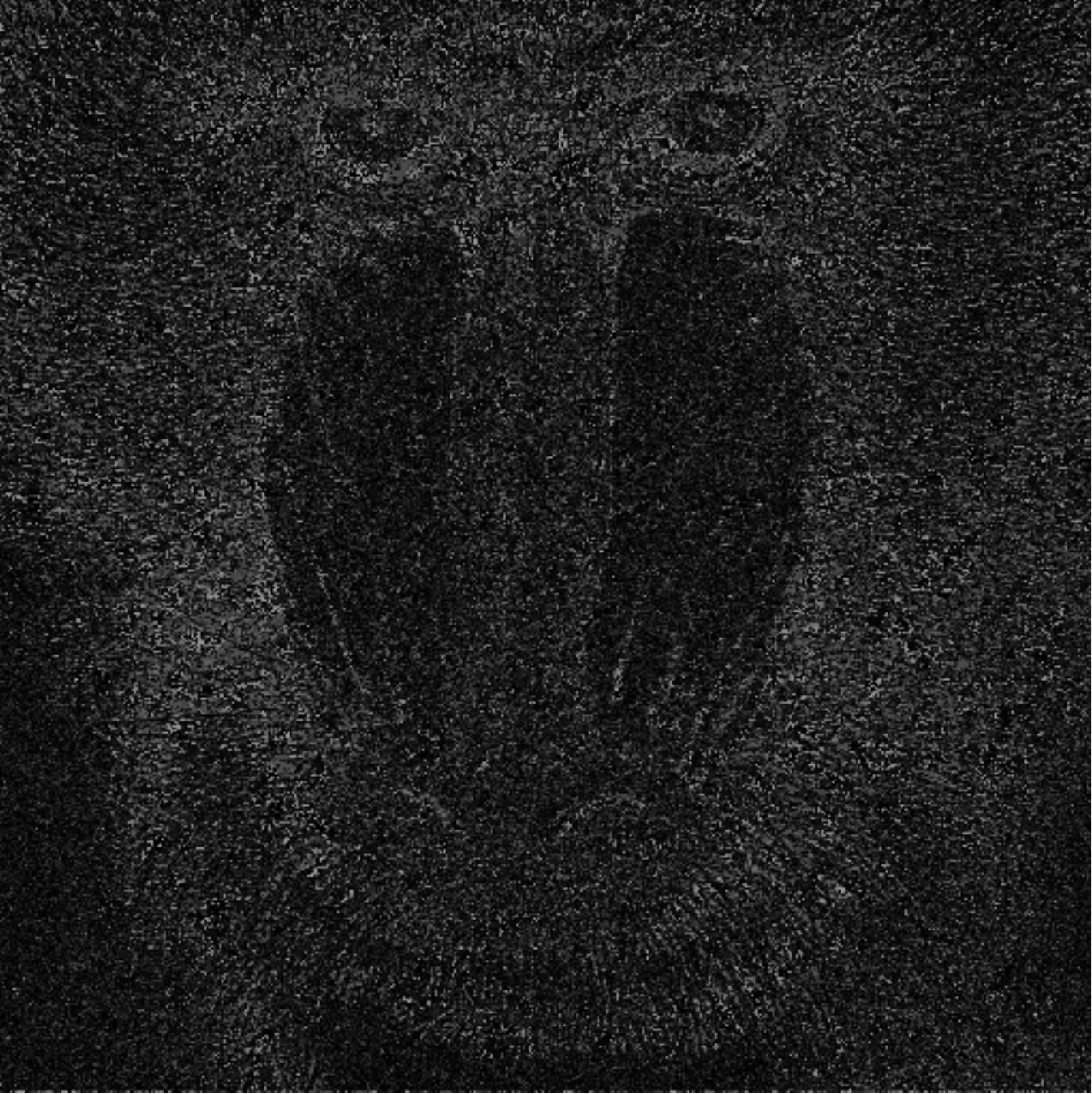}}
          \subfigure[$r(\hat f,\hat \mu) \approx 7.87\%$]{
        \includegraphics[width = 5.3cm]{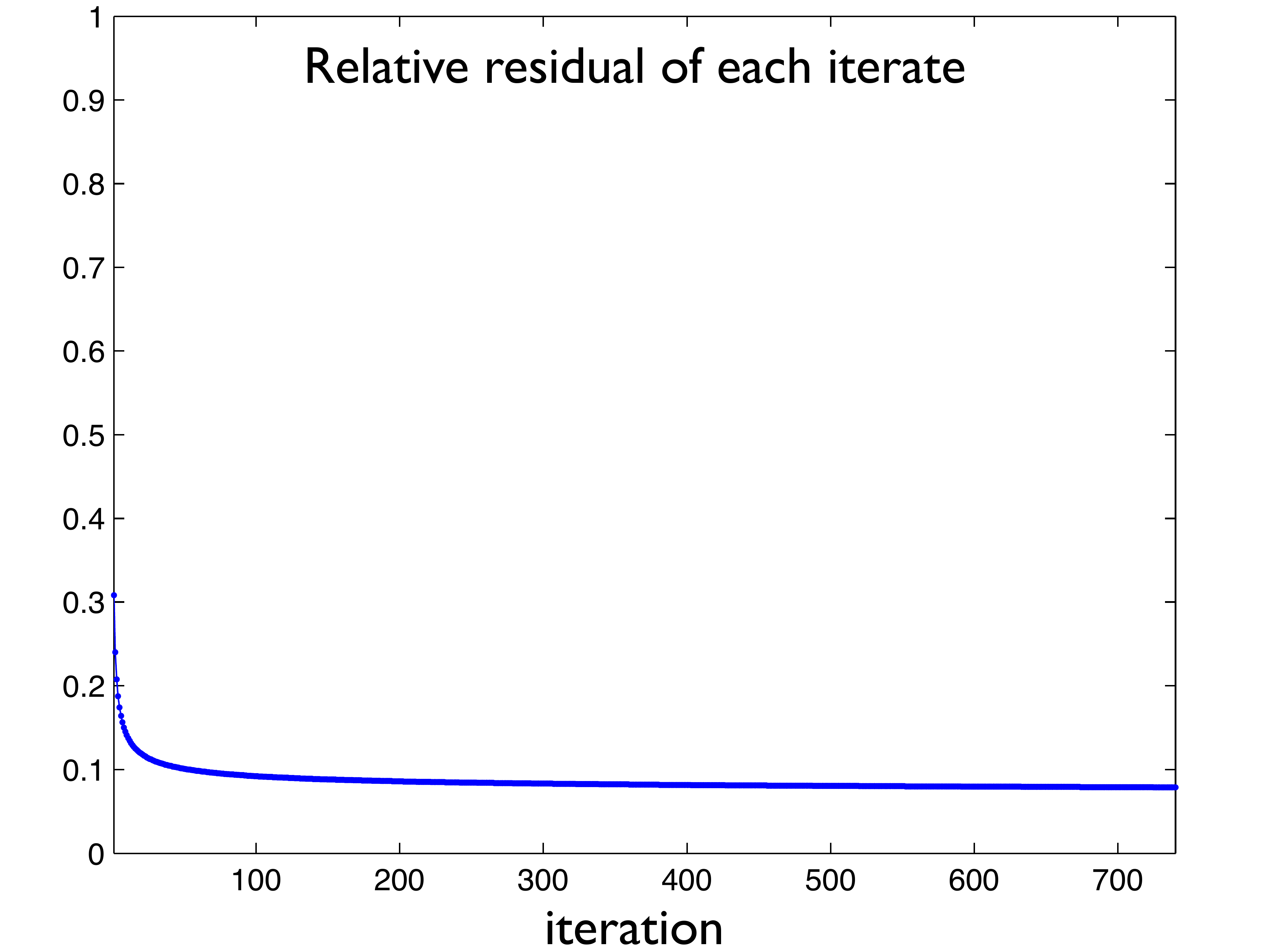}}
                         \subfigure[$e(\hat f) \approx 59.50\%$]{
         \includegraphics[width = 4cm]{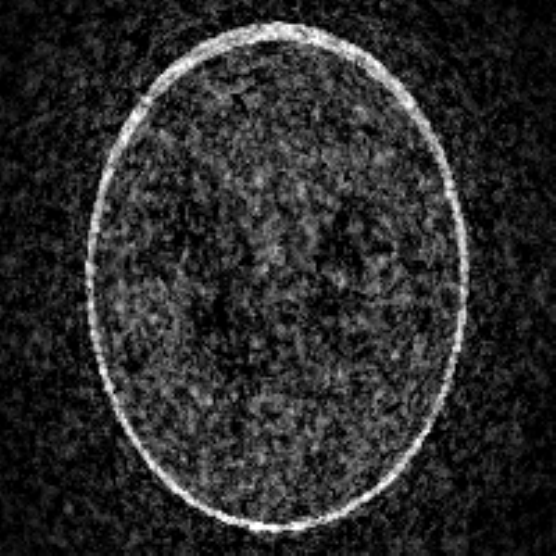}}     
         \subfigure[]{
        \includegraphics[width = 4cm]{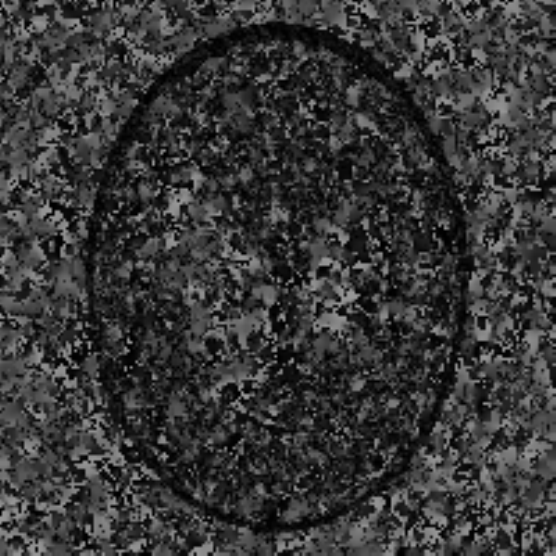}}
          \subfigure[$r(\hat f,\hat \mu) \approx 10.74\%$]{
         \includegraphics[width = 5.3cm]{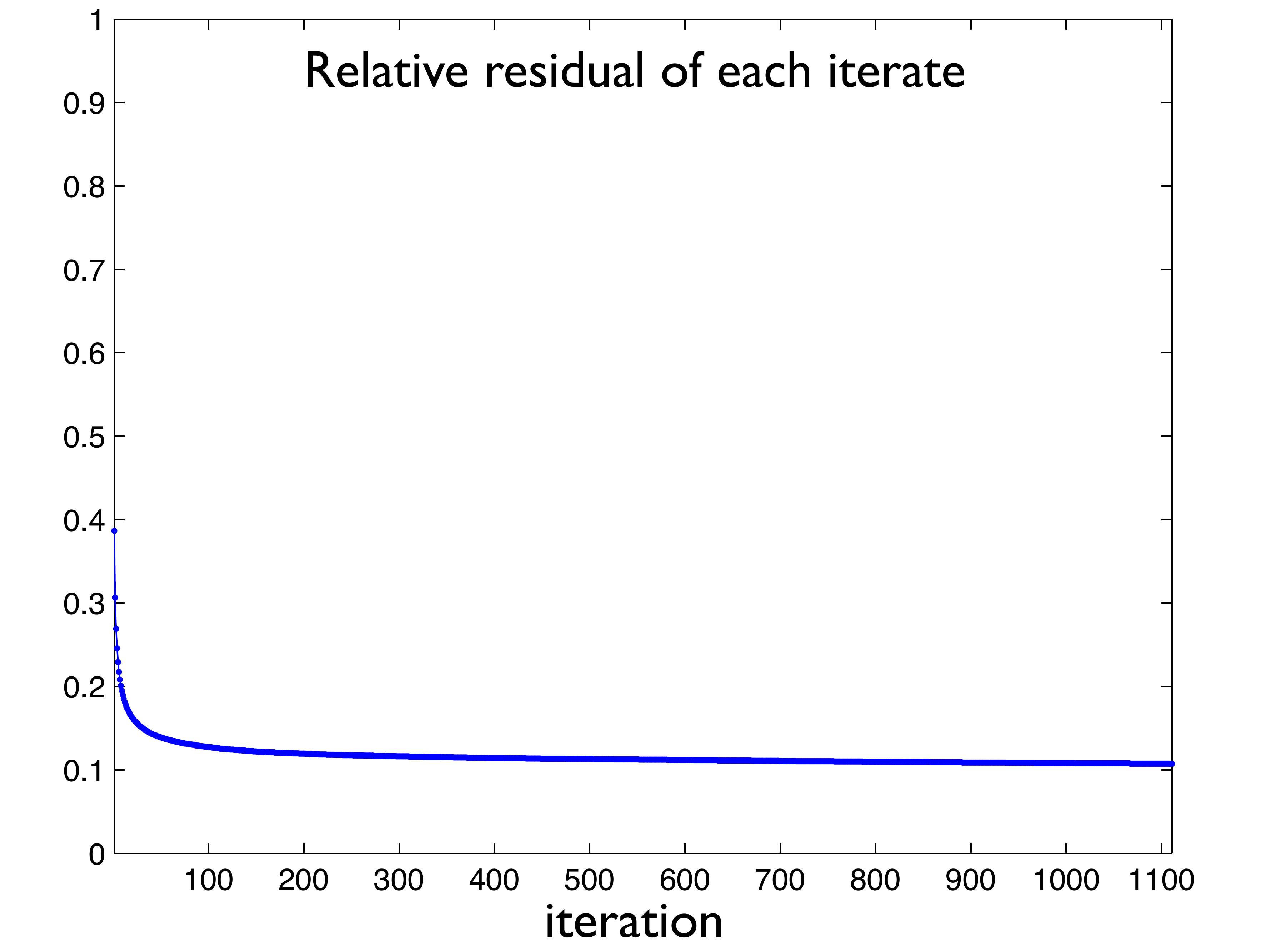}}

             \caption{Recovery of non-negative images by AER with one LRM of $\delta=0.3$.
                The middle column shows the absolute phase differences between $\mu$ and $\hat\mu$.            The right column shows the  relative residual at each iteration.  
             }
             \label{fig:aer}
\end{figure}

First we use AER (\ref{aer})  to recover the non-negative images with the stopping rule
$\|f_{k+1}-f_k\|/\|f_k\| < 0.05\%$ and one LRM of uncertainty $\delta=0.3$. 
The results, shown in Fig. \ref{fig:aer}, are noisy and inaccurate with 36.56\% error for
the cameraman, 44.28\% for the mandrill and 59.50\% error for the phantom. Consistent with the residual reduction property (Theorem \ref{Thm:er}), the residual curves in Fig. \ref{fig:aer} are monotonically decreasing.

\begin{figure}[hthp]
\centering
         \subfigure[$e(\hat{f}) \approx 1.26\%$]{
         \includegraphics[width = 4cm]{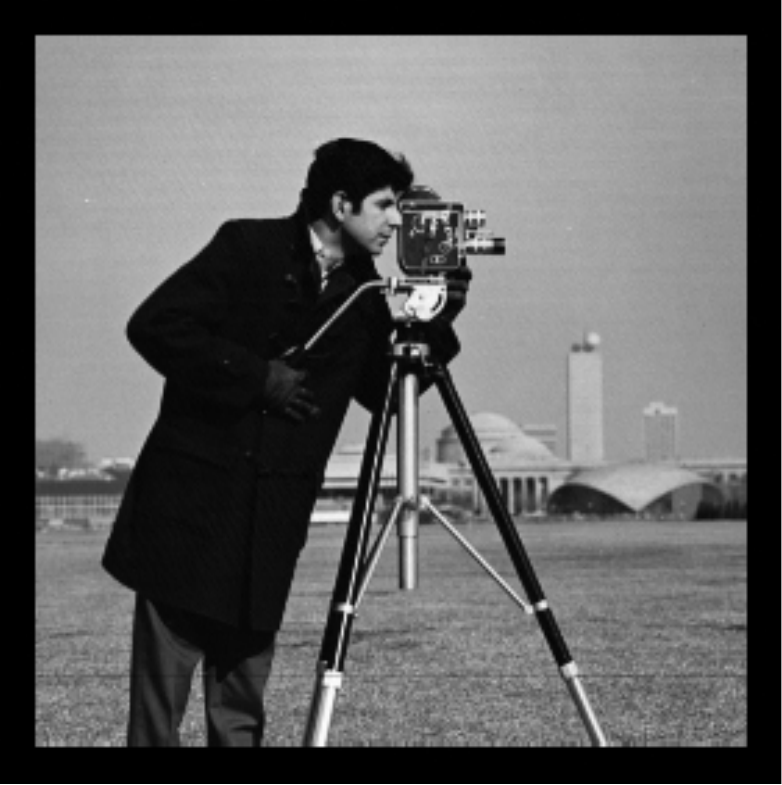}}
                 \subfigure[]{
         \includegraphics[width = 4cm]{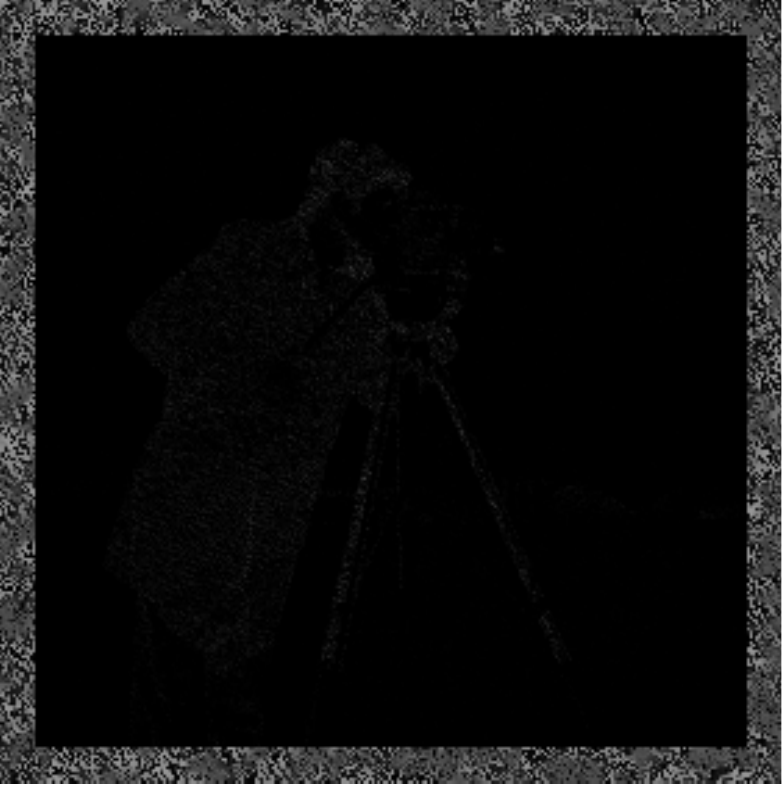}}
         \subfigure[$\rho(\hat{f},\hat{\mu}) \approx 0.25\%$]{
         \includegraphics[width = 5.3cm]{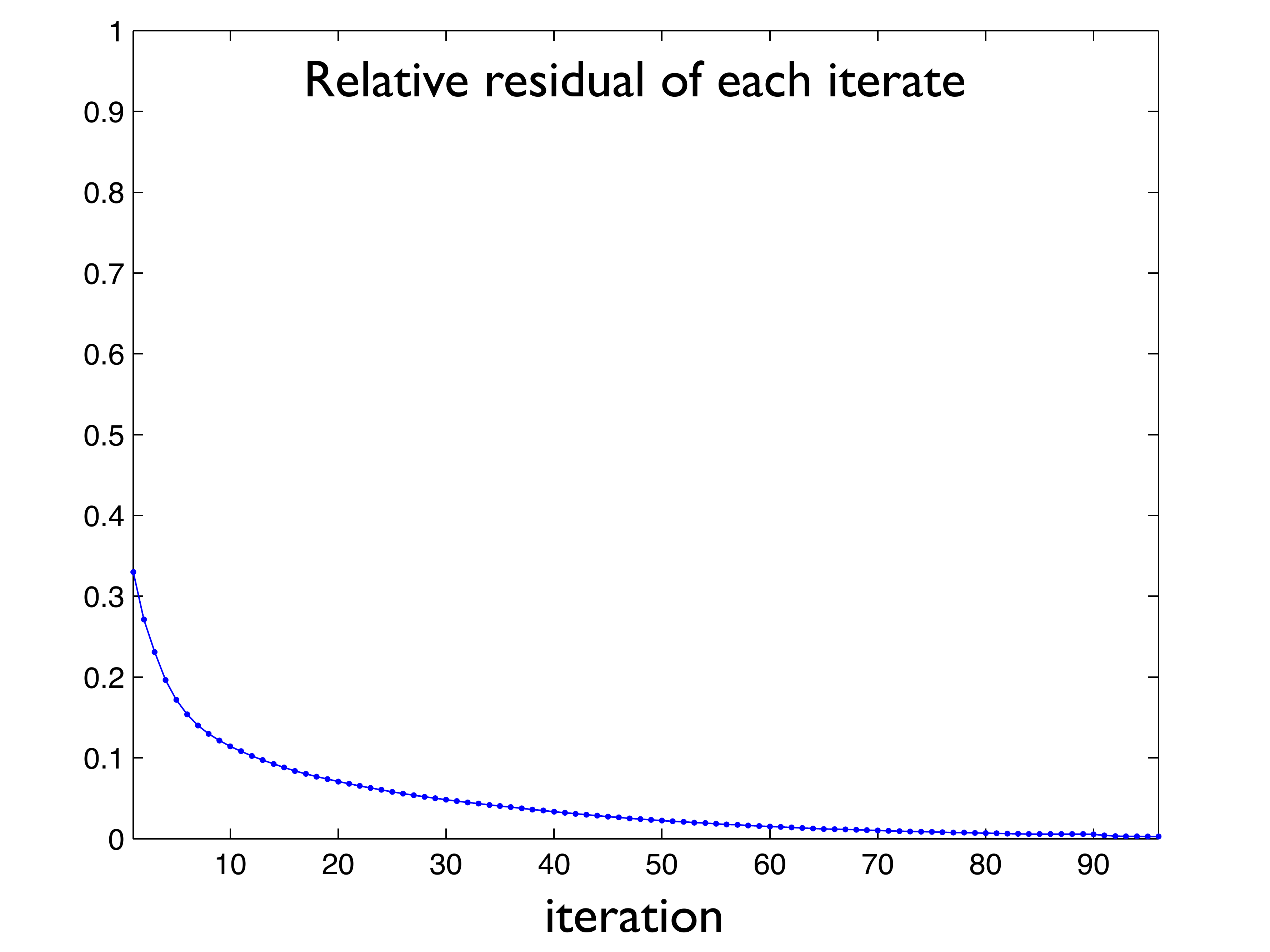}}
         \subfigure[$e(\hat{f}) \approx 0.96\%$]{
         \includegraphics[width = 4cm]{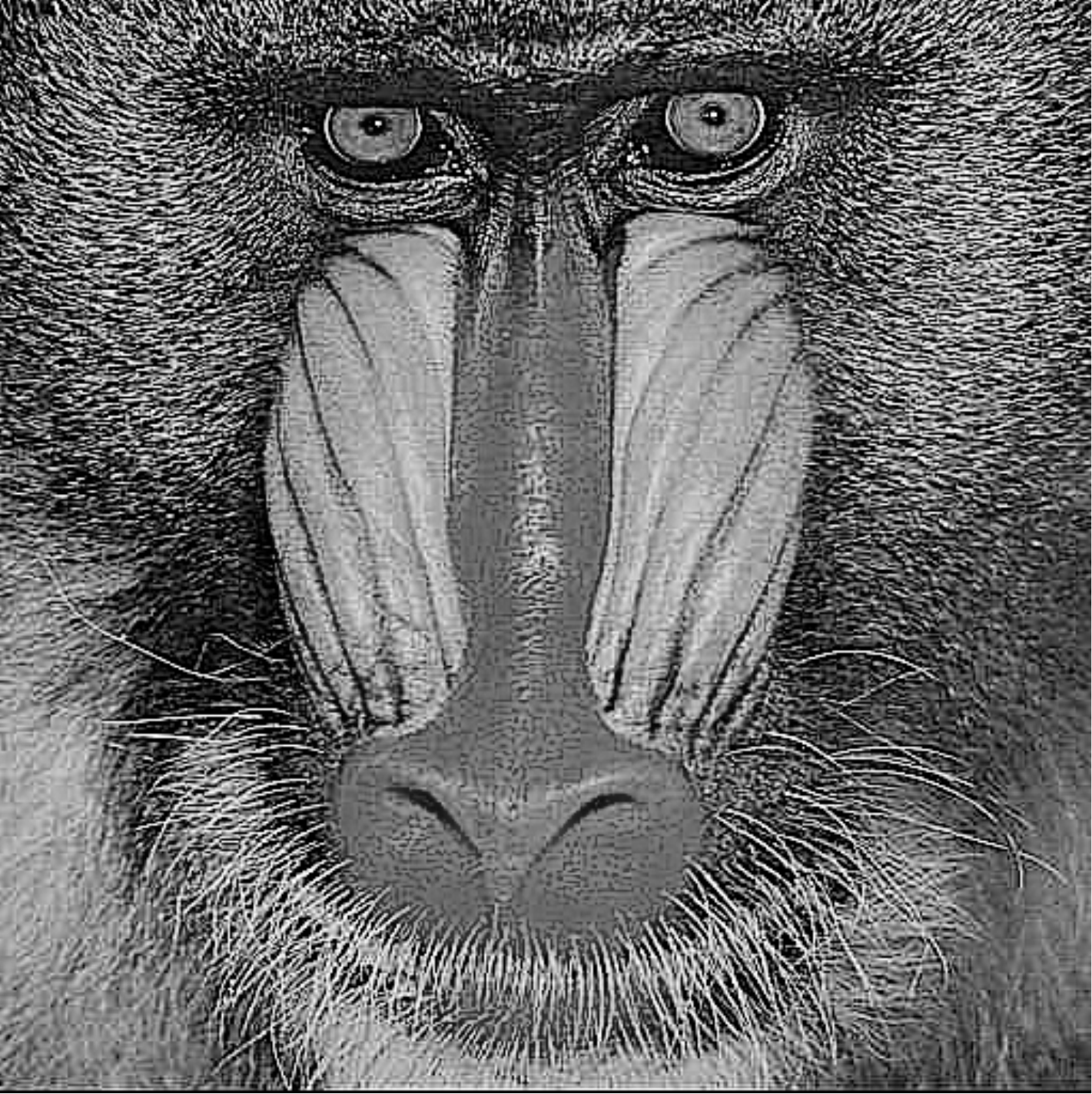}}                \subfigure[]{
         \includegraphics[width = 4cm]{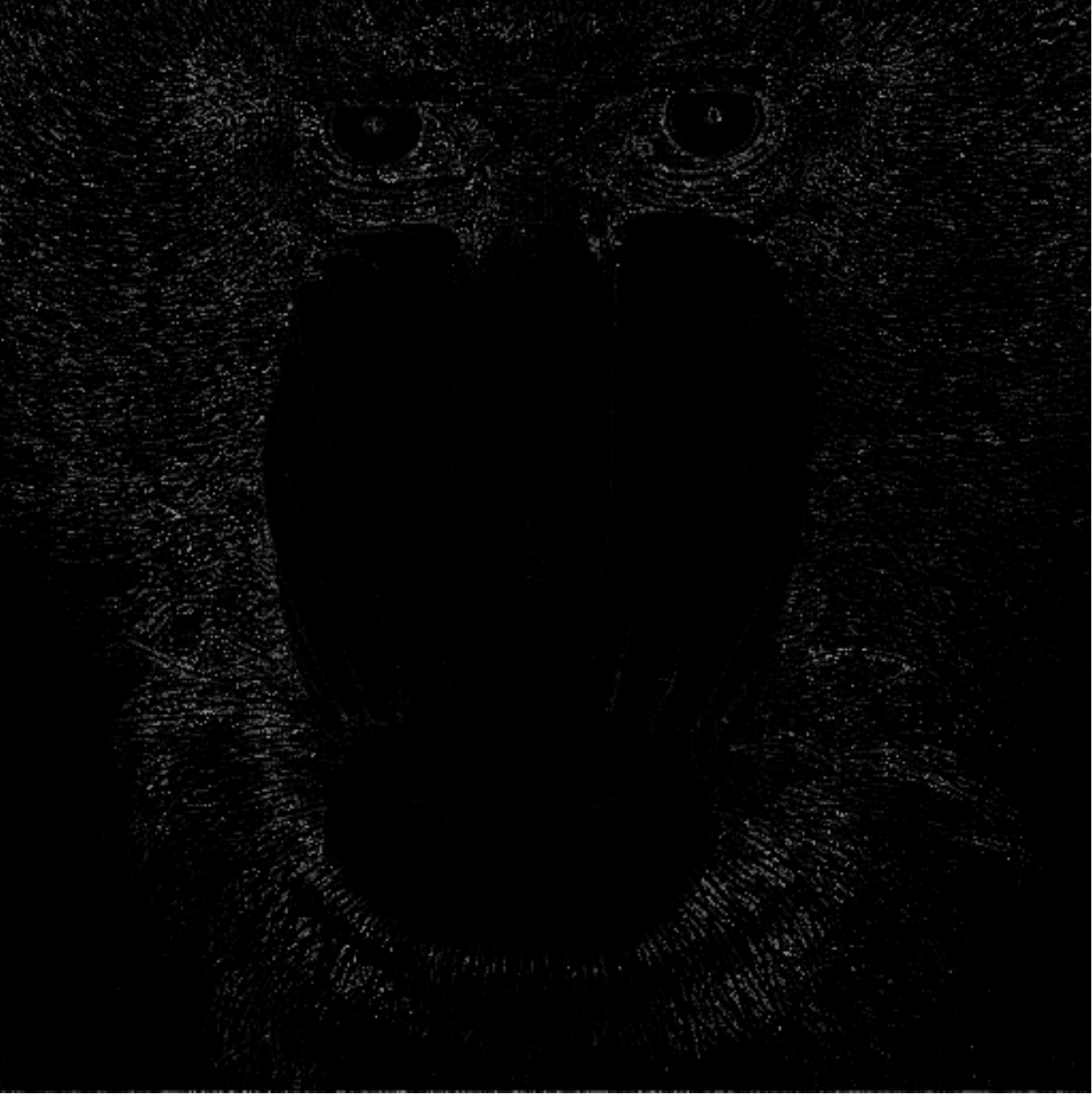}}
         \subfigure[$\rho(\hat{f},\hat{\mu}) \approx 0.23\%$]{
         \includegraphics[width = 5.3cm]{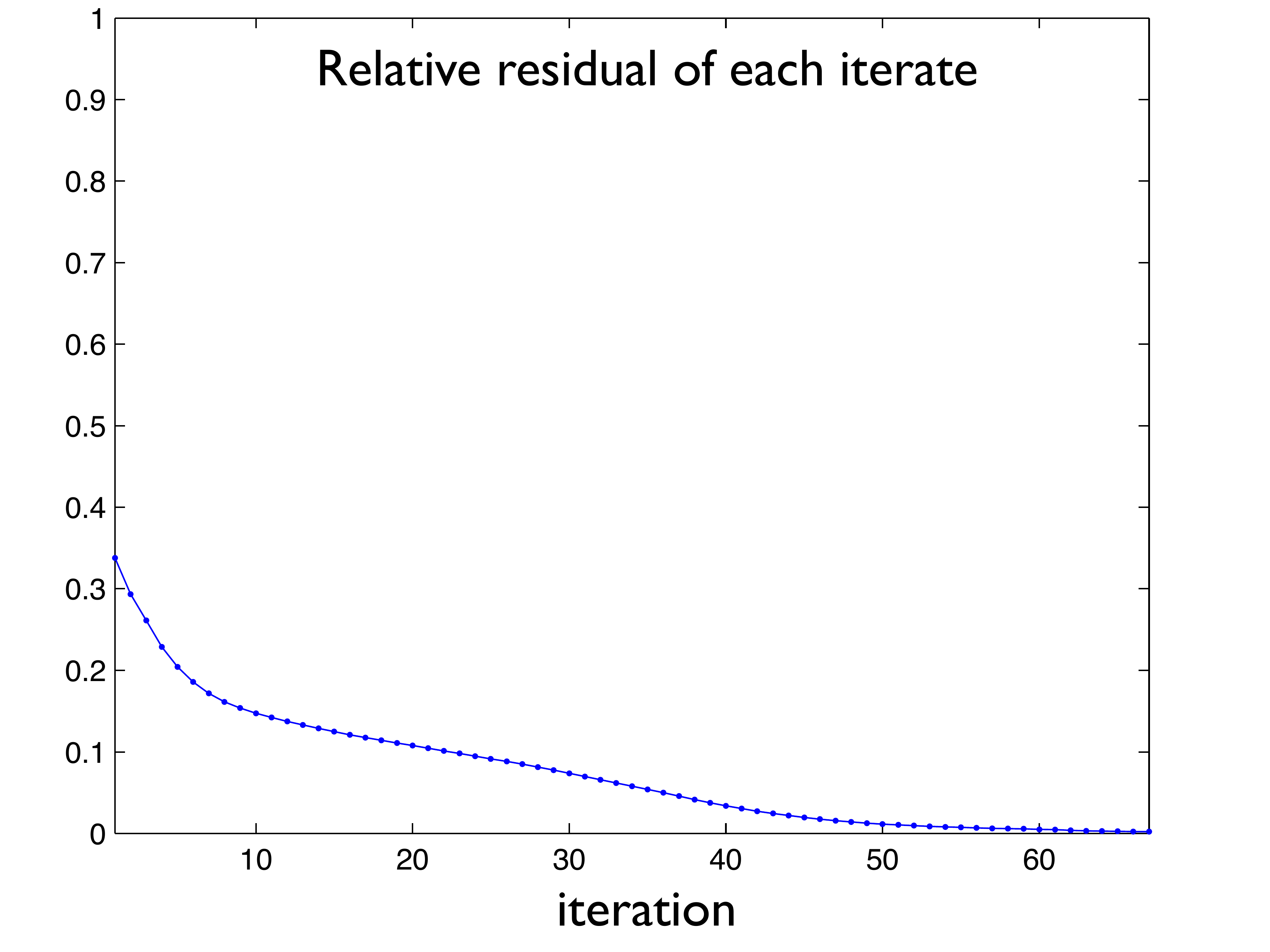}}
          \subfigure[$e(\hat{f}) \approx 0.37\%$]{
         \includegraphics[width = 4cm]{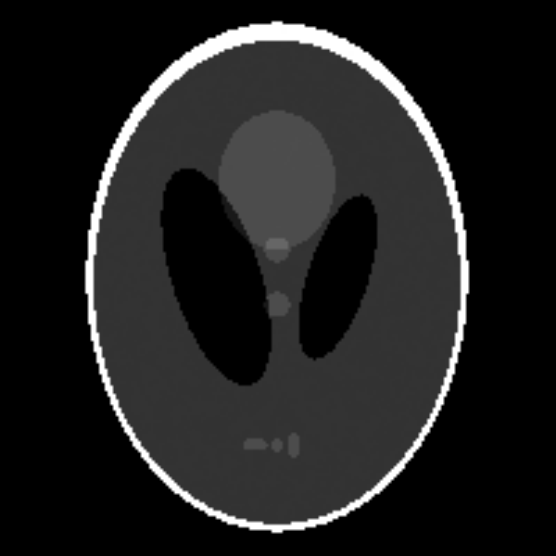}}
                 \subfigure[]{
         \includegraphics[width = 4cm]{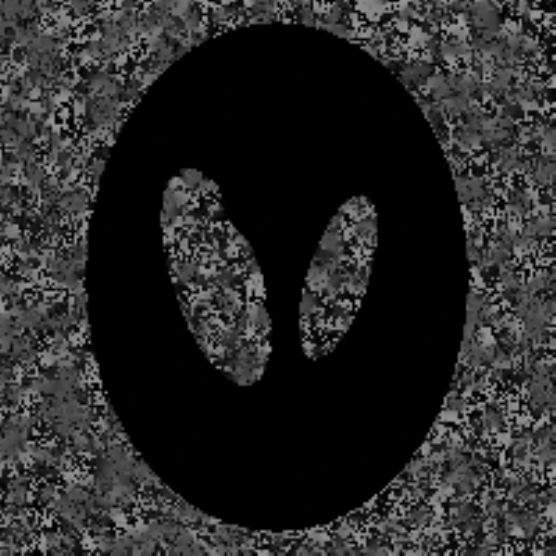}}
         \subfigure[$\rho(\hat{f},\hat{\mu}) \approx 0.12\%$]{
         \includegraphics[width = 5.3cm]{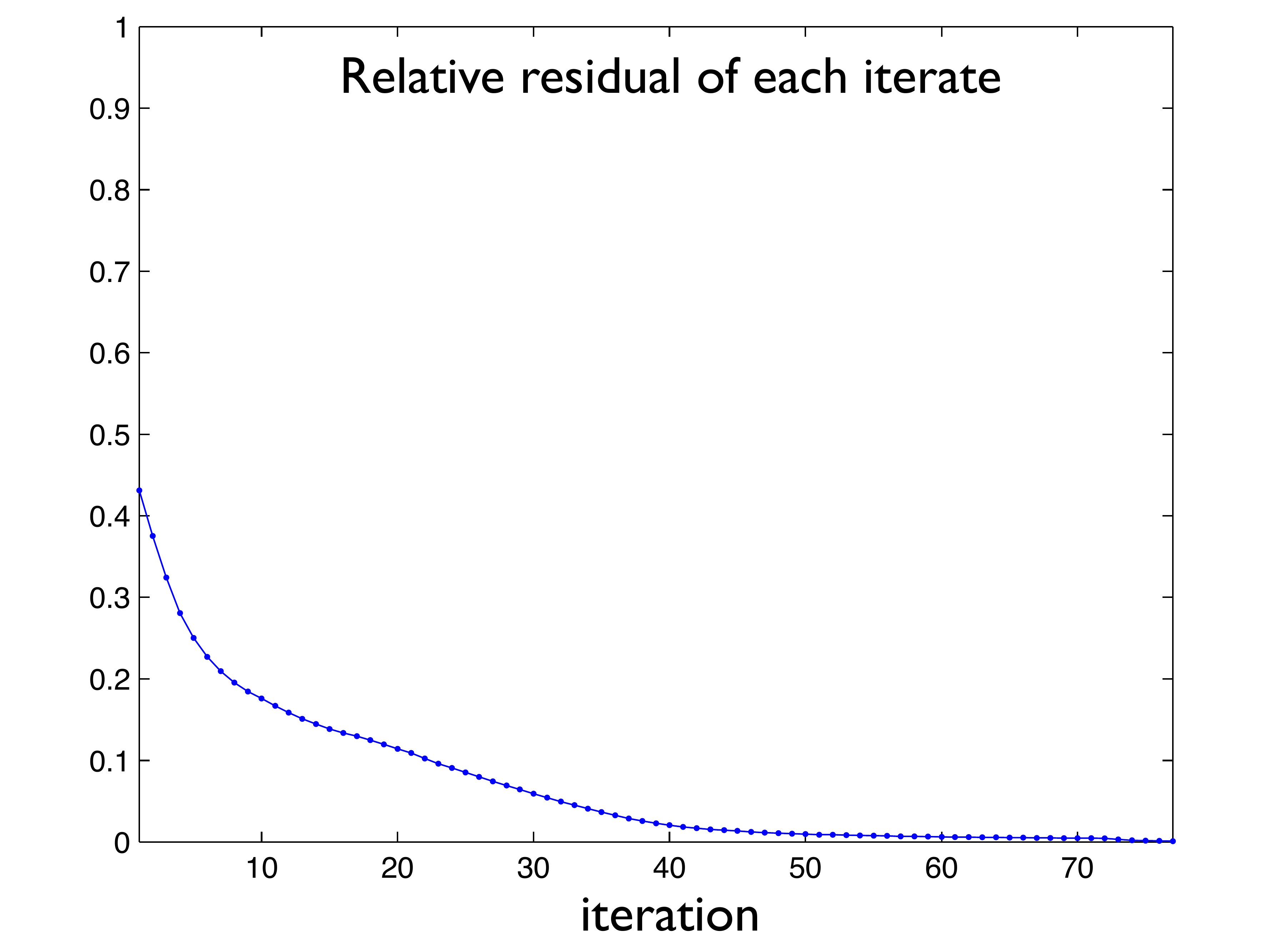}}
                  \caption{Recovery of non-negative images  with one  LRM of $\delta = 0.3$. 
         (a) the recovered cameraman $\hat f$    by $90$ DRER $+$ $6$ AER steps.        
           (d) the recovered mandrill  $\hat f$    by $61$ DRER $+$ $6$ AER steps.
                      (g) the recovered phantom  $\hat f$    by $72$ DRER $+$ $5$ AER steps.
          The middle column shows the absolute phase differences between $\mu$ and $\hat\mu$. 
         The right column shows the relative residual at each iteration. 
         } 
          \label{FigureRealPositive}
\end{figure}

\begin{figure}[hthp]
\centering
         \subfigure[$e(\hat{f}) \approx 6.43\%$]{
         \includegraphics[width = 4cm]{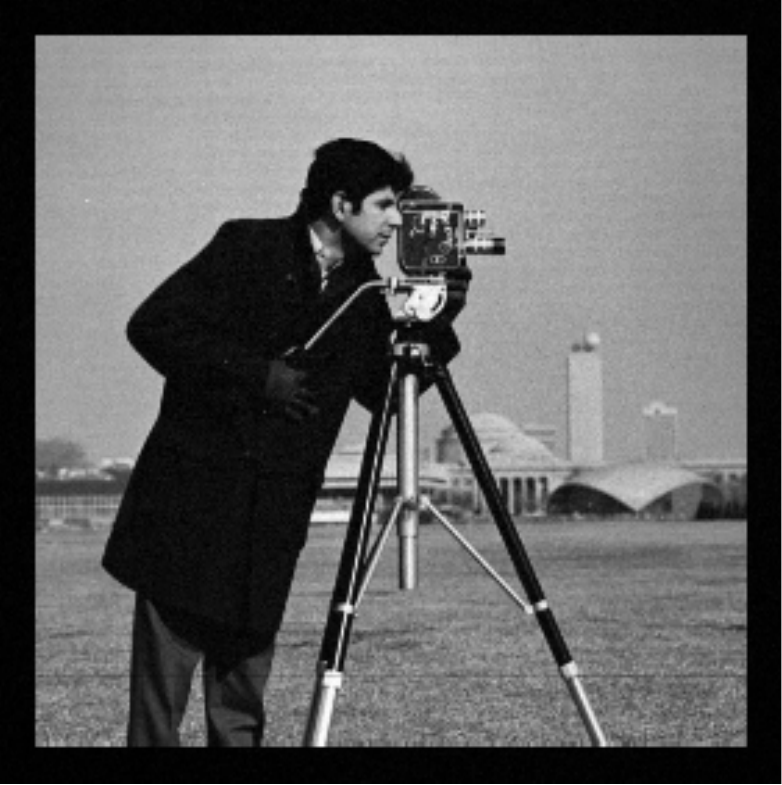}}     
         \subfigure[]{
         \includegraphics[width = 4cm]{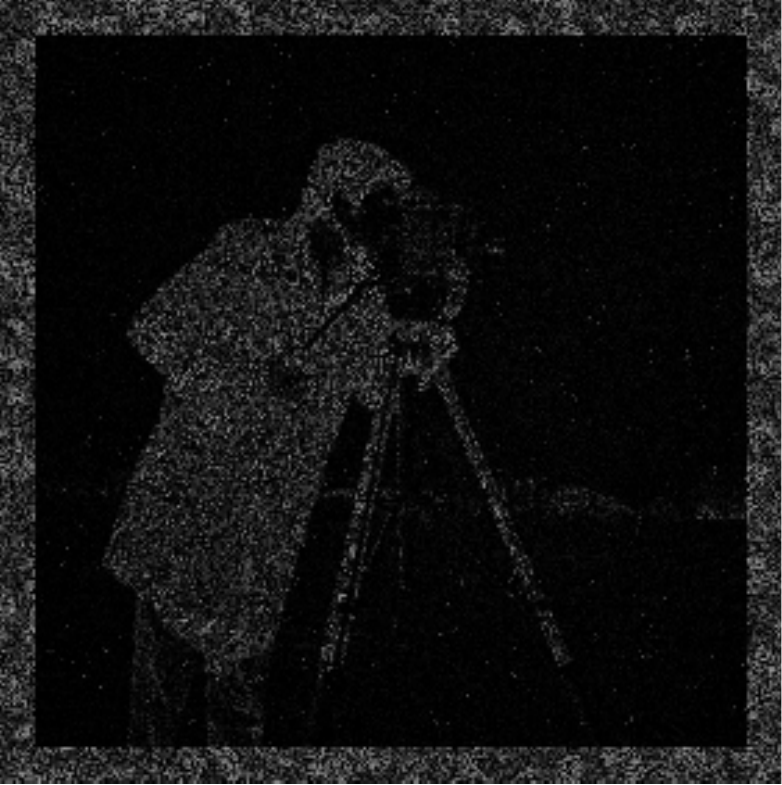}}
          \subfigure[$\rho(\hat{f},\hat{\mu}) \approx 2.66\%$]{
         \includegraphics[width = 5.3cm]{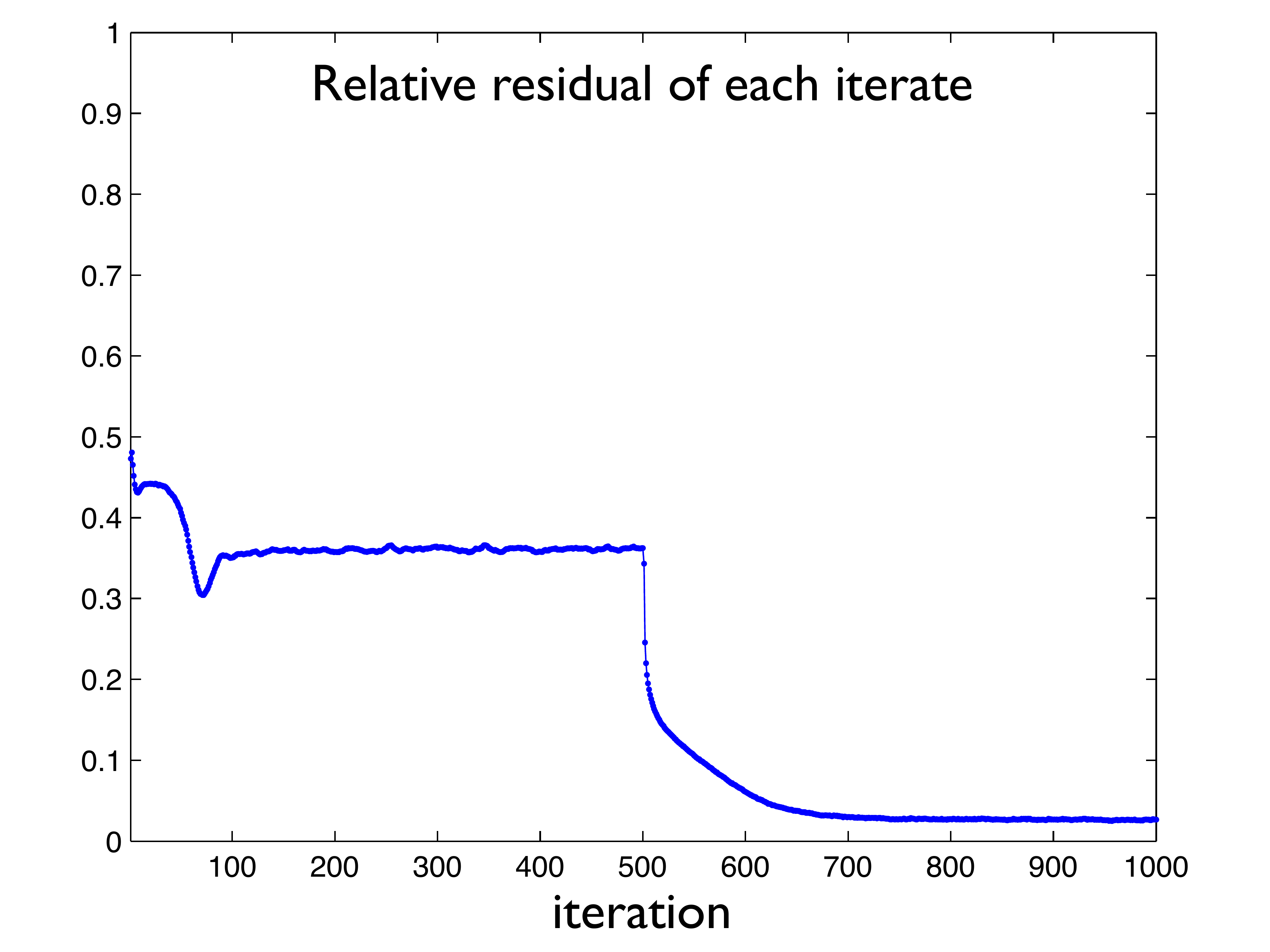}}
            \subfigure[$e(\hat{f}) \approx 4.62\%$]{
         \includegraphics[width = 4cm]{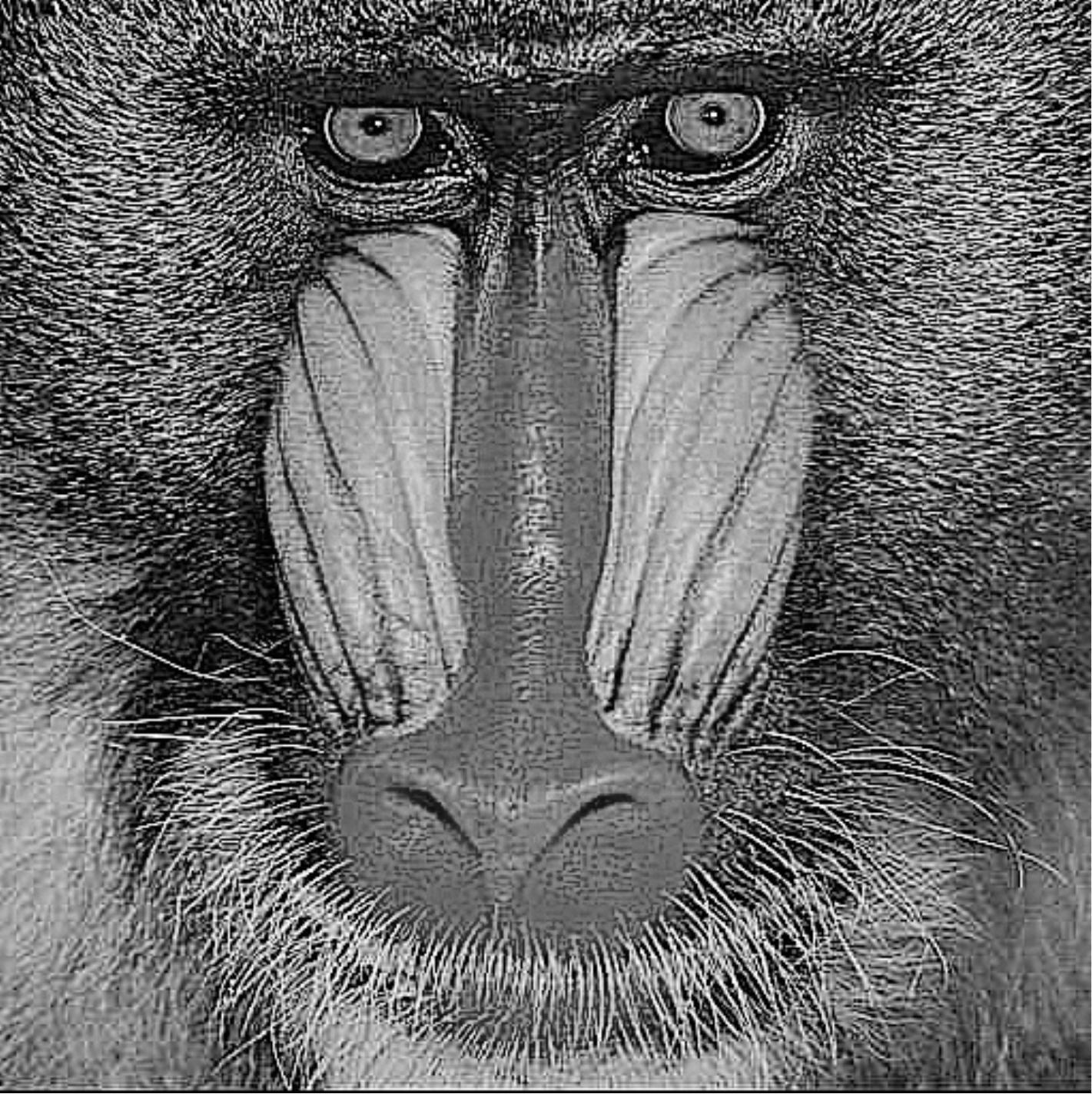}}
                 \subfigure[]{
         \includegraphics[width = 4cm]{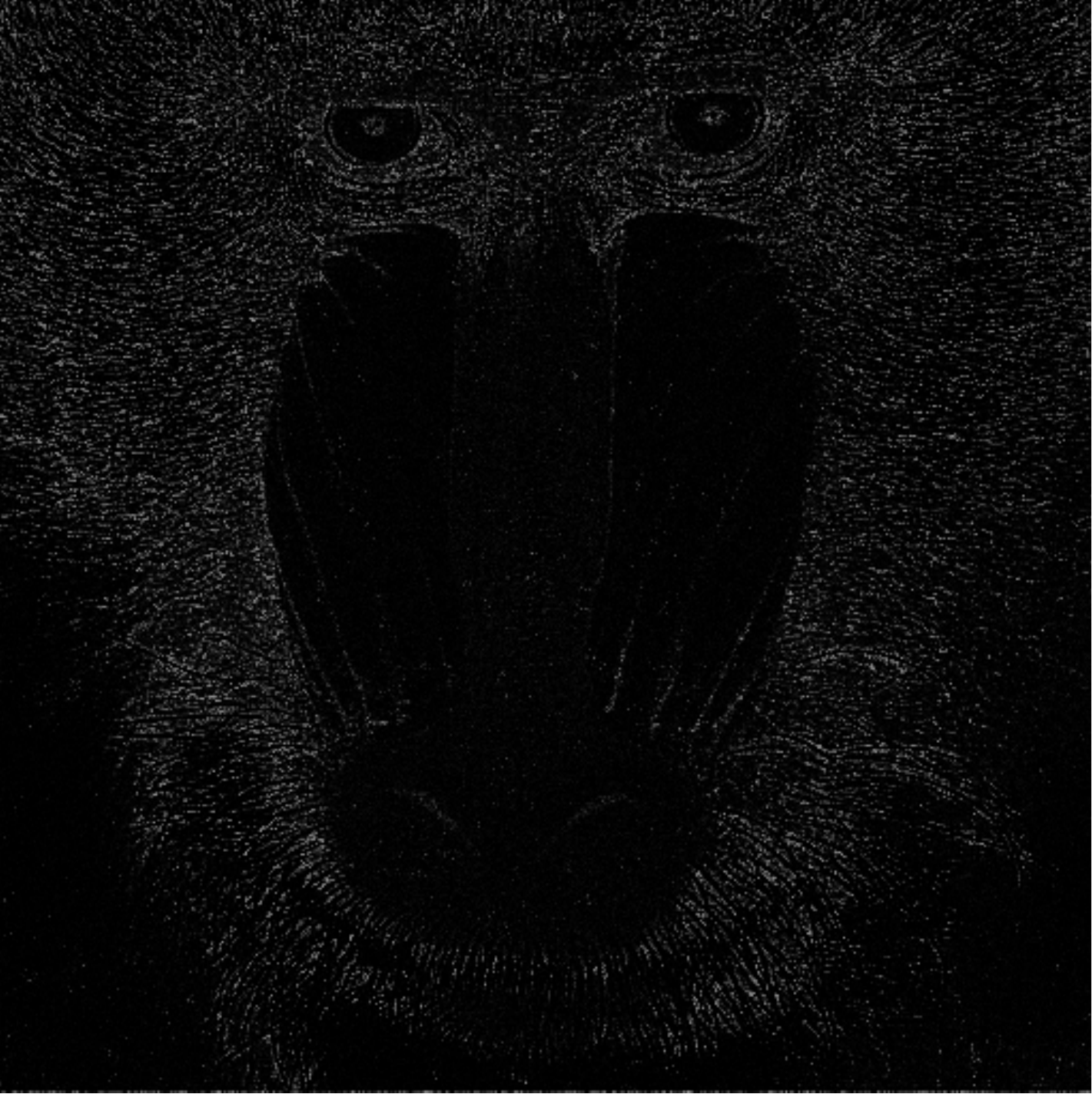}}
         \subfigure[$\rho(\hat{f},\hat{\mu}) \approx 2.04\%$]{
         \includegraphics[width = 5.3cm]{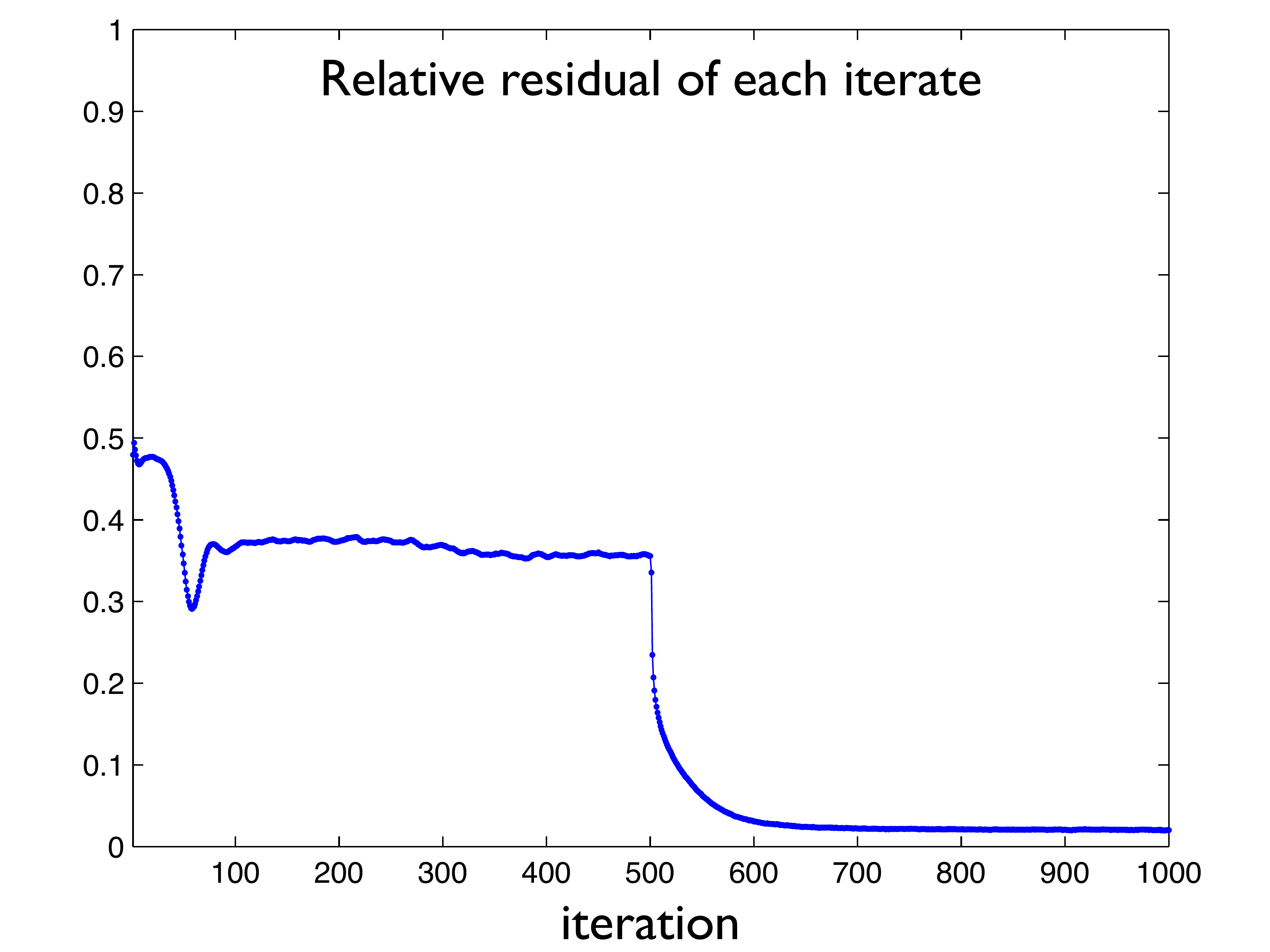}}  
            \subfigure[$e(\hat{f}) \approx 2.20\%$]{
         \includegraphics[width = 4cm]{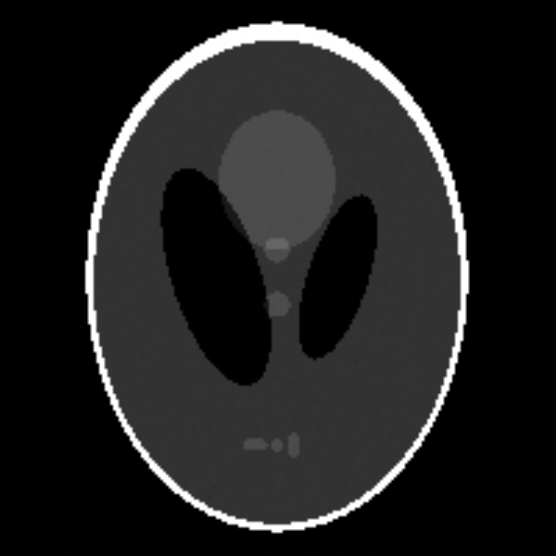}}
                 \subfigure[]{
         \includegraphics[width = 4cm]{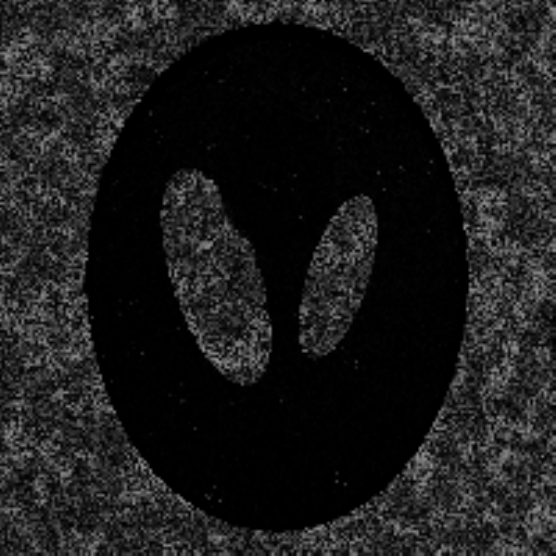}}
         \subfigure[$\rho(\hat{f},\hat{\mu}) \approx 1.31\%$]{
         \includegraphics[width = 5.3cm]{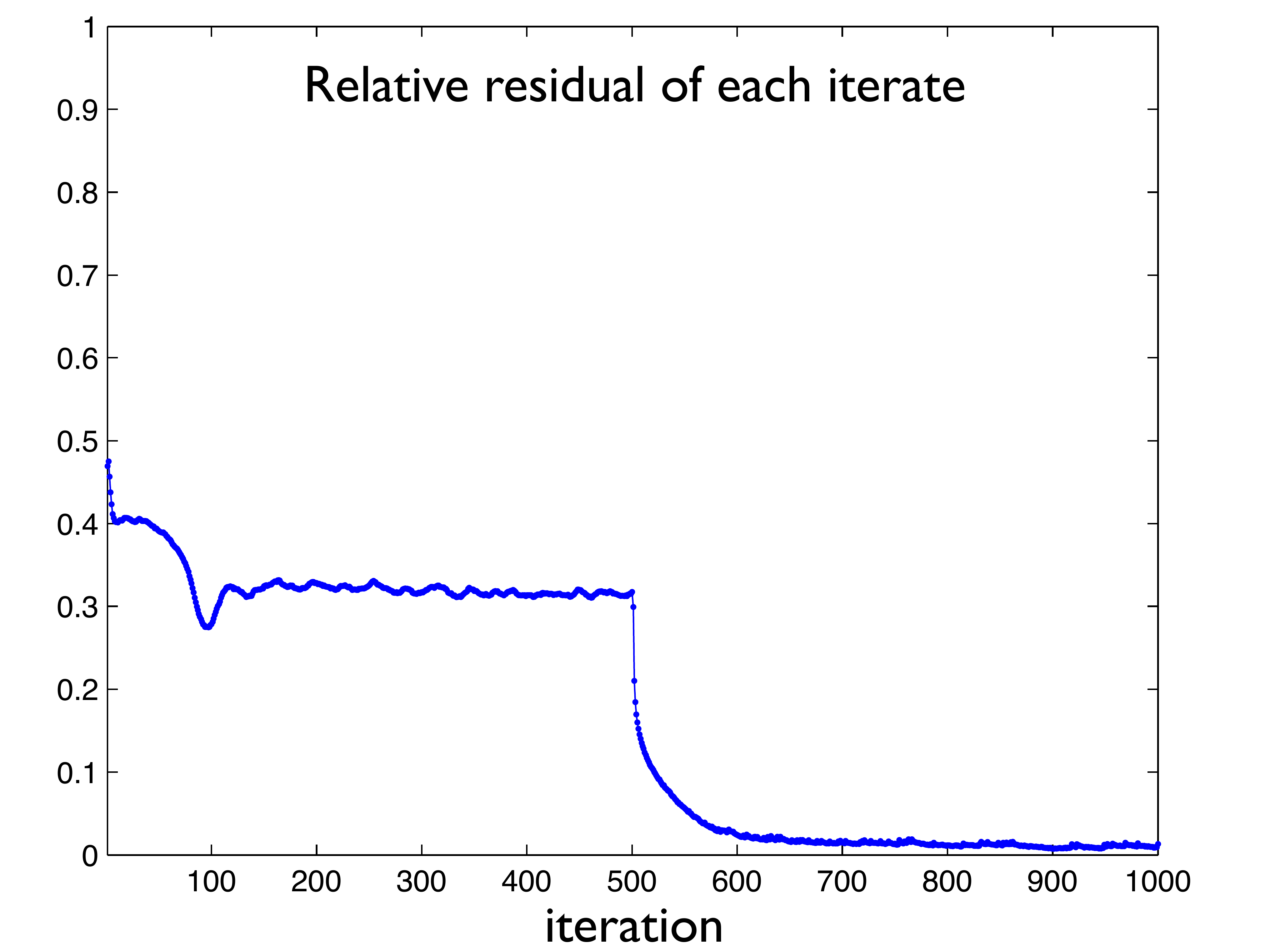}}
    \caption{Recovery of unconstrained complex-valued images  with one UM and  one LRM of $\delta=0.3$.
           (a) absolute values of the recovered cameraman  $\hat f$    by $500$ DRER $+$ $500$ AER steps.           
           (d) absolute values of the recovered mandrill $\hat f$    by $500$ DRER $+$ $500$ AER steps.   
                         (g) absolute values of the recovered phantom $\hat f$    by $500$ DRER $+$ $500$ AER steps.      
          The middle column shows the absolute phase differences between $\mu$ and $\hat\mu$.            The right column shows the relative residual at each iteration.  
                         } 
 \label{FigureComplex}
\end{figure}

\begin{figure}[hthp]
\centering
         \subfigure[$e(\hat{f}) \approx 2.62\%$]{
         \includegraphics[width = 4cm]{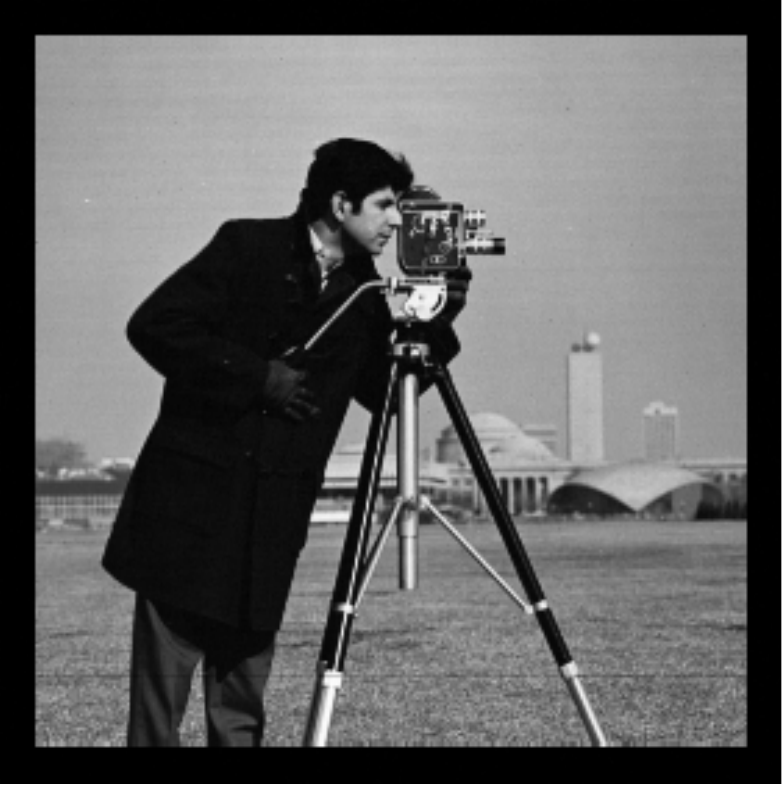}}    
         \subfigure[]{
         \includegraphics[width = 4cm]{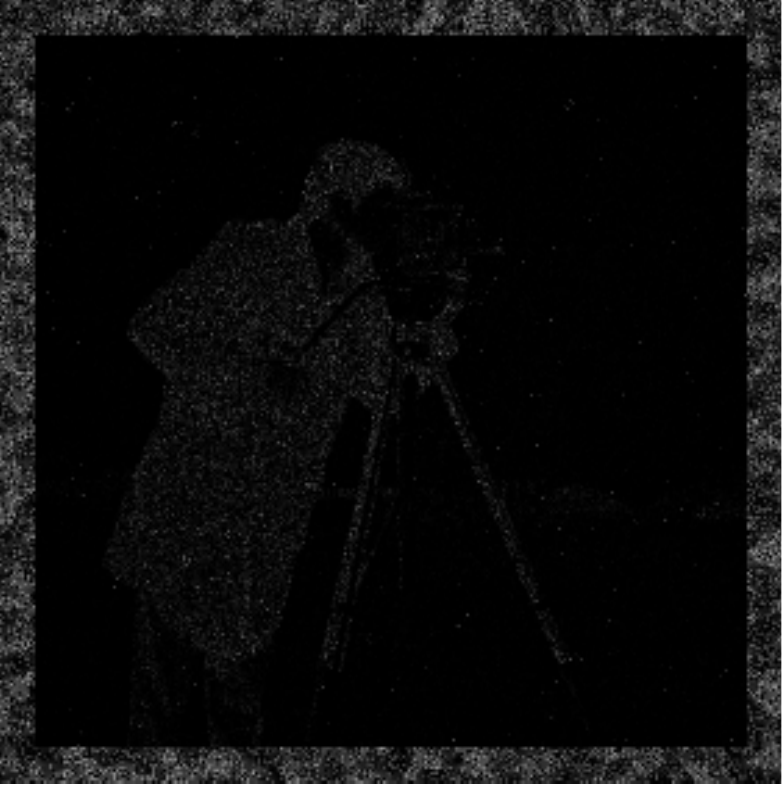}}
          \subfigure[$\rho(\hat{f},\hat{\mu}) \approx 1.12\%$]{
         \includegraphics[width = 5.3cm]{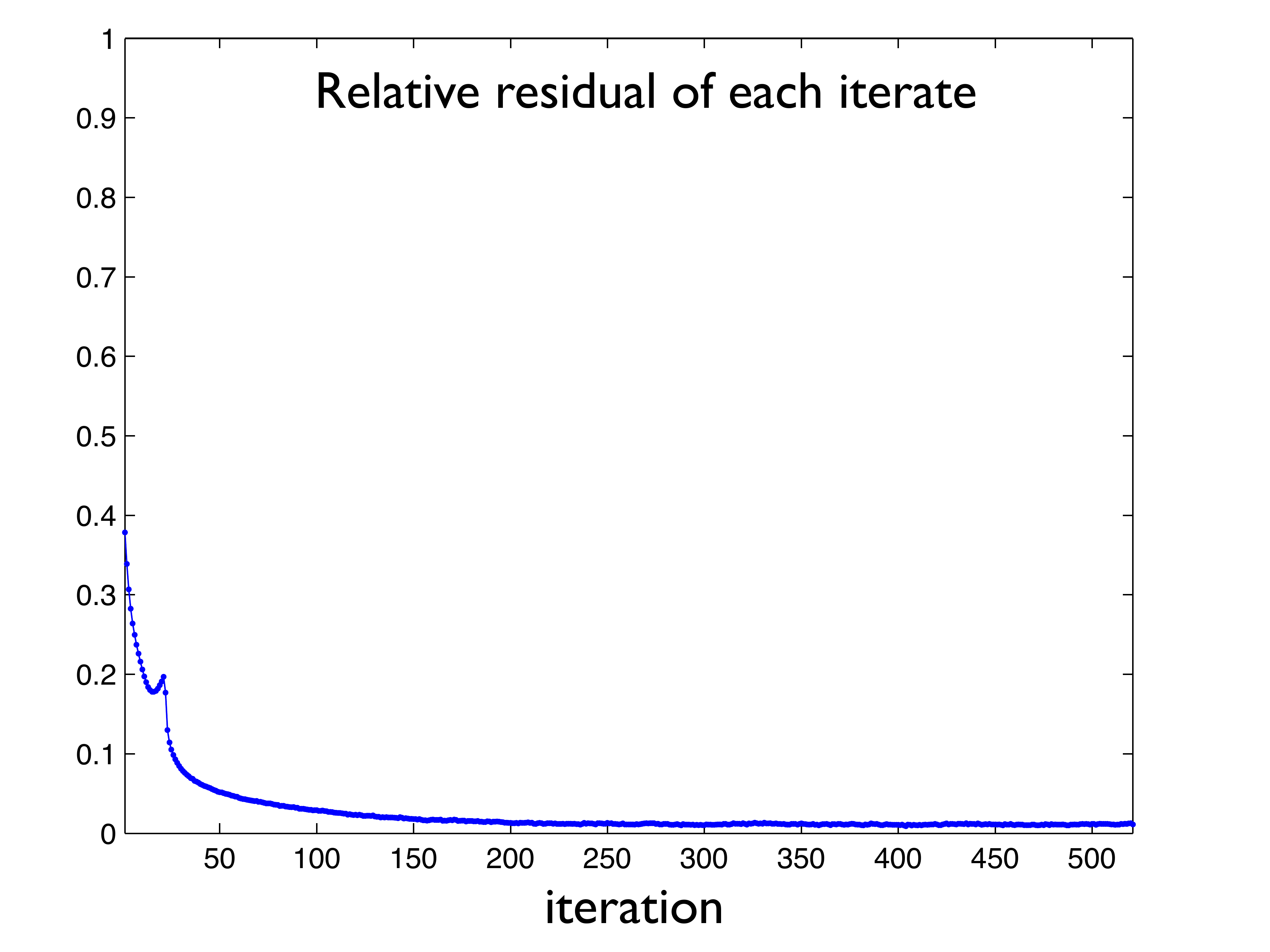}}
          \subfigure[$e(\hat{f}) \approx 2.16\%$]{
         \includegraphics[width = 4cm]{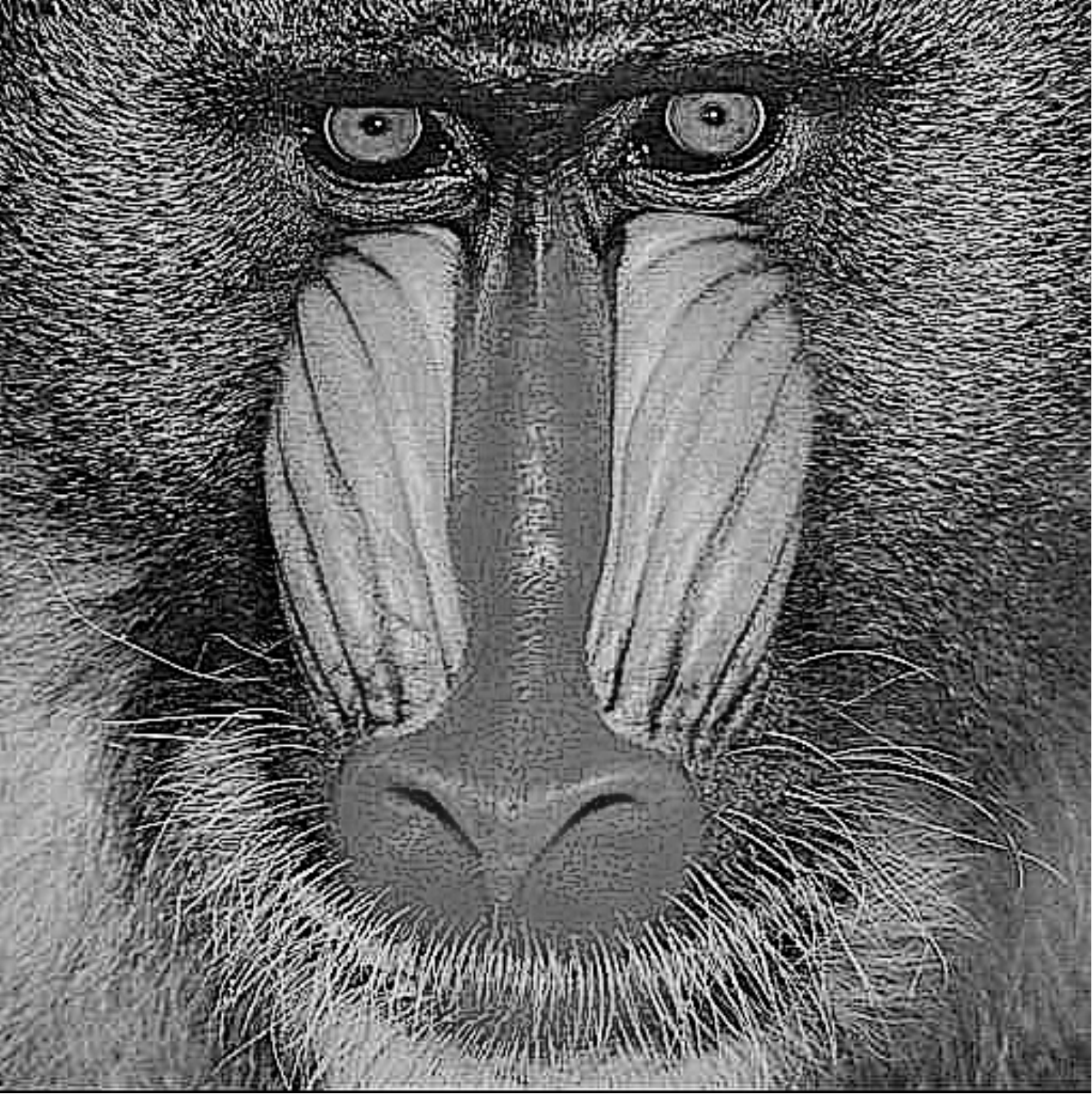}}
                 \subfigure[]{
         \includegraphics[width = 4cm]{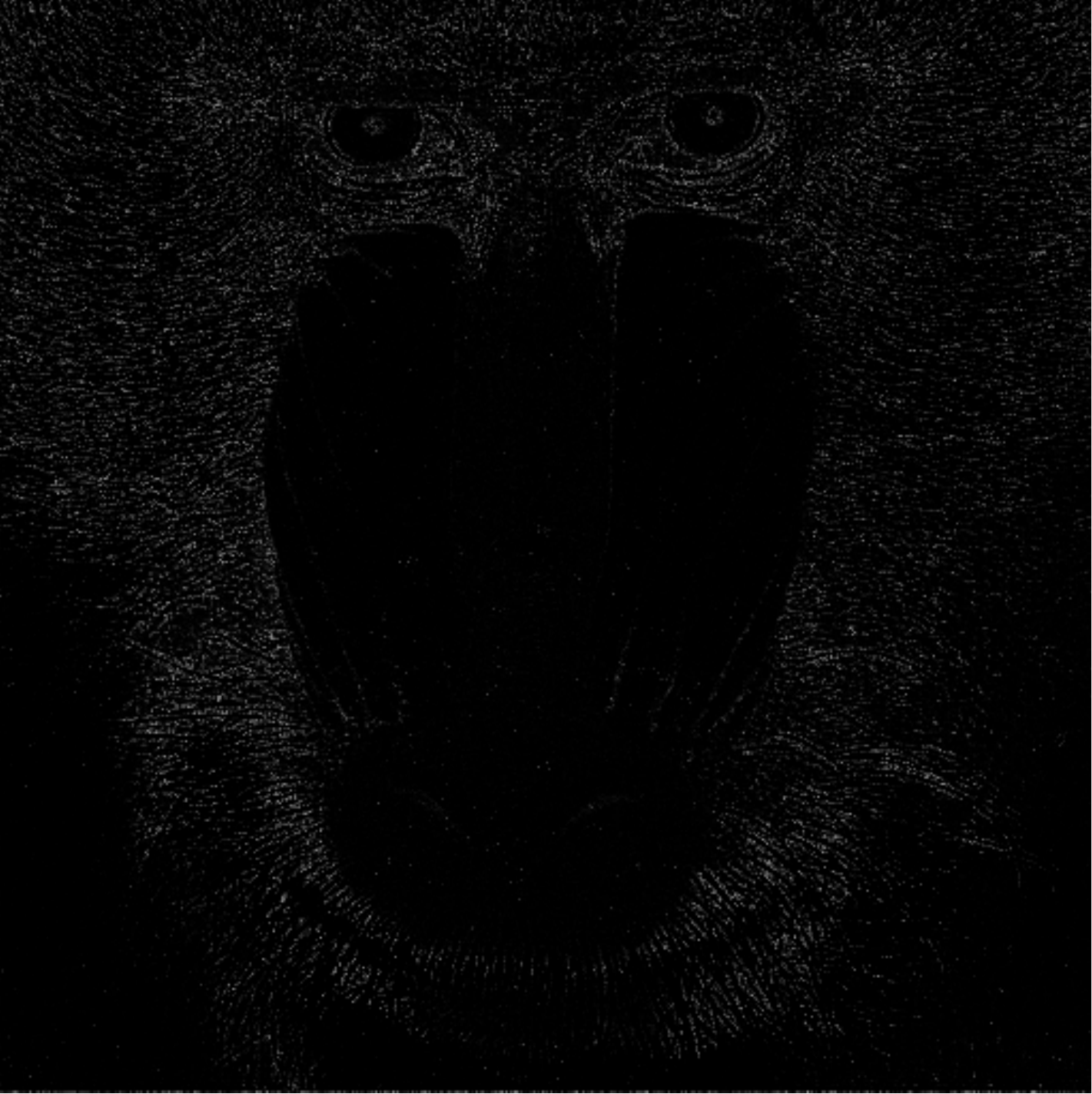}}
         \subfigure[$\rho(\hat{f},\hat{\mu}) \approx 1.03\%$]{
         \includegraphics[width = 5.3cm]{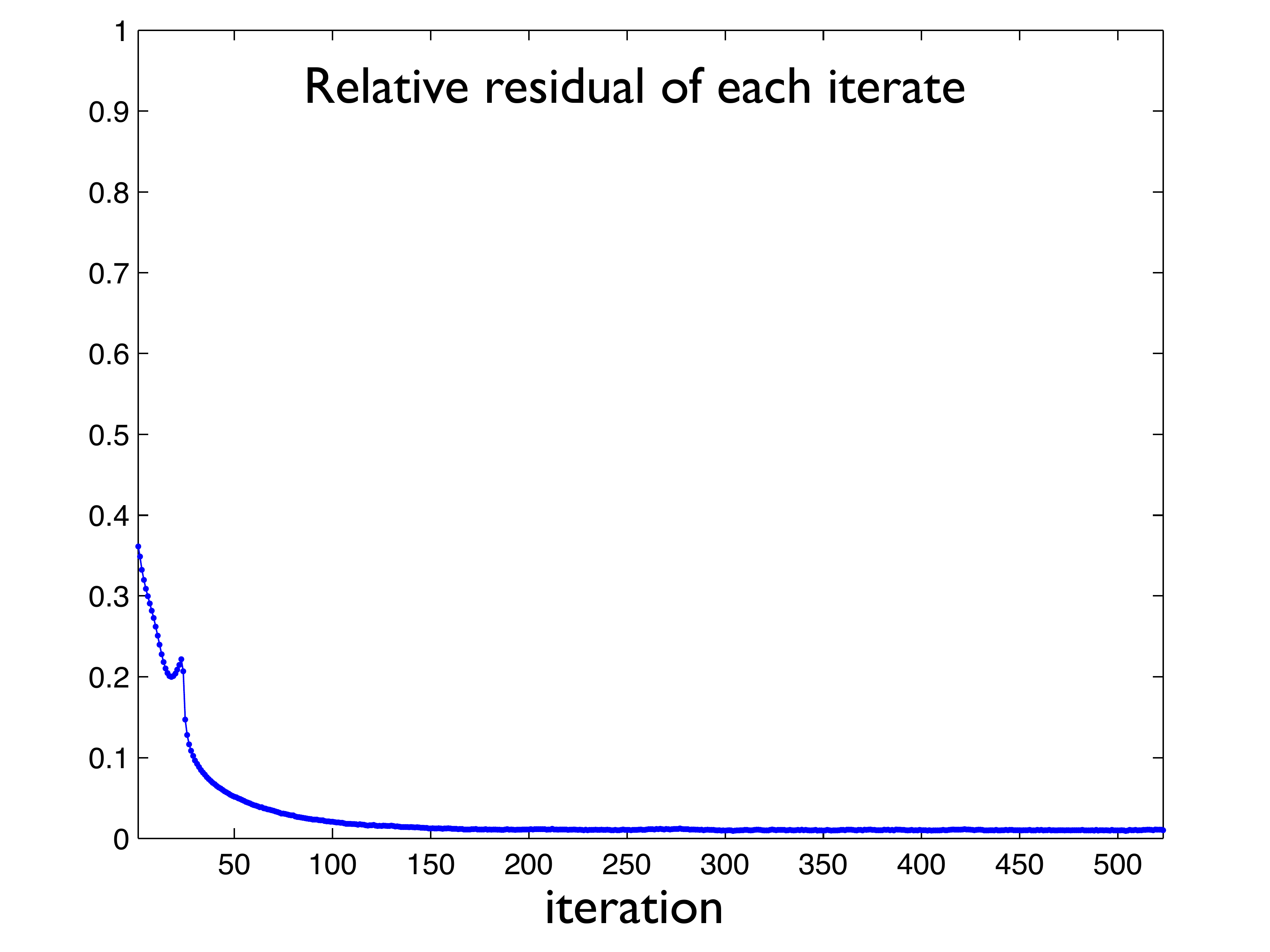}}        
          \subfigure[$e(\hat{f}) \approx 1.47\%$]{
         \includegraphics[width = 4cm]{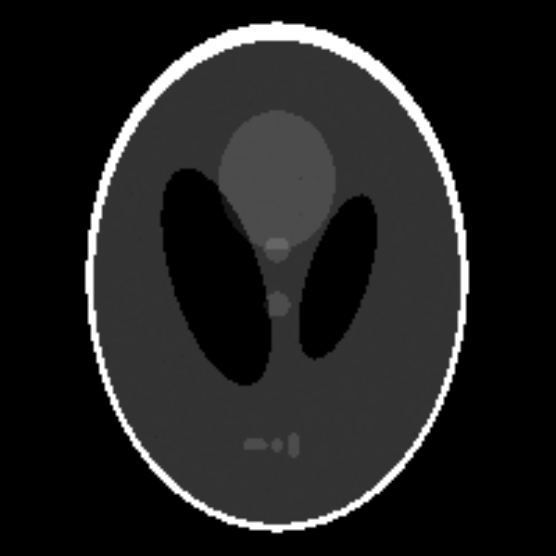}}
                 \subfigure[]{
         \includegraphics[width = 4cm]{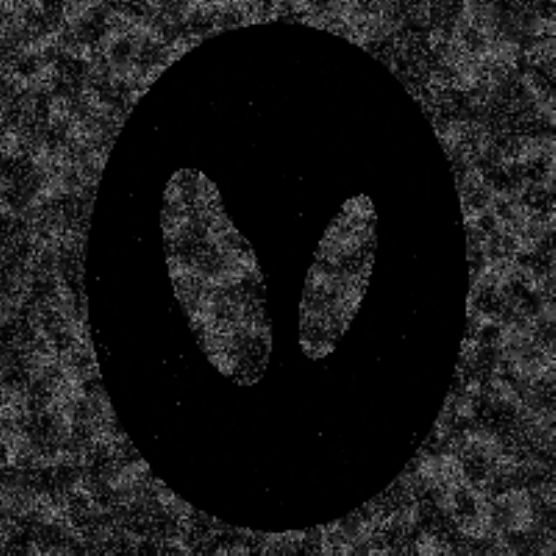}}
         \subfigure[$\rho(\hat{f},\hat{\mu}) \approx 0.80\%$]{
         \includegraphics[width = 5.3cm]{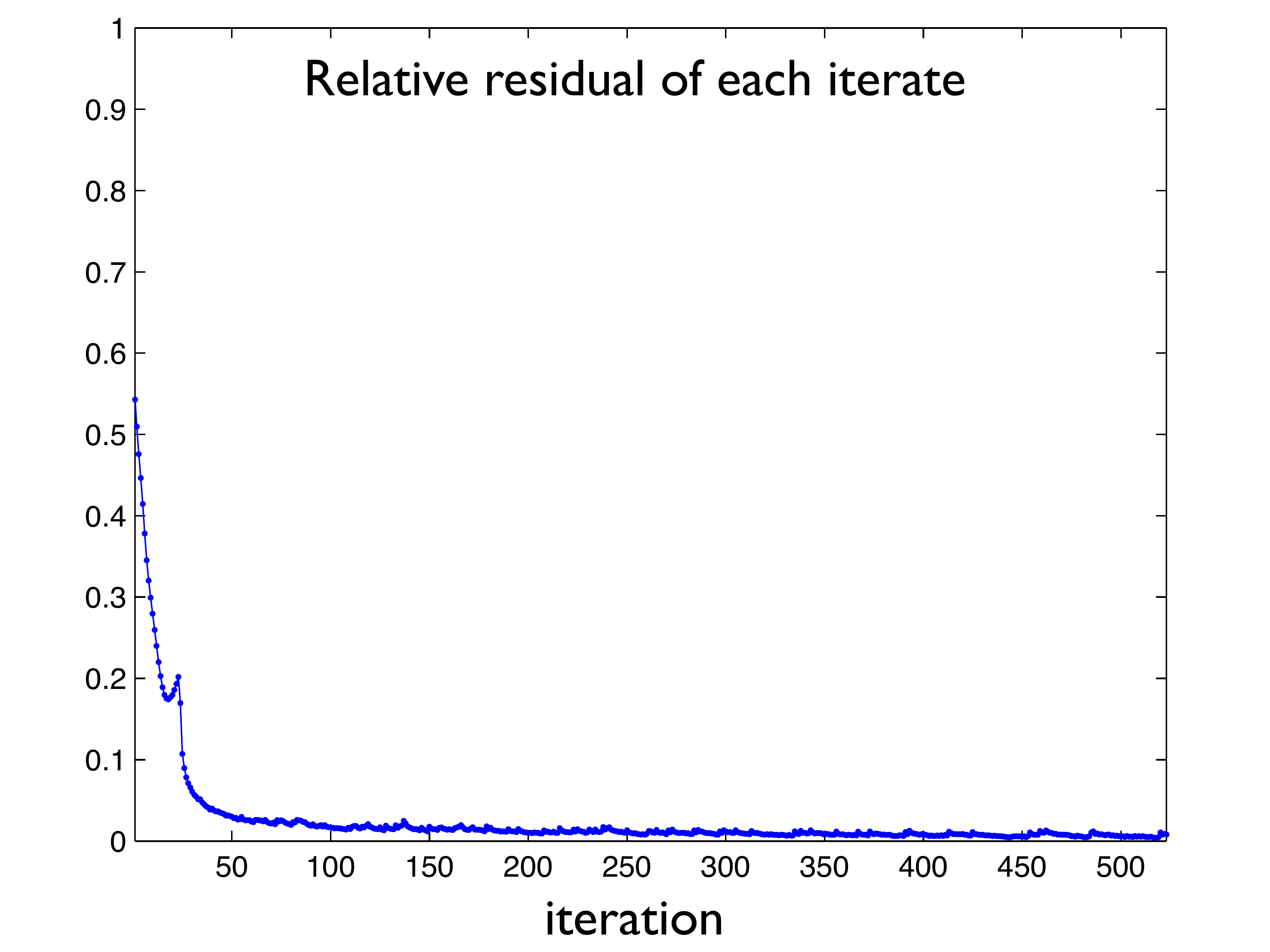}}
     \caption{Recovery  of  the $\pi/2$-sector constrained images with one UM and  one LRM of $\delta=0.3$.
           (a) absolute values of the recovered cameraman $\hat f$    by $21$ DRER $+$ $500$ AER steps.             
             (d) absolute values of the recovered mandrill$\hat f$    by $23$ DRER $+$ $500$ AER steps.                    (g) absolute values of the recovered phantom $\hat f$    by $23$ DRER $+$ $500$ AER steps.              
            The middle column shows the absolute phase differences between $\mu$ and $\hat\mu$. 
           The right column shows the relative residual at each iteration. 
            } 
 \label{FigureComplexPositive}
\end{figure}

\begin{figure}[hthp]
\centering
\subfigure[]{
         \includegraphics[width = 5.4cm]{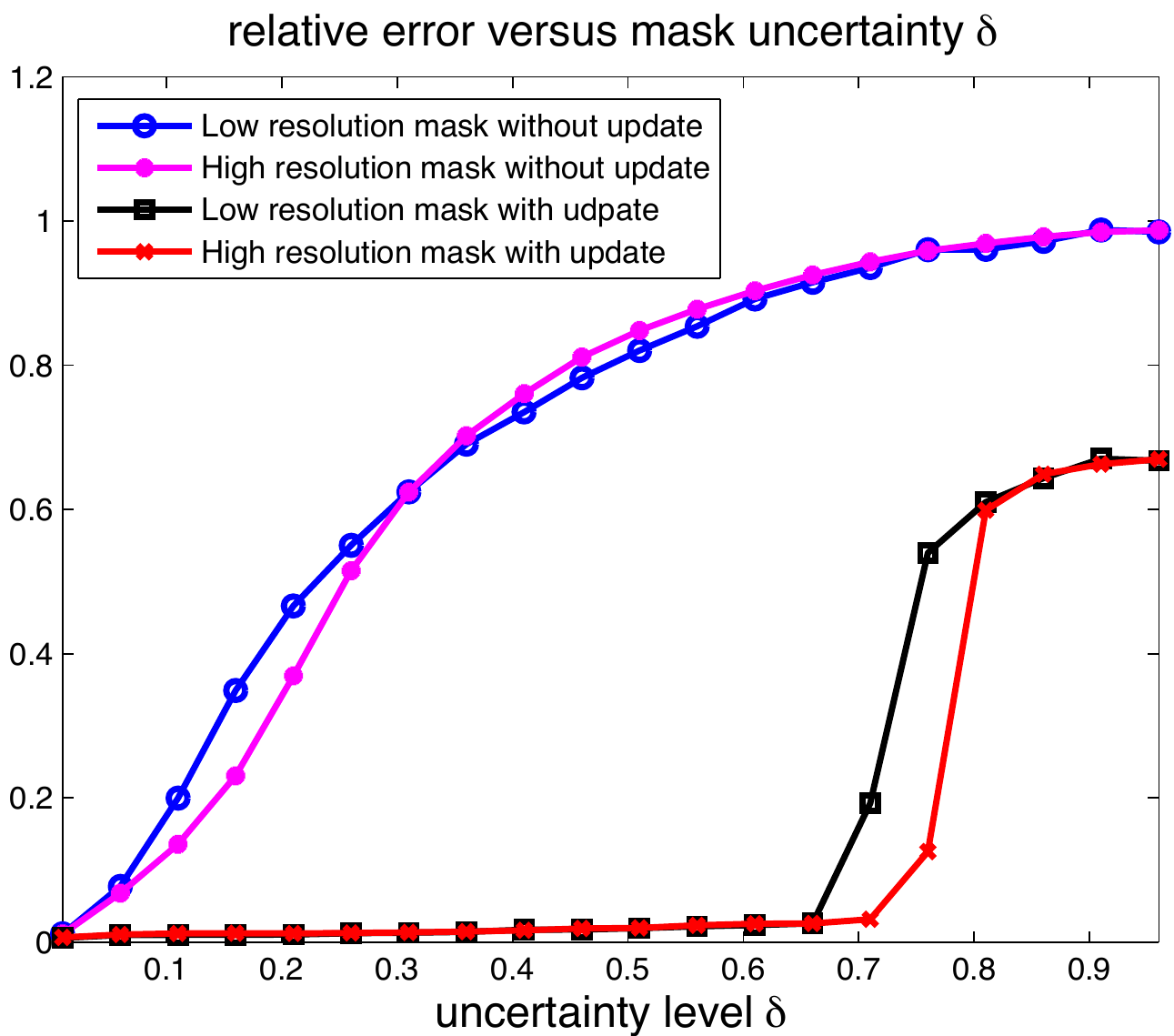}}   
         \hspace{-0.4cm}
         \subfigure[]{
                \includegraphics[width = 5.4cm]{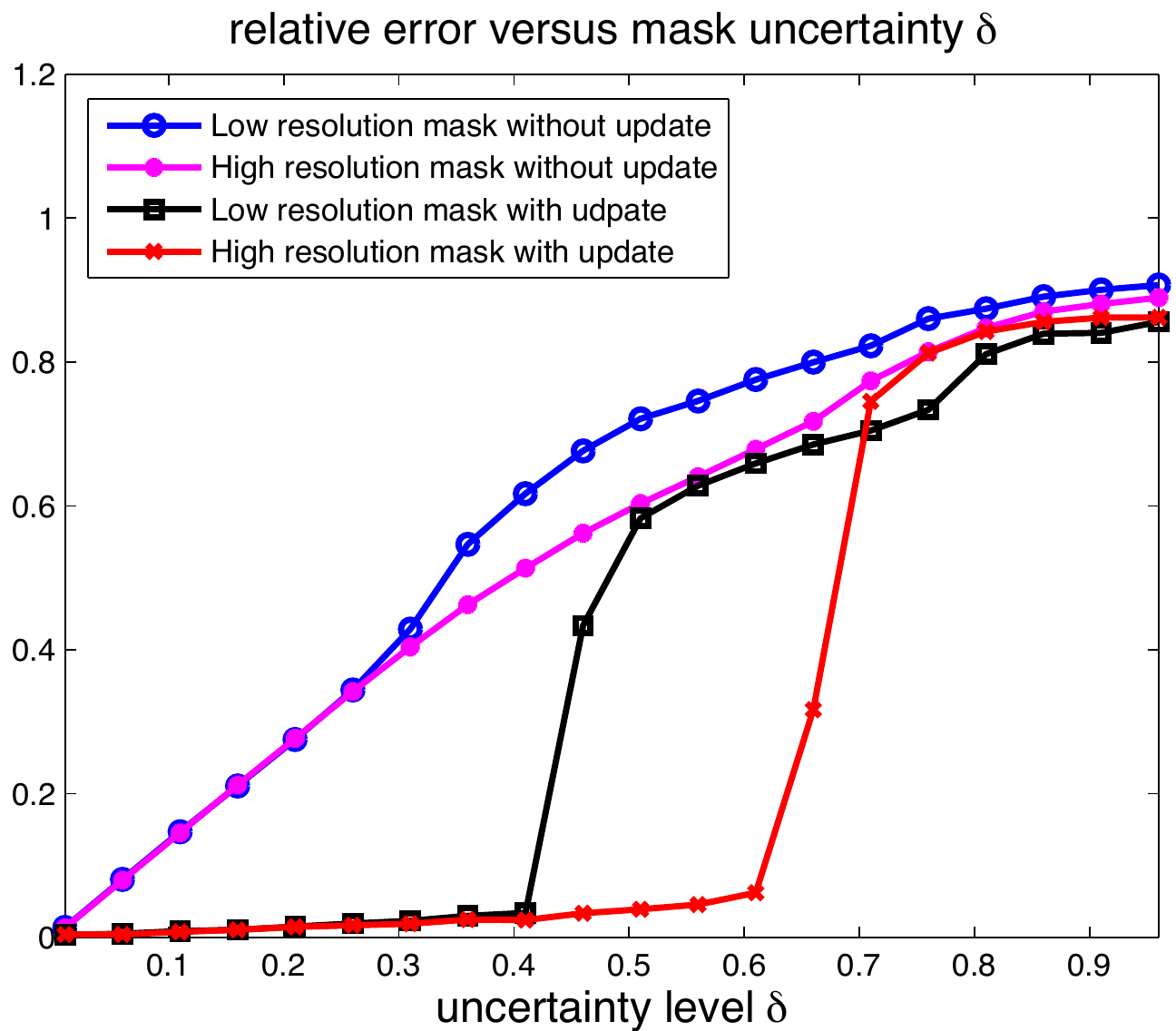}} \hspace{-.4cm}
                            \subfigure[]{
                \includegraphics[width = 5.4cm]{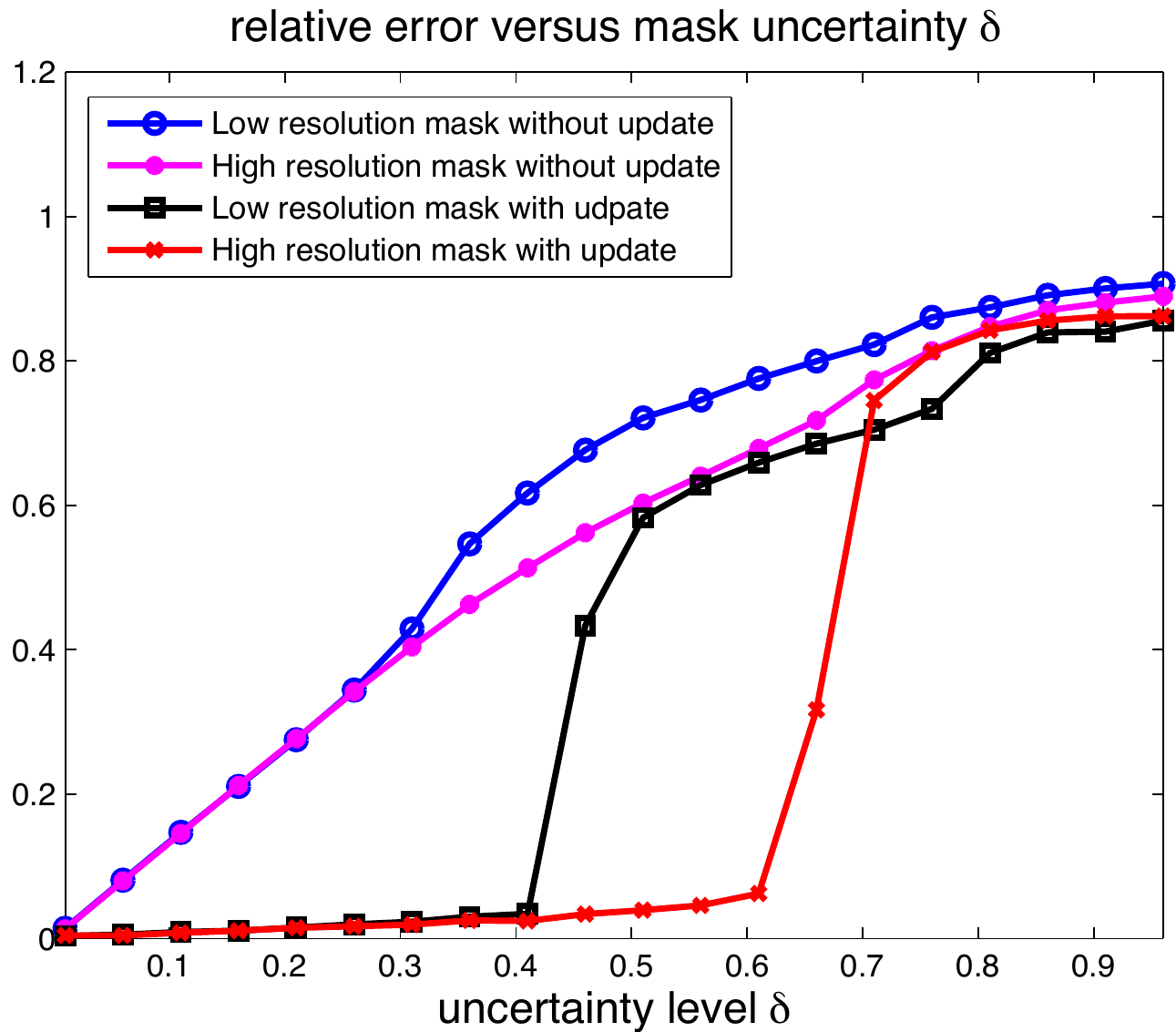}}  
      \subfigure[]{ \includegraphics[width = 5.4cm]{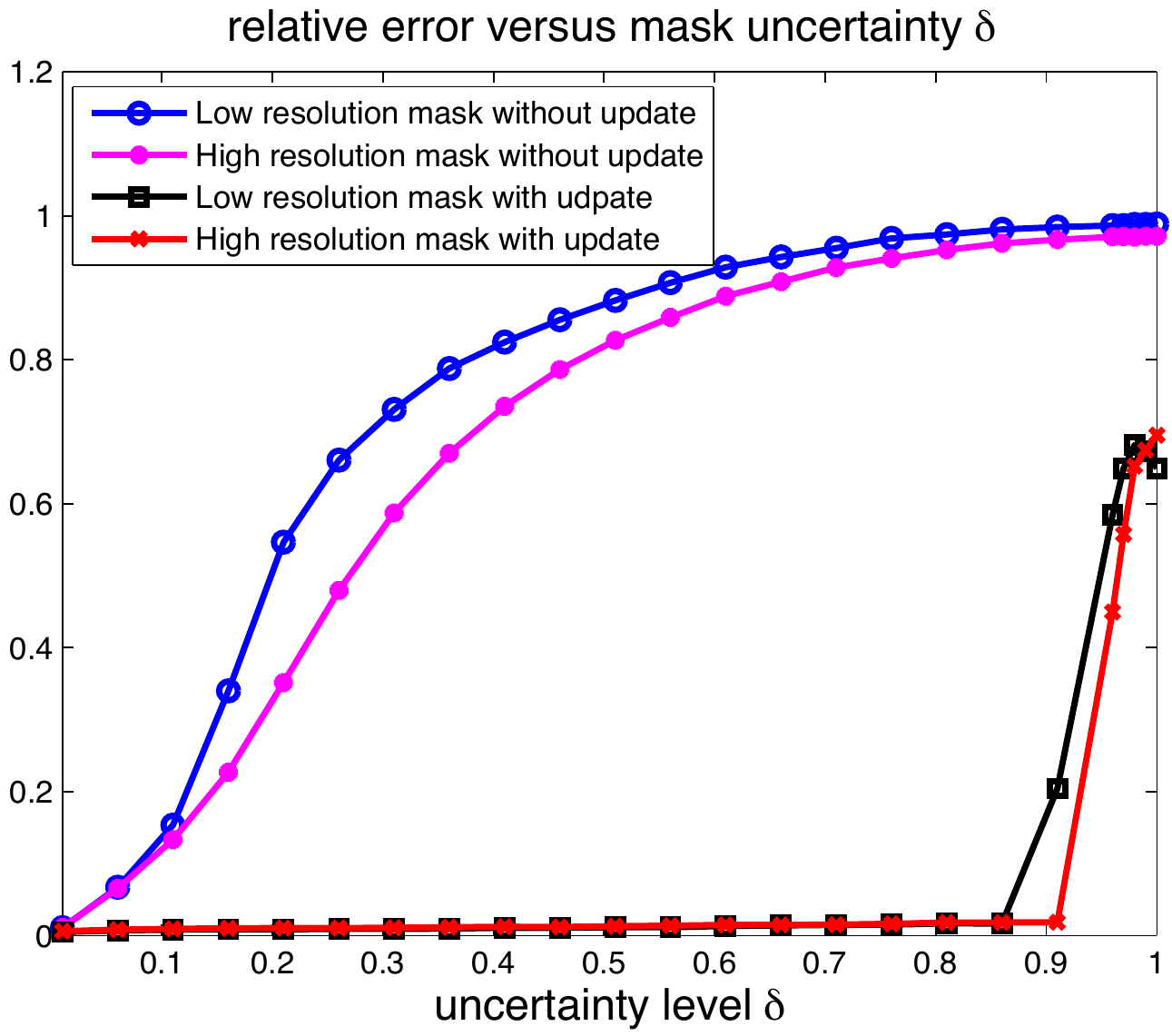}} \hspace{-0.4cm} 
         \subfigure[]{
                \includegraphics[width = 5.4cm]{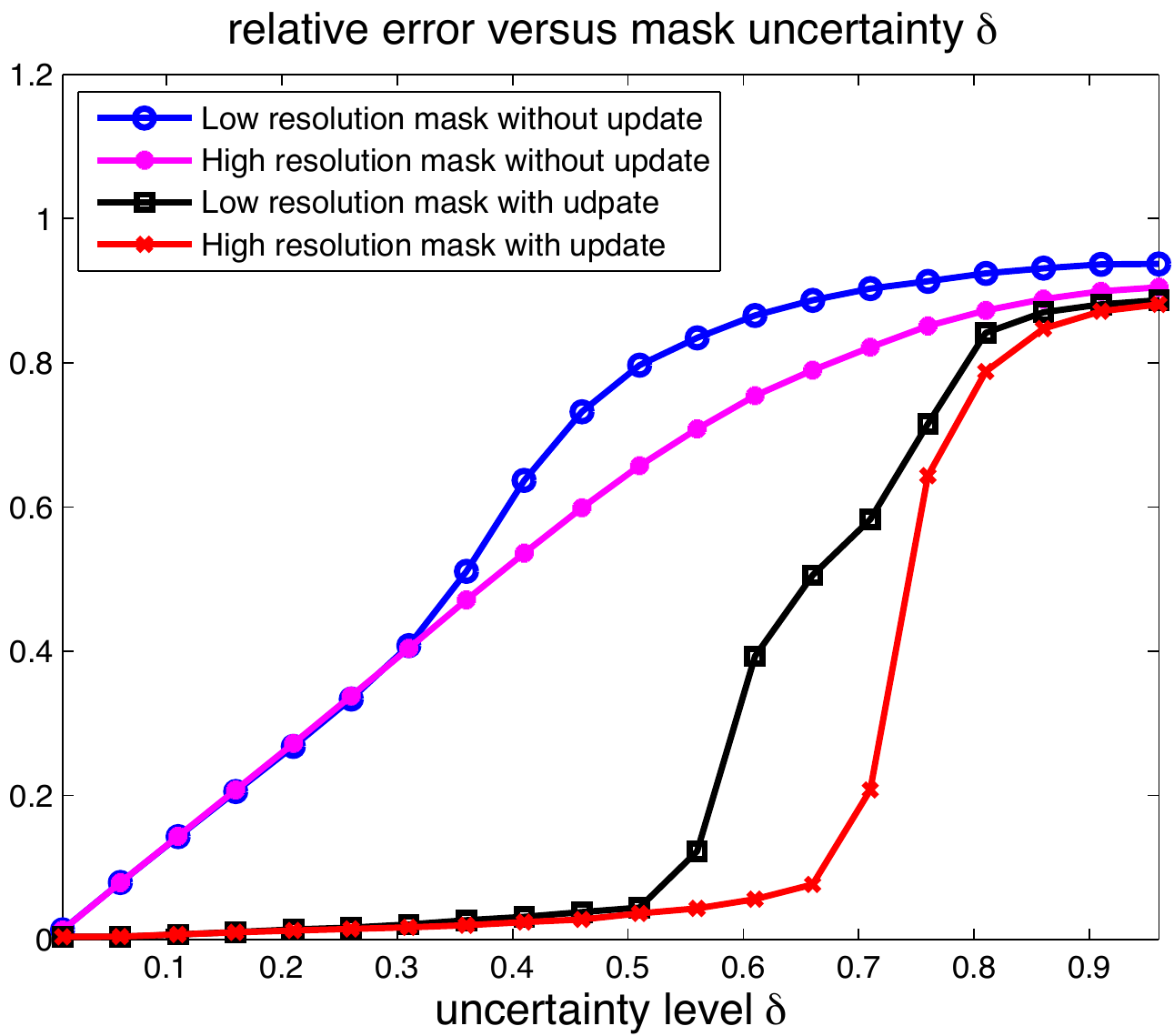}}    \hspace{-.4cm}  
                           \subfigure[]{
                \includegraphics[width = 5.4cm]{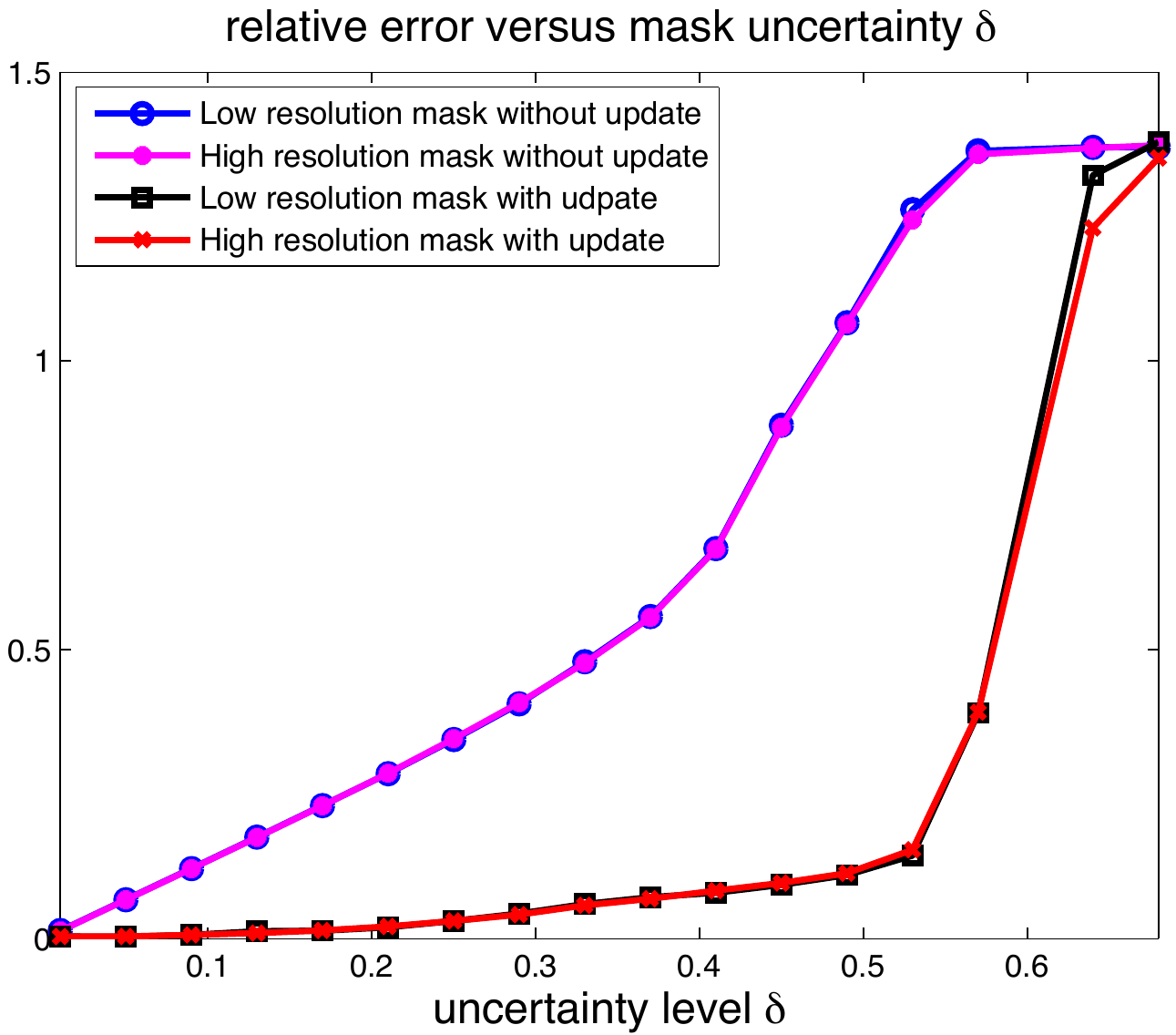}} 
  \subfigure[]{
         \includegraphics[width = 5.4cm]{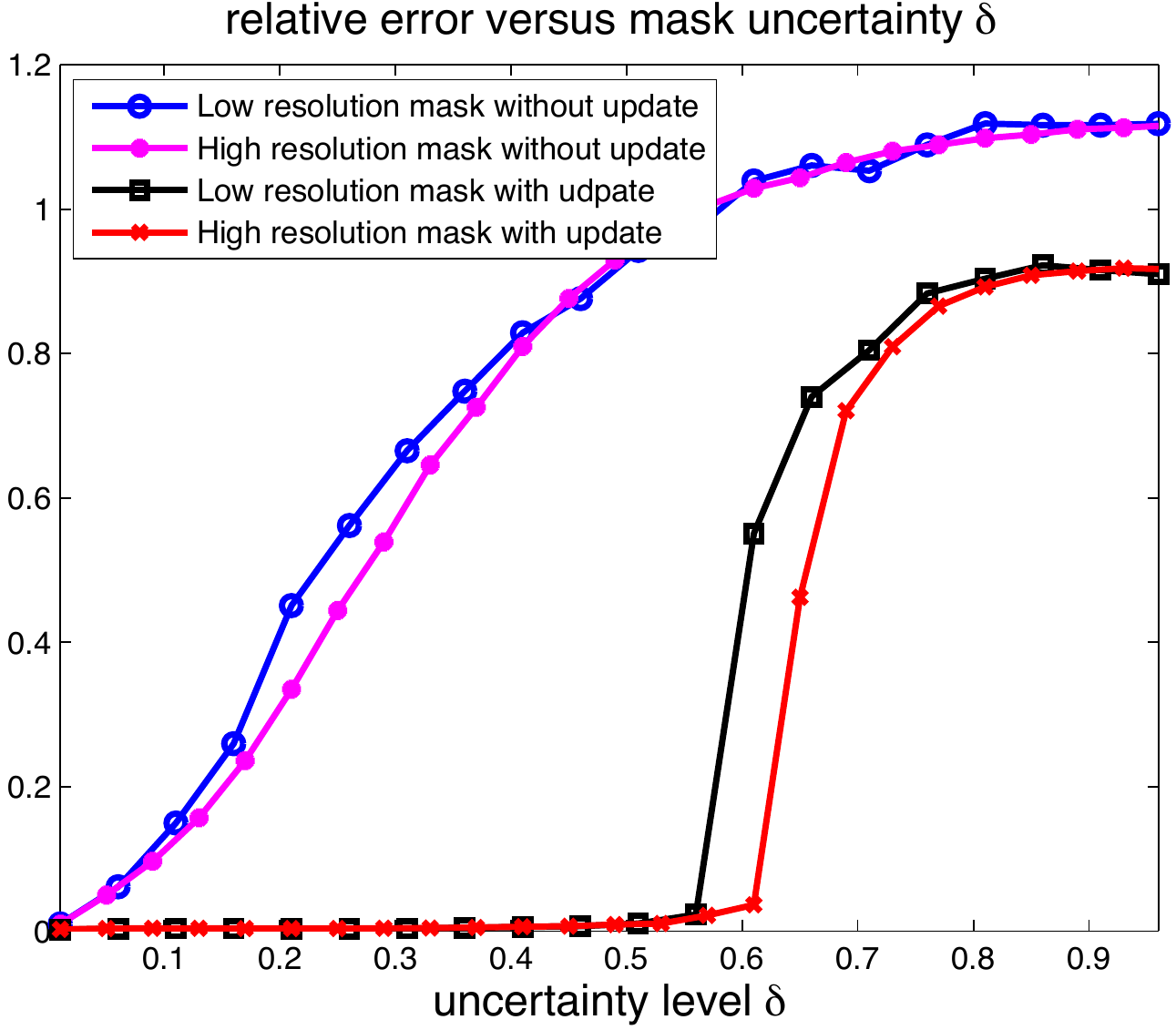}}   \hspace{-.4cm} 
               \subfigure[]{
                \includegraphics[width = 5.4cm]{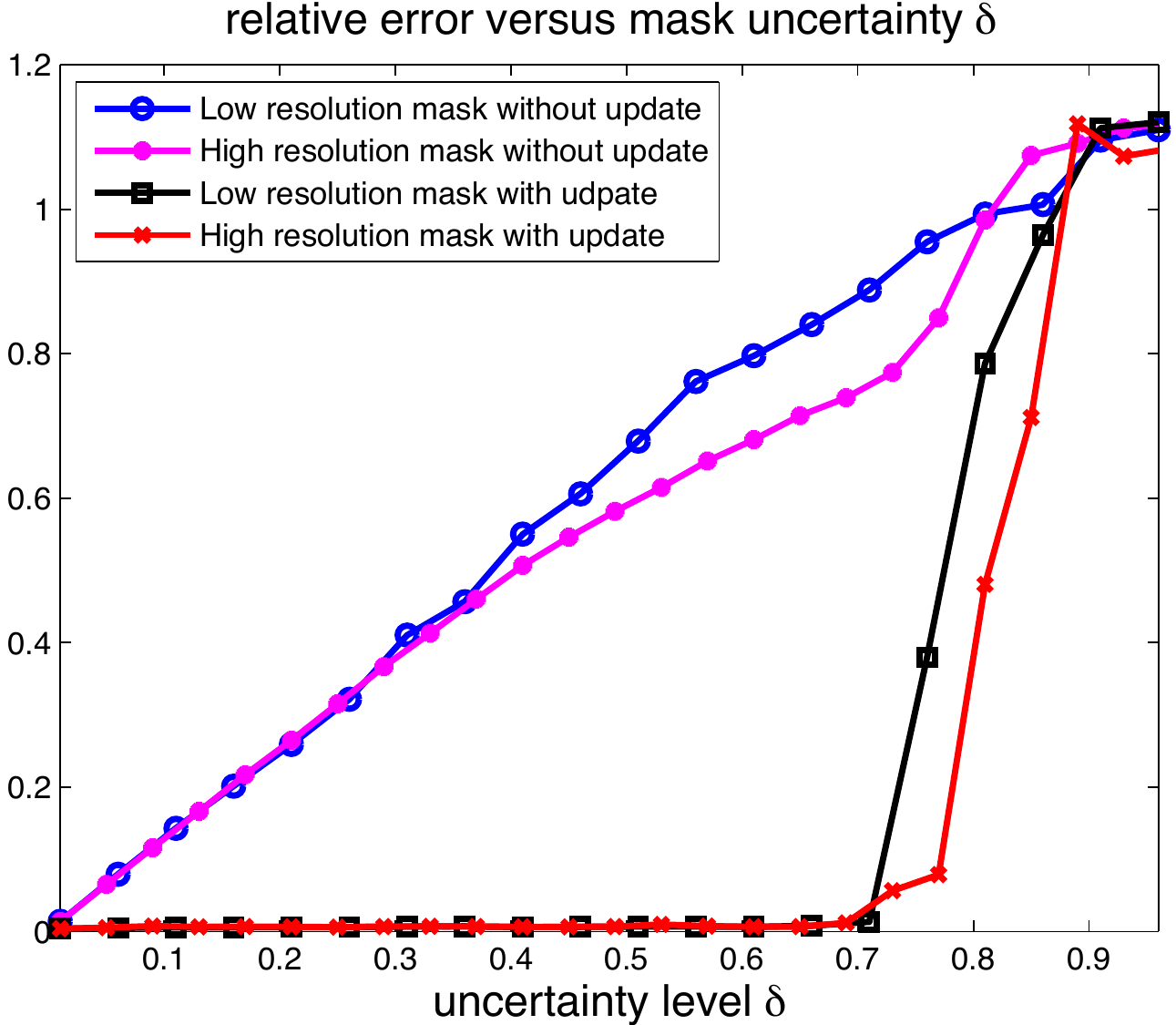}}     \hspace{-.4cm}  
             \subfigure[]{
                \includegraphics[width = 5.4cm]{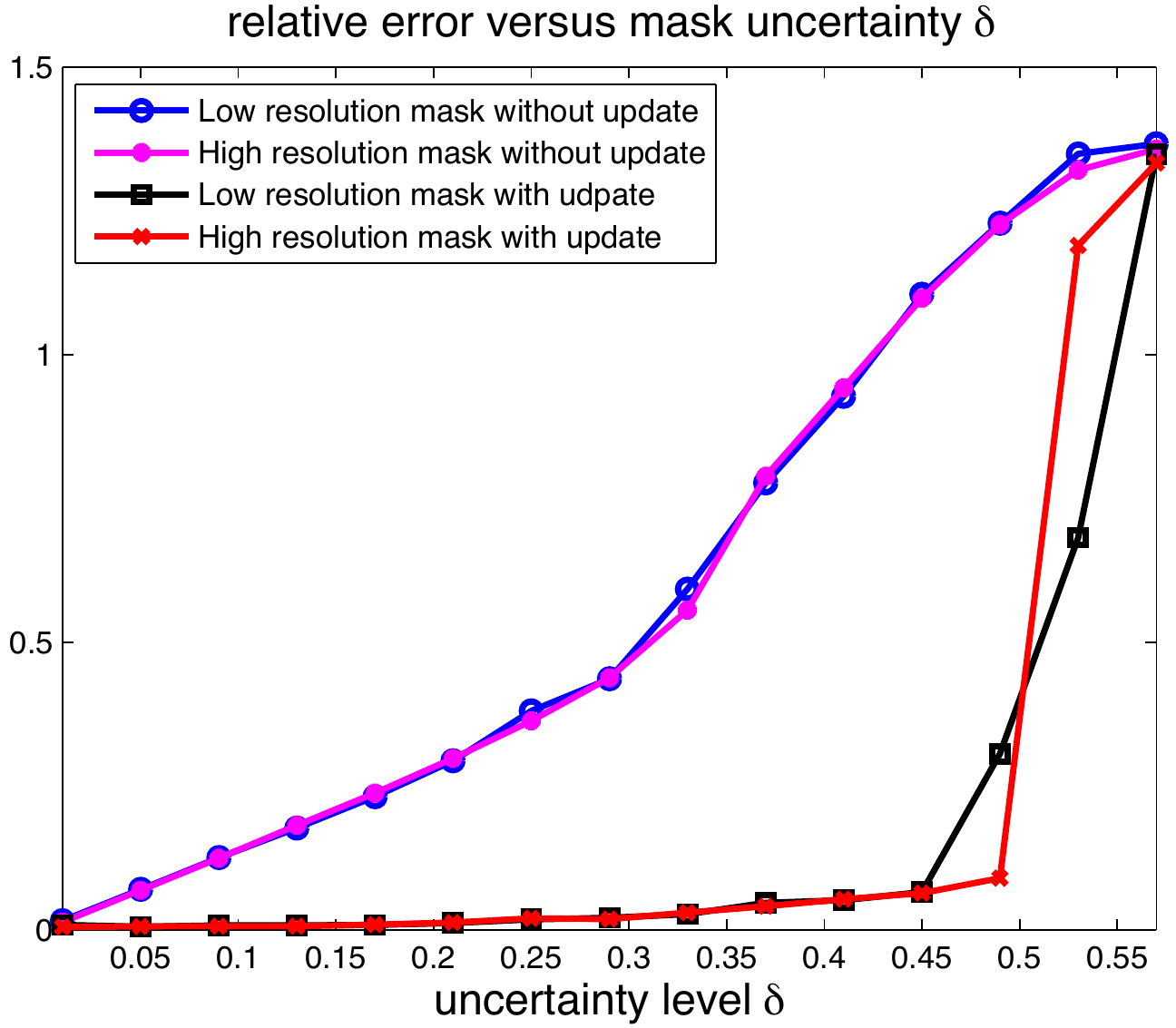}} 
                      \caption{Noiseless reconstruction error with or without mask update. Averaged  relative error $e(\hat{f})$ of 5 independent runs  versus the percentage of mask uncertainty for nonnegative images (left column), $\pi/2$-sector images (middle column) and  unconstrained images (right column)  in the order of cameraman, mandrill and phantom (top to bottom).
                   The stopping rules and mask updating rules are the same  as described      in the main text for each case
with  the maximum of  $200+1000\cdot\delta$ steps for  DRER and AER separately. 
                  }
            \label{FigureErrVersusNoise}
\end{figure}

Much improvement can be gained 
by running DRER first, followed by AER. 
For real-valued objects, we use the version of DRER (\ref{drer}). DRER (\ref{drer})  is stopped when $\|f_{k+1}-f_k\|/\|f_k\| < 1\%$, with the maximum of 500 steps, and AER (\ref{aer}) is terminated when $\|f_{k+1}-f_k\|/\|f_k\| < 0.05\%$, with the maximum of 500 steps. As shown in Figure \ref{FigureRealPositive}, the results are 
 90 DRER and 6 AER steps with
1.26\% error for the cameraman, 61 DRER and 6 AER with 0.96\% error for the mandrill and 72 DRER and 5 AER 
with 0.37\% error for the phantom. Consistent with Theorem \ref{TheoremReal}, the mask errors occur
only outside the object supports. 

\subsection{Unconstrained complex images}
Next  we consider the case of the complex-valued objects without phase constraint and  with one UM and one LRM
of uncertainty  $\delta=0.3$.  We apply the alternative versions of  DRER (\ref{drer2b}) and AER (\ref{aer2b}) which tend to outperform  (\ref{drer}) and (\ref{aer})
for complex-valued objects. 
DRER (\ref{drer2b}) 
 is stopped when $\|f_{k+1}-f_k\|/\|f_k\| < 1\%$, with the maximum of 500 steps, and AER (\ref{aer2b}) is terminated when $\|f_{k+1}-f_k\|/\|f_k\| < 0.05\%$, with the maximum of 500 steps.
 Fig. \ref{FigureComplex} shows the results for object phases randomly distributed on $[0,2\pi)$. 
 Both algorithms  ran their full course of 500 steps with 6.43\% error for the cameraman, 4.62\% for the mandrill and
 2.20\% error for the phantom. The mask errors occur only outside the object supports,
 consistent with Theorem \ref{TheoremComplex2}. 

\subsection{$\pi/2$-sector constrained complex images}
Here  we  consider $\pi/2$-sector constrained complex images with  randomly distributed phases in $[0,\pi/2]$.  
 
With the   sector constraint, we found that the following stopping rule
can significantly reduce the number of iterations: 
DRER (\ref{drer2b})  is stopped if the residual increases in five consecutive steps, with the maximum of 500 steps,  and AER (\ref{aer2b}) is terminated when $\|f_{k+1}-f_k\|/\|f_k\| < 0.05\%$,
with the maximum of 500 steps.  Fig. \ref{FigureComplexPositive} shows the results with  one UM  and one LRM of uncertainty $\delta=0.3$. With the new stopping rule and the sector constraint,
21 DRER and 500 AER steps took place with 2.62\% error for the cameraman,  23 DRER and 500 AER steps with  2.16\% for the mandrill and 23 DRER  and 500 AER steps with 1.47 \% error for the phantom.

\subsection{Reconstruction error versus mask uncertainty}
Fig. \ref{FigureErrVersusNoise} shows the averaged relative error $e(\hat{f})$,  after 5  runs of independently chosen initial guesses for the object,  with  or without   mask update,  as a function of  the mask uncertainty of HRM or LRM  for non-negative images (a)(d)(g),
complex-valued images under the $\pi/2$-sector  condition (b)(e)(h) and
 complex-valued images images with unconstrained random phases (c)(f)(i). We use the same stopping rules and
updating rules as above for each case, except that the maximum number of steps is changed to
 $200+\delta\cdot 1000$ for DRER and AER separately to adapt to 
 variable uncertainty.   

Without mask update the error curves are roughly linear with the noise amplification factor
roughly  2 (top two curves), consistent with our previous results reported in \cite{FL}.
With mask update, the results (bottom two curves) are drastically improved 
in all cases.

         \subsection{Reconstruction error versus Gaussian and Poisson noises}
                 
\begin{figure}[hthp]
\centering 
  \subfigure[]{
         \includegraphics[width = 5.4cm]{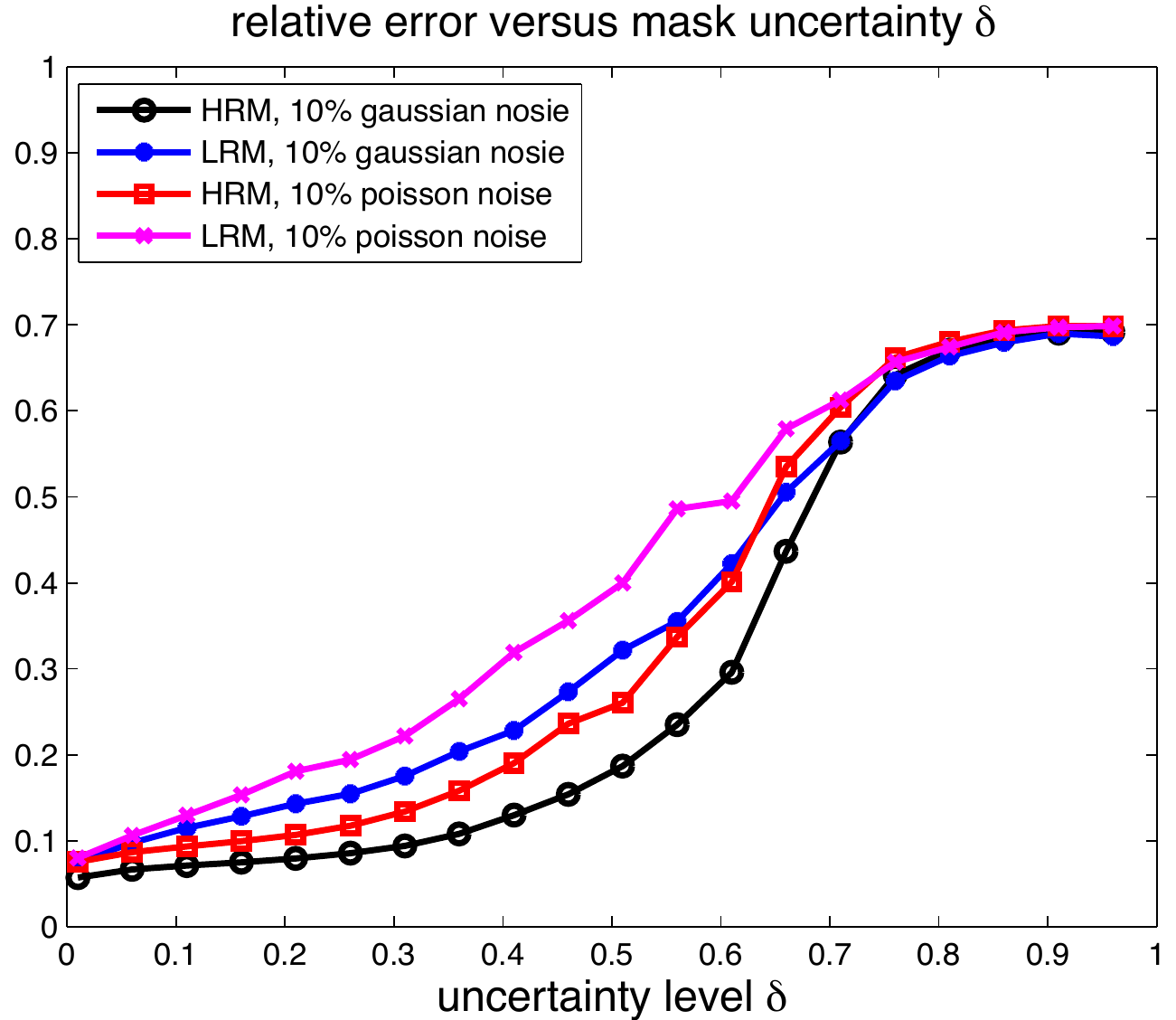}}   \hspace{-.4cm} 
               \subfigure[]{
                \includegraphics[width = 5.2cm]{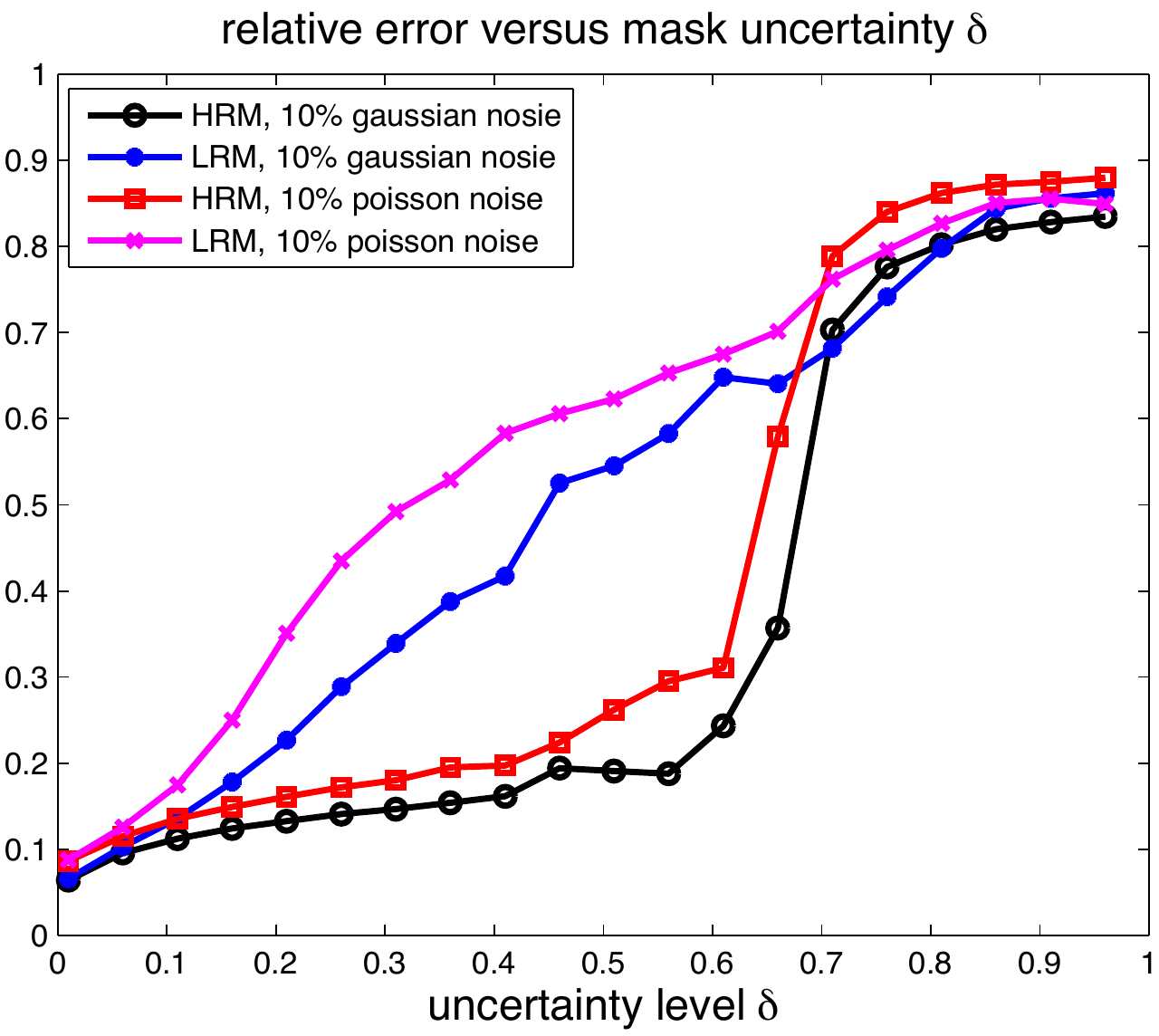}}     \hspace{-.4cm}  
             \subfigure[]{
                \includegraphics[width = 5.4cm]{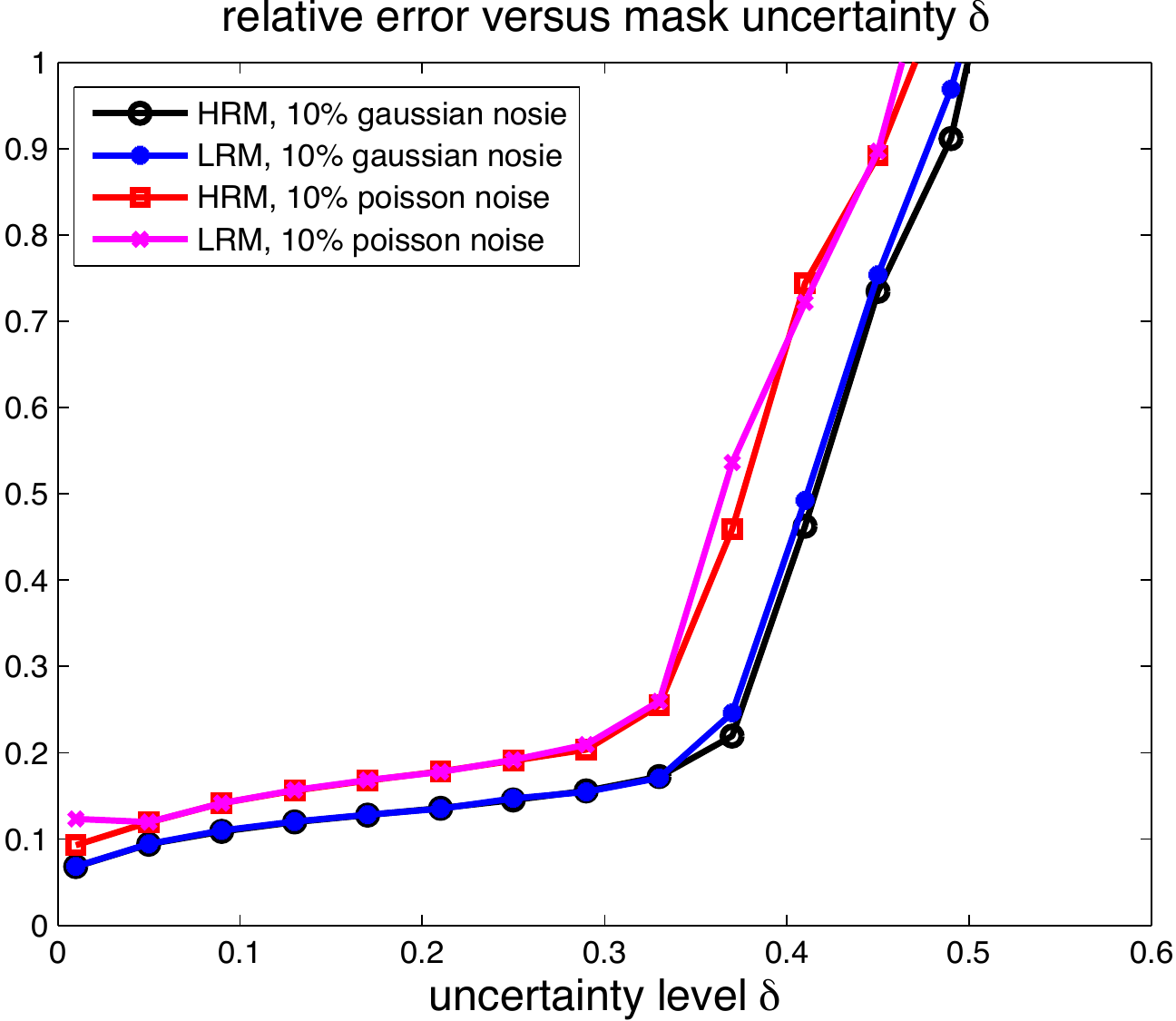}} 
  \subfigure[]{
         \includegraphics[width = 5.4cm]{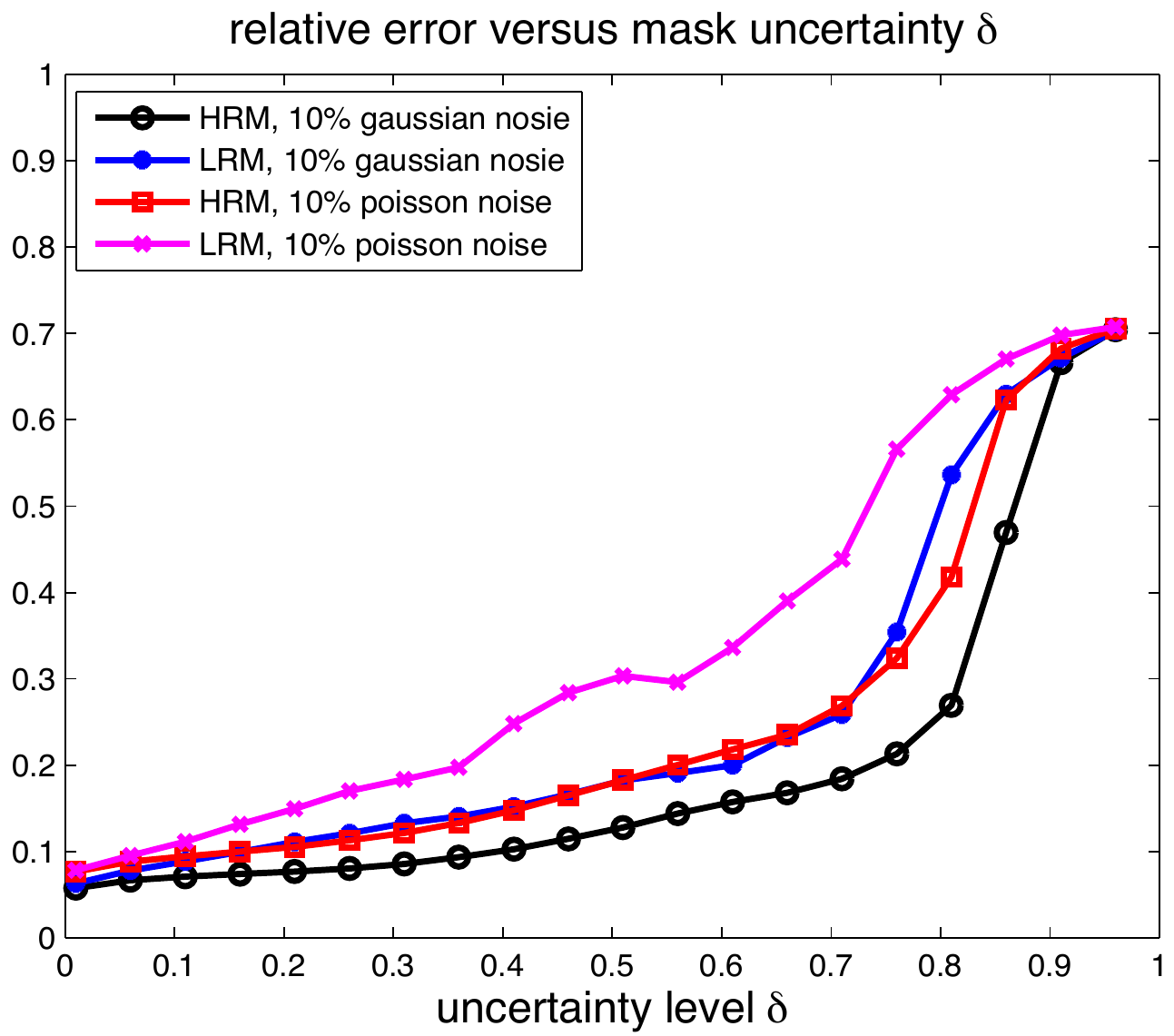}}   \hspace{-.4cm} 
               \subfigure[]{
                \includegraphics[width = 5.4cm]{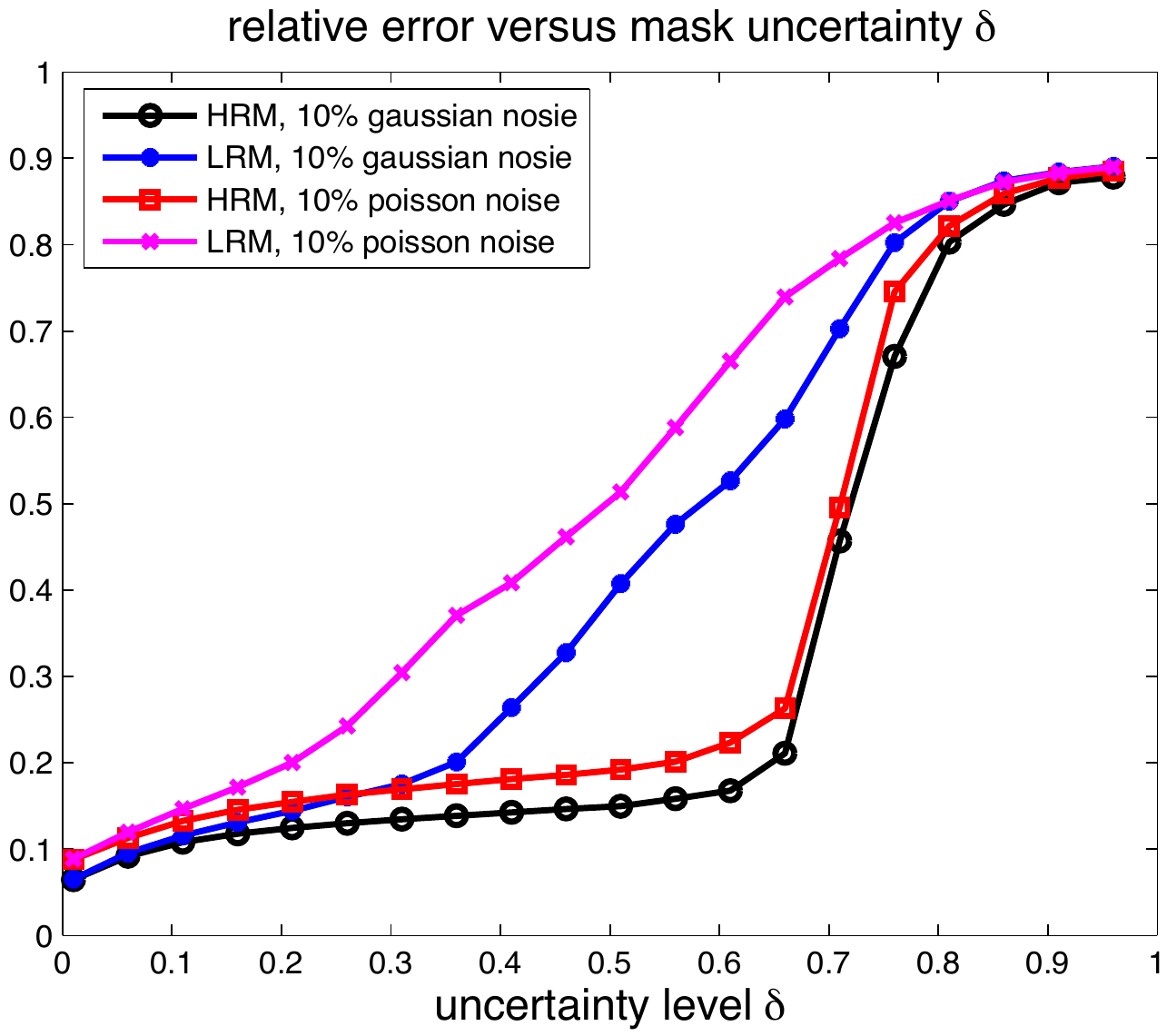}}     \hspace{-.4cm}  
             \subfigure[]{
                \includegraphics[width = 5.4cm]{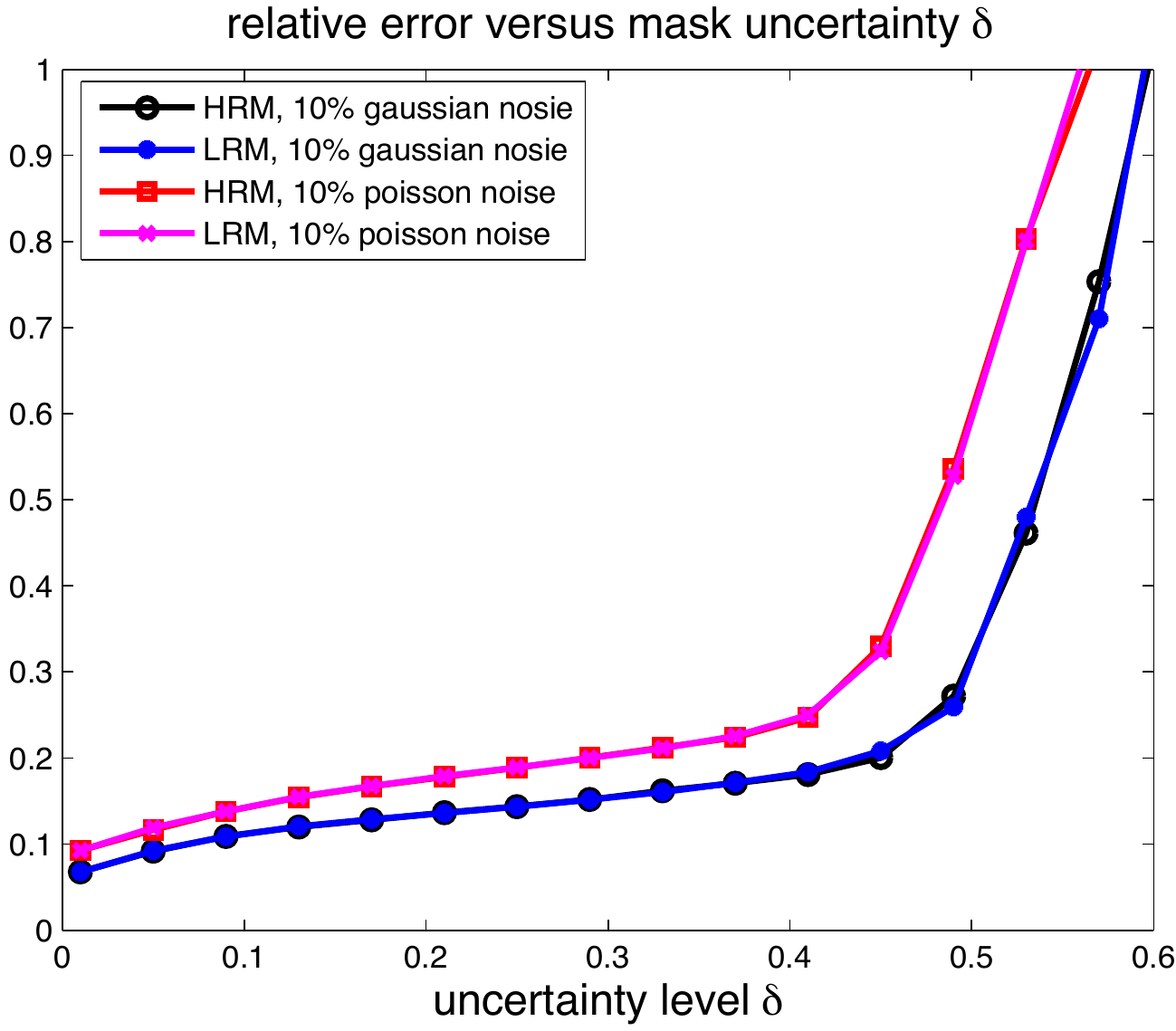}} 
  \subfigure[]{
         \includegraphics[width = 5.4cm]{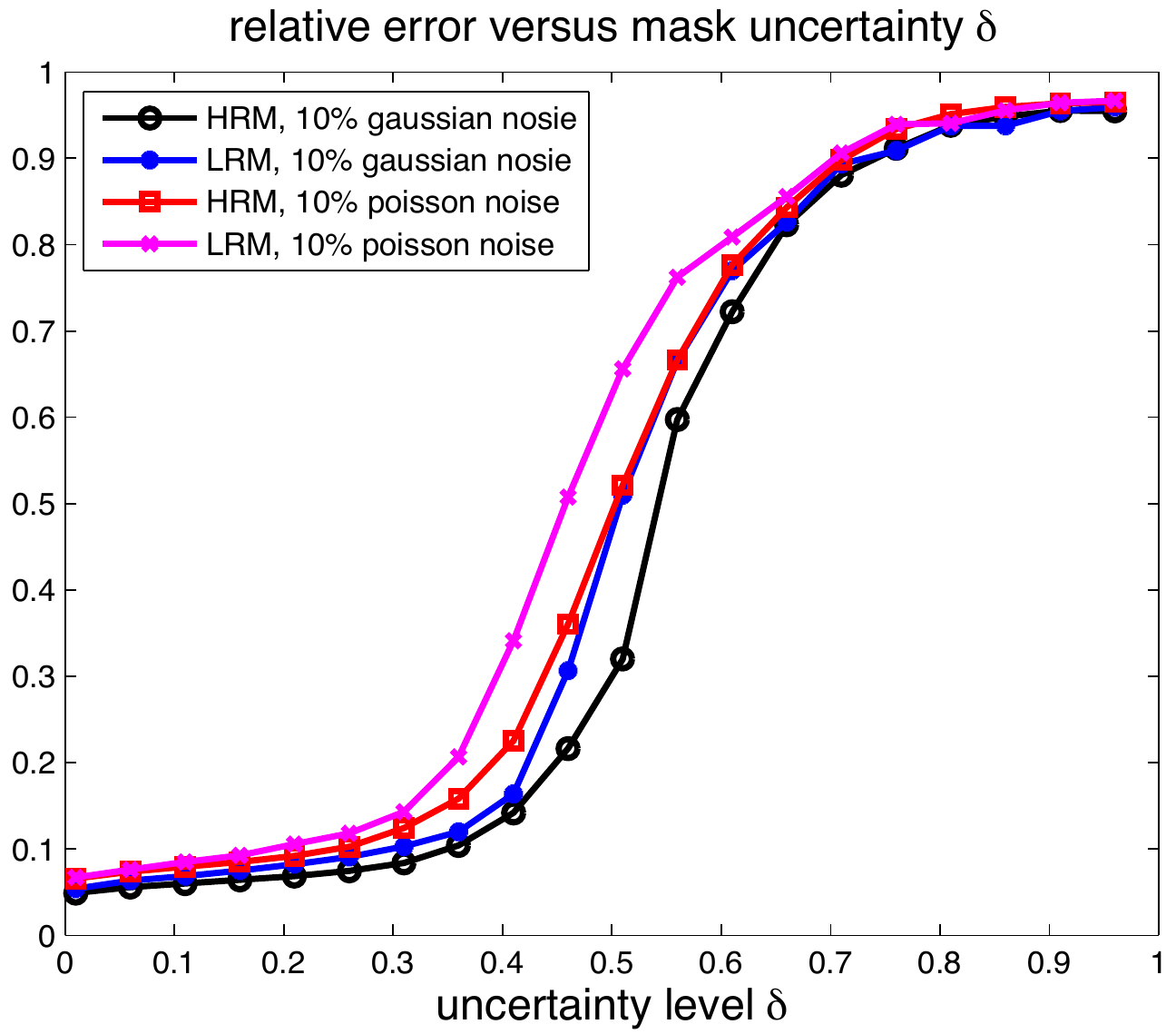}}   \hspace{-.4cm} 
               \subfigure[]{
                \includegraphics[width = 5.4cm]{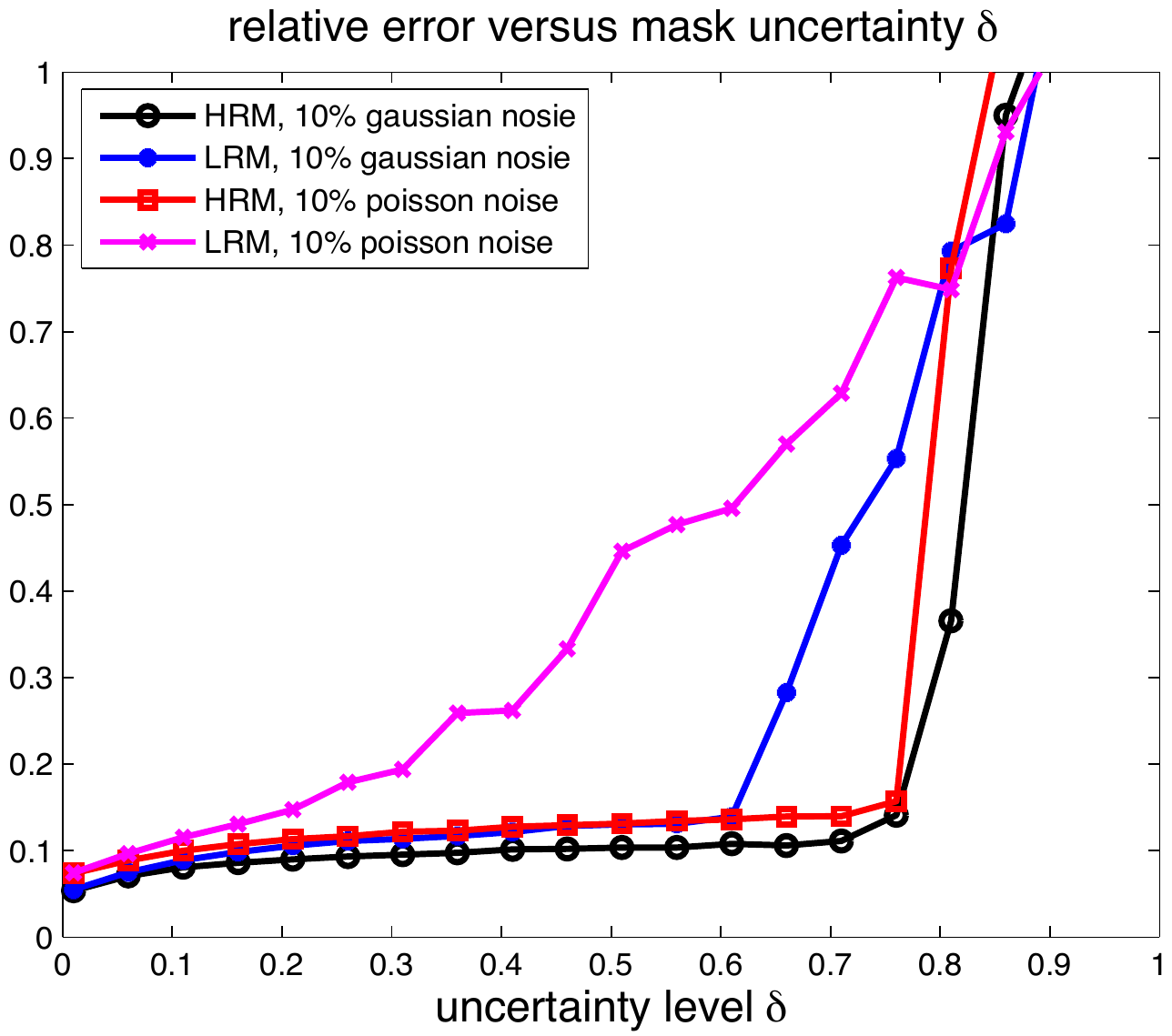}}     \hspace{-.4cm}  
             \subfigure[]{
                \includegraphics[width = 5.4cm]{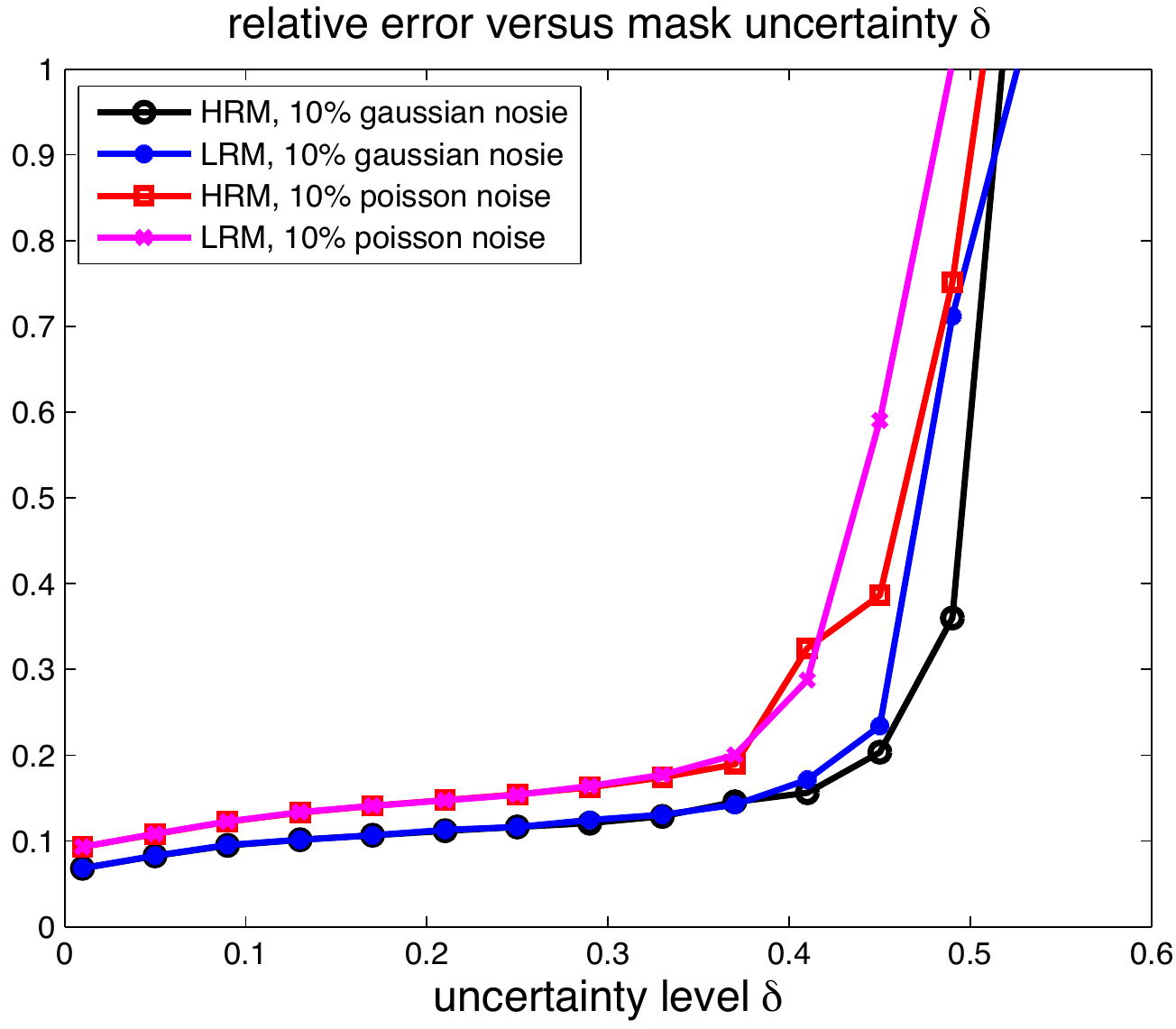}} 
                                         \caption{Noisy reconstruction error with Gaussian or Poisson noise. Averaged  relative error $e(\hat{f})$  versus mask uncertainty $\delta$ for non-negative images (left column), $\pi/2$-sector images (middle column) and  unconstrained images (right column) in the order of cameraman, mandrill and phantom (top to bottom).
                   The stopping rules and mask updating rules are the same  as described      in the main text for each case
with  the maximum of  $200+1000\cdot\delta$ steps for  DRER and AER separately.                }
                                         \label{Noisy1}
\end{figure}
We demonstrate the stability of our method with respect to
additional Gaussian and Poisson noises (10\%).

For the Gaussian noise let  $E = E_1 + i E_2$ be a complex Gaussian vector where  $E_1$ and $E_2$ consist of $|\cL|$ independent Gaussian random variables with zero mean and variance $\sigma^2$.  The noisy Fourier intensity data are given by $Y^2_{\rm noisy} = |\bPhi\Lambda f+E|^2$. We set $\sqrt{2|\cL|\sigma^2}/\|\bPhi\Lambda f\| = 10\%.$

For the Poisson noise, let the noisy data $Y_{\rm noisy}=X_{\rm noisy}/a$  where $X_{\rm noisy}$ 
 consists of  $|\cL|$ independent Poisson random variables with mean $a|\bPhi\Lambda f|$ where the scaling factor $a>0$ is chosen so that the overall noise-to-signal ratio $\|\sqrt{a |\bPhi\Lambda f|}\|/\|a |\bPhi\Lambda f|\| = 10\%$.

Figure ~\ref{Noisy1}  shows the averaged relative error $e(\hat{f})$, over 5 runs of independent random initial guesses, as a function of the mask uncertainty $\delta$ of HRM or LRM, in the presence of $10\%$ gaussian or $10\%$ poisson noise. Not surprisingly, the reconstruction with the Poisson noise
is generally worse than that with the Gaussian noise. The presence of (Gaussian or Poisson) noise amplifies the difference in performance between HRM and LRM especially in the case
of $\pi/2$-sector images  (the middle column). The reconstruction with HRM 
is stable across the board. 

\subsection{Reconstruction error versus uncertainty-to-diversity ratio (UDR)}
\begin{figure}[hthp]
\centering
\subfigure[Nonnegative images ]{ \includegraphics[width = 5.4cm]{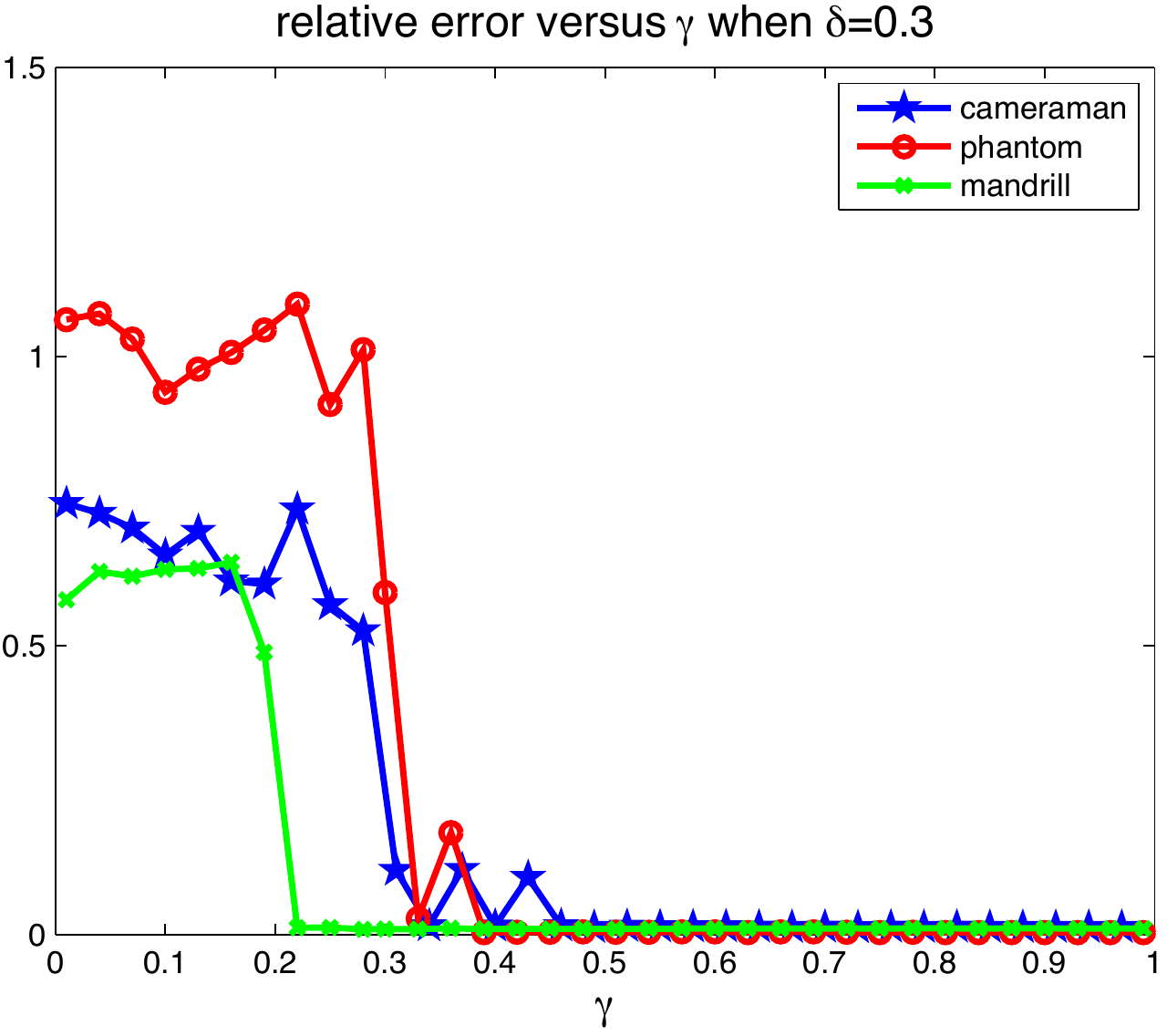}} \hspace{-.4cm} 
          \subfigure[$\pi/2$-sector images]{
         \includegraphics[width = 5.3cm]{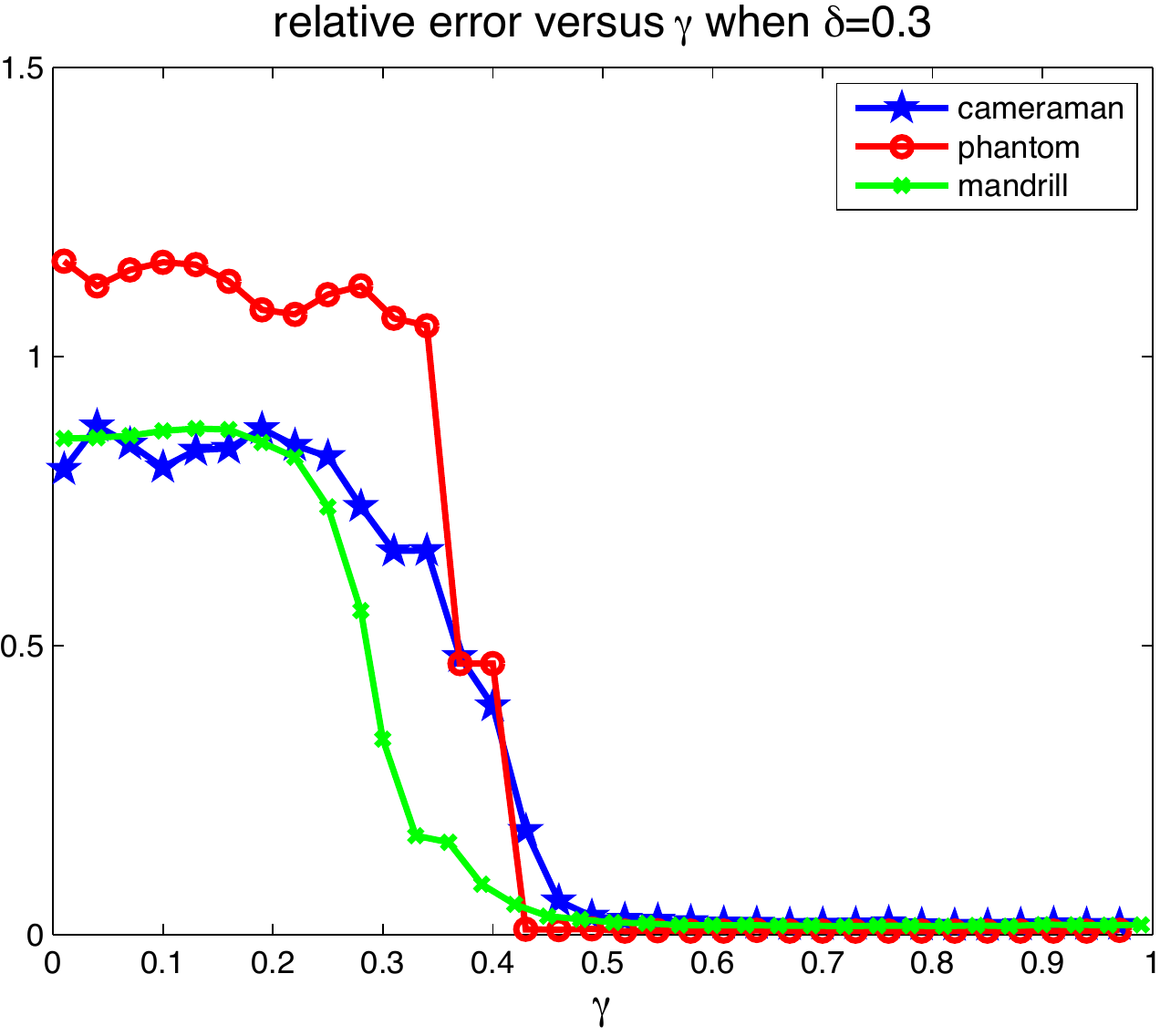}} \hspace{-.4cm} 
                 \subfigure[Unconstrained images]{
         \includegraphics[width = 5.4cm]{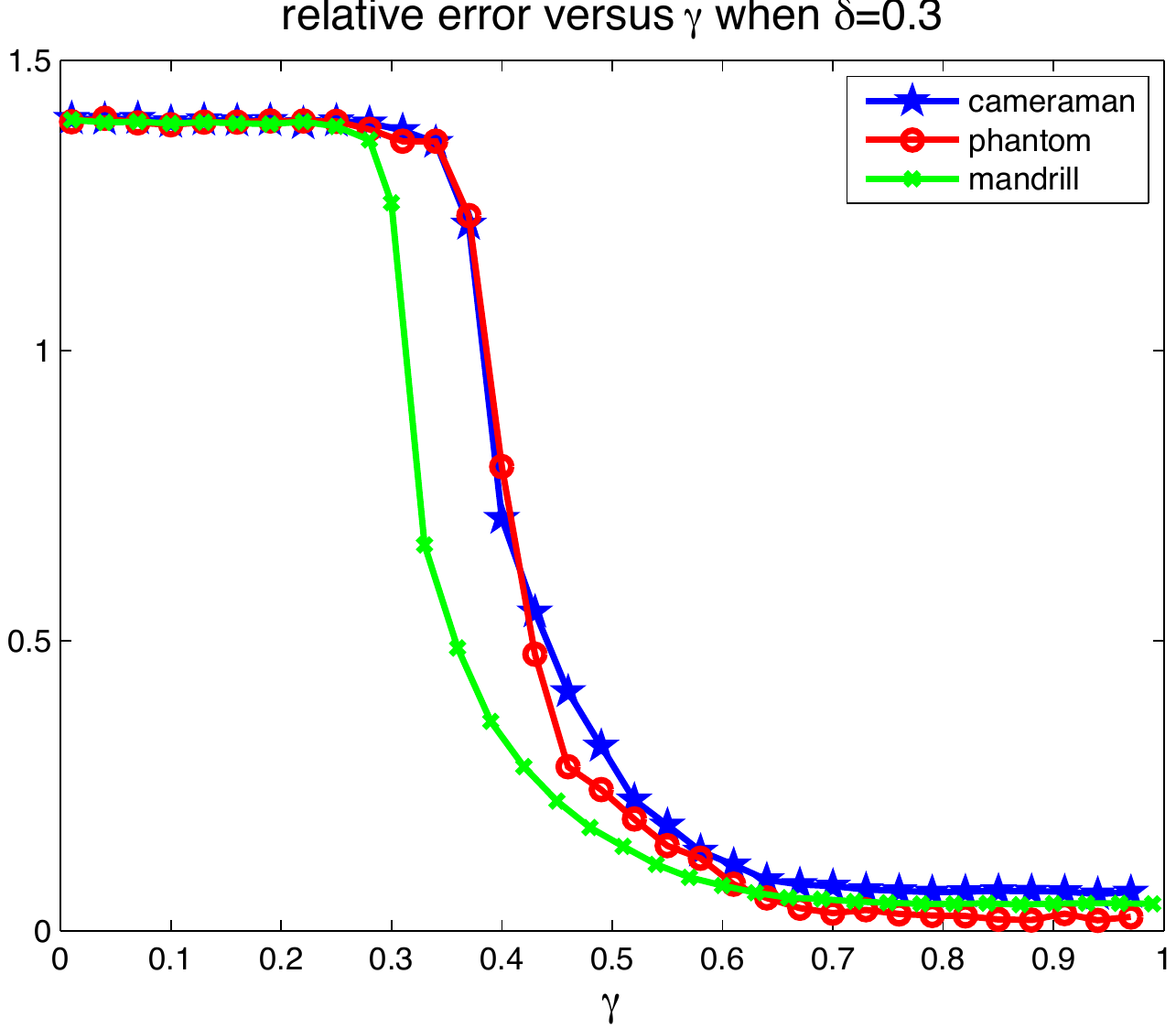}}
         \caption{Reconstruction error with variable $\gamma$ with $\delta=0.3$  for (a) nonnegative, (b) $\pi/2$-sector constrained and (c) unconstrained images.}
         \label{fig:gamma}
\end{figure}
Fig. \ref{fig:gamma} shows the relative error versus the range of mask phases for $\delta=0.3$. 
The error starts to change  precipitously around UDR $\approx 1$ consistent with the threshold predicted by  the probability bound 
$1-|\cN|\hbox{\rm UDR}^{\lfloor S/2\rfloor}$ in Theorems \ref{TheoremReal} (for non-negative images) and \ref{TheoremComplex2}. 

The real or complex mandrill has the best performance near the threshold UDR $\approx 1$  probably due to its highest sparsity $S$ among the tested images. Surprisingly  the non-negative mandrill image can be accurately recovered
with $\gamma$ just slightly greater than 0.2 (Fig. \ref{fig:gamma}(a)). By contrast,
the image with the lowest sparsity (i.e. phantom) also has the worst performance.

\section{Conclusion}\label{sec:last}

We proved the uniqueness, up to a global phase, for phasing with PUM 
with probability exponentially close to one, depending on the object sparsity and
the uncertainty-to-diversity ratio (UDR) of the mask. We 
designed algorithms that achieve nearly perfect recovery for mask uncertainty 
up to half of that promised by the uniqueness results. Additional 
object constraints such as the sector condition help mitigate the mask uncertainty.   As a by-product of
object recovery
the unknown mask can be recovered accurately within the object support.
The numerical performance is robust  with respect to the correlation in the mask as well
as external noises.

Our method  can be easily extended to general masks with phase and amplitude modulation  if
the mask amplitudes are known exactly. 
 If the mask amplitudes are  also uncertain, the proposed method will have to be substantially modified.
This will be a topic of future study.

\appendix
\begin{appendix}
\section{Proof of Theorem \ref{TheoremReal}}

\begin{proof}
As a  consequence of Theorem 2 \cite{UniqueRI}
the global ambiguities are the only  ambiguities possible as far as
the masked object (\ref{mo}) is concerned \cite{Hayes82}.  
As a consequence, 
there exist some $\mbm$ and $\theta \in [0,2\pi)$ such that either 
\begin{equation}
\tilde\mu(\bn)\tilde f (\bn) = \exp{(i \theta)} \mu(\mbm+\bn)f(\mbm+\bn)
\label{Theorem1Tem1}
\end{equation}
or 
\begin{equation}
 \tilde\mu(\bn)\tilde f (\bn)= \exp{(i \theta)} \overline{\mu(\mbm-\bn)f(\mbm-\bn)}.
\label{Theorem1Tem2}
\end{equation}
In the case of   \eqref{Theorem1Tem1} with any $\mbm \neq \mathbf{0}$ and any $\theta \in [0,2\pi)$,
$$\tilde f(\bn) = \exp{(i \theta)} \frac{|\mu(\mbm+\bn)|\exp{(i \phi(\bn+\mbm))}}{|\tilde\mu(\bn)|\exp{(i \measuredangle\tilde\mu(\bn))}} f(\bn+\mbm).$$
Consider the $\lfloor S/2\rfloor $ independently distributed r.v.s. of $\mu(\bn+\mbm)$ where $f(\bn+\mbm) \neq 0$ corresponding to $\lfloor S/2\rfloor $ nonoverlapping pairs of points $\{\bn, \bn+\mbm\}$.
For every $\bn$ where $f(\bn+\mbm) \neq \mathbf{0}$, a proper choice of $\measuredangle{\tilde\mu(\bn)}$ makes $\tilde f(\bn)$ real-valued if and only if either
\[
 \phi(\bn+\mbm)\in \lb \tphi(\bn) - \theta  -\measuredangle{f(\bn+\mbm)} \pm \delta\pi\rb
 \]
 
or
\[
\phi(\bn+\mbm)\in \lb (\tphi(\bn) - \theta -\measuredangle{f(\bn+\mbm)} +\pi) \pm \delta\pi \rb
 \]

However, $\phi(\bn+\mbm)$ is independently and uniformly distributed in $[-\gamma \pi,\gamma\pi]$, so it  falls in these two regions with probability at most $2\delta/\gamma$. The probability for every such $\tilde f(\bn)$ to be real-valued is at most $2\delta/\gamma$ and hence the probability for all $\tilde f(\bn)$ with $\mbm \neq 0$ to be real-valued is at most $(2\delta/\gamma)^{\lfloor S/2\rfloor }$. 

The union over $\mbm \neq 0$ of these events has probability at most $| \mathcal{N}| (2\delta/\gamma)^{\lfloor S/2\rfloor }$. Therefore, with probability at least $1-| \mathcal{N}| (2\delta/\gamma)^{\lfloor S/2\rfloor }$, $\mbm = \mathbf{0}$ and 
$\exp{(i \theta)} \mu(\bn) f(\bn) = \tilde\mu(\bn)\tilde f(\bn)\ \forall \bn$
which further implies that
$\tilde f(\bn) = \pm f(\bn) \ \forall \bn$
and 
$\tilde\mu(\bn) = \pm \exp{(i\theta)} \mu(\bn) \text{ on } \bn\text{ where } f(\bn) \neq 0.$

Likewise the probability for all $\tilde f(\bn)$ given by   \eqref{Theorem1Tem2} to be real-valued for any $\mbm$ is at most $| \mathcal{N}| (2\delta/\gamma)^{\lfloor S/2\rfloor }$.
\end{proof}

\section{Proof of Theorem \ref{TheoremComplex2}}
\begin{proof}As a  consequence of Theorem 2 \cite{UniqueRI}
the global ambiguities are the only  ambiguities possible as far as
the masked object (\ref{mo}) is concerned \cite{Hayes82}. Consequently,
 for some $\mbm_1, \mbm_2$ and $\theta_1 , \theta_2 \in [0,2\pi)$  either
\begin{equation}
\exp{(i \theta_1)} \mu(\bn+\mbm_1) f(\bn+\mbm_1) = \tilde\mu(\bn)\tilde f(\bn)
\label{Theorem3Tem1}
\end{equation}
or 
\begin{equation}
\exp{(i \theta_1)} \overline{\mu(\mbm_1-\bn)f(\mbm_1-\bn)} = \tilde\mu(\bn)\tilde f(\bn)
\label{Theorem3Tem2}
\end{equation}
as well as
\begin{equation}
\exp{(i \theta_2)} \mu^{(2)}(\bn+\mbm_2)f(\bn+\mbm_2) = \mu^{(2)}(\bn)\tilde f(\bn)
\label{Theorem3Tem3}
\end{equation}
or
\begin{equation}
\exp{(i \theta_2)} \overline{\mu^{(2)}(\mbm_2-\bn)f(\mbm_2-\bn)} = \mu^{(2)}(\bn)\tilde f(\bn).
\label{Theorem3Tem4}
\end{equation}
There are four possible combinations of   \eqref{Theorem3Tem1},   \eqref{Theorem3Tem2},   \eqref{Theorem3Tem3} and   \eqref{Theorem3Tem4}.

In the case of \eqref{Theorem3Tem1}\&\eqref{Theorem3Tem3}, we have
\begin{equation}
\exp{(i \theta_1)} \mu(\bn+\mbm_1) \mu^{(2)}(\bn) f(\bn+\mbm_1)
= \exp{(i \theta_2)} \mu^{(2)}(\bn+\mbm_2)\tilde\mu(\bn) f(\bn+\mbm_2).
\label{Theorem3Tem5}
\end{equation}

For any $\mbm_1 \neq \mathbf{0}$ and any $\theta_1,\theta_2 \in [0,2\pi)$, consider the $\lfloor S/2\rfloor $ pairs of independently distributed r.v.s. of $\mu(\bn+\mbm_1)$ where $f(\bn+\mbm_1) \neq 0$ corresponding to $\lfloor S/2\rfloor $ non overlapping sets of points $\{\bn, \bn+\mbm_1\}$. For every $\bn$, a proper choice of $\tilde\mu(\bn)$ makes   \eqref{Theorem3Tem5} true if and only if
\beq
\lefteqn{\phi(\bn+\mbm_1)}\nn\\
&\in& \lb ( \tphi(\bn) + \theta_2 - \theta_1 +\phi^{(2)}(\bn+\mbm_2)  -\phi^{(2)}(\bn) + \measuredangle{f(\bn+\mbm_2)} - \measuredangle{f(\bn+\mbm_1)})\pm\delta\pi\rb \label{Theorem3Tem6} 
\eeq
where $\phi^{(2)}(\bn) = \measuredangle{\mu^{(2)}(\bn)}$.

Since $\phi(\bn+\mbm_1)$ are independently and uniformly distributed in $[-\gamma\pi,\gamma\pi]$, \eqref{Theorem3Tem6} holds  for each $\bn$ with probability at most $\delta/\gamma$ and hence \eqref{Theorem3Tem5} 
holds for all $\bn$ at once with probability  at most $(\delta/\gamma)^{\lfloor S/2\rfloor }$. 

The union over $\mbm_1 \neq \mathbf{0}$ of these events has probability at most $|\mathcal{N}|(\delta/\gamma)^{\lfloor S/2\rfloor }$. Therefore, with probability at least $1-|\mathcal{N}|(\delta/\gamma)^{\lfloor S /2\rfloor }$, $\mbm_1 = \mathbf{0}$ and   \eqref{Theorem3Tem5} becomes
\begin{equation*}
\frac{\mu(\bn)}{\tilde\mu(\bn)} = \exp{(i\theta_2-i\theta_1)} \frac{\mu^{(2)}(\bn+\mbm_2)f(\bn+\mbm_2)}{\mu^{(2)}(\bn)f(\bn)}.
\end{equation*}
Moreover, if $\mu^{(2)} f$ satisfies the non-degeneracy condition, then $\mbm_2 = \mathbf{0}$, 
$\tilde f(\bn) = \exp{(i \theta_2)} f(\bn), \ \forall \bn, $ and
$\tilde\mu(\bn) = \exp{(i\theta_1-i\theta_2)}\mu(\bn), \text{ if }  f(\bn) \neq 0, $
with probability at least $1-|\mathcal{N}|(\delta/\gamma)^{\lfloor S /2\rfloor }$.

In the case of \eqref{Theorem3Tem1}\&\eqref{Theorem3Tem4}, we have 
\begin{equation}
\exp{(i \theta_1)} \mu(\bn+\mbm_1) \mu^{(2)}(\bn) f(\bn+\mbm_1) = \exp{(i \theta_2)}\tilde\mu(\bn) \overline{\mu^{(2)}(\mbm_2-\bn)f(\mbm_2-\bn)}.
\label{Theorem3Tem7}
\end{equation}
The same argument applies and $\mbm_1 = 0$ with probability at least $1-|\mathcal{N}|(\delta/\gamma)^{\lfloor S /2\rfloor }$, and  \eqref{Theorem3Tem7} becomes
\begin{equation*}
\frac{\mu(\bn)}{\tilde\mu(\bn)} = \exp{(i\theta_2-i\theta_1)} \frac{\overline{\mu^{(2)}(\mbm_2-\bn)f(\mbm_2-\bn)}}{\mu^{(2)}(\bn)f(\bn)},
\end{equation*}
which violates the non-degeneracy condition. In other words,   \eqref{Theorem3Tem1}\&\eqref{Theorem3Tem4} holds with probability at most $|\mathcal{N}|(\delta/\gamma)^{\lfloor S /2\rfloor }$.

Similar conclusions follow in the case of   \eqref{Theorem3Tem2}\&\eqref{Theorem3Tem3} and   \eqref{Theorem3Tem2}\&\eqref{Theorem3Tem4}.
\end{proof}
\section{Proof of Lemma \ref{lemmaer2}}
\begin{proof} Since the operator $\cT$ enforces the measured  Fourier intensities $Y^2$ 
\beq
{r(f_{k+1},\mu_{k+1}) }
& =& \|  \   |\bPhi\Lambda_{k+1}f_{k+1}| - Y  \  \|= \|  \   \bPhi\Lambda_{k+1}f_{k+1} - \cT\bPhi\Lambda_{k+1}f_{k+1}  \  \| \nn\\
& \le& \| \bPhi \Lambda_{k+1} f_{k+1} - \PT \bPhi \Lambda_{k} f_{k+1} \|  = \| \Lambda_{k+1} f_{k+1} - \bPhi^{-1}\PT \bPhi \Lambda_{k} f_{k+1} \| \nonumber 
\eeq
by the unitarity of the Fourier transform. By splitting the summation and using the definition (\ref{QF}), the rightmost term becomes
\beq
 & & 
\left(\sum_{f_{k+1}(\bn) \neq 0} | f_{k+1}|^2(\bn)\Big| \mu_{k+1}(\bn)- \mu_k'(\bn) \Big|^2
 +
\sum_{f_{k+1}(\bn) = 0}  \Big|\bPhi^{-1}\PT\bPhi\Lambda_k f_{k+1}(\bn) \Big|^2\right)^{1/2}.
 \label{c.2}
\eeq
Now since $\mu_{k+1}(\bn)=\QM \mu'_k(\bn)$ is a pixel-wise
projection of $\mu_k'(\bn)$,  $| \mu_{k+1}(\bn)- \mu_k'(\bn) | \leq
| \mu_{k}(\bn)- \mu_k'(\bn)|$ and hence (\ref{c.2}) is less than or equal to 
\beq
& \le &\left( 
\sum_{f_{k+1}(\bn) \neq 0}  |f_{k+1}|^2(\bn)\Big|\mu_{k}(\bn)- \mu_k'(\bn) \Big |^2
 +
\sum_{f_{k+1}(\bn) = 0}  \Big | \bPhi^{-1}\PT\bPhi\Lambda_k f_{k+1}(\bn) \Big |^2
\right)^{1/2}  \nonumber  \\
& = &\| \Lambda_k f_{k+1} - \bPhi^{-1}\PT\bPhi \Lambda_k f_{k+1} \| = \| \bPhi \Lambda_k f_{k+1} - \PT\bPhi \Lambda_k f_{k+1} \| \nonumber \\
& =& \| \ |\bPhi\Lambda_k f_{k+1}  | - Y  \ \| = r(f_{k+1},\mu_k)  \nonumber
\eeq
which is the desired result. 
\end{proof}

\end{appendix}


\begin{thebibliography}{plain}
\bibitem{BCL02} 
H. H. Bauschke, P. L. Combettes and D. R. Luke, \lq\lq Phase retrieval, error reduction algorithm, and Fienup variants: a view from convex optimization,"  {J. Opt. Soc. Am. A} {\bf 19},  13341-1345 (2002).  

\bibitem{BCL04}
H. H. Bauschke,
P. L. Combettes,
and
D. R. Luke,
``Finding best approximation pairs relative to two
closed convex sets in Hilbert spaces,"  J. Approx. Th. {\bf 127}, 178Ð192 (2004)
\bibitem{ptycho10}
M. Dierolf, A. Menzel, P. Thibault, P. Schneider, C. M. Kewish, R. Wepf, O. Bunk, and F. Pfeiffer, `` Ptychographic x-ray computed tomography at the nanoscale,'' { Nature} {\bf 467}, 436-439 (2010).

\bibitem{DR}
J. Douglas and H.H. Rachford, "On the numerical solution of heat conduction problems in two and three space variables," {\em Trans.  Am.  Math.  Soc.}{\bf 82}, 421-439 (1956).



\bibitem{UniqueRI} 
A. Fannjiang, \lq\lq Absolute uniqueness of phase retrieval with random illumination," { Inverse Problems} {\bf 28}, 075008(2012). 

\bibitem{FL} 
A. Fannjiang and W. Liao, \lq\lq Phase retrieval with random phase illumination," {J. Opt. Soc. A}, {\bf 29},   1847-1859(2012).  

\bibitem{RKM}
A. Fannjiang and W. Liao, ``Phase retrieval with roughly known mask," arXiv:1212.3858. 

\bibitem{Fie82} 
J. R. Fienup, \lq\lq Phase retrieval algorithms: a comparison,"  {Appl. Opt.} {\bf 21},   2758-2769 (1982).  

\bibitem{Fie2}
J.R. Fienup, ``Reconstruction of a complex-valued object from the modulus of its Fourier transform using a support constraint,'' {\em J. Opt. Soc. Am. A } {\bf 4}, 118 -123 (1987).

\bibitem{FW}
J.R. Fienup and C.C. Wackerman,
``Phase-retrieval stagnation problems and solutions," {\em J. Opt. Soc. Am. A} {\bf 3} 1897-1907 (1986).


\bibitem{GS72} 
R. W. Gerchberg and W. O. Saxton, \lq\lq A practical algorithm for the determination of the phase from image and diffraction plane pictures,"  { Optik } {\bf 35},  237-246, 1972.  



\bibitem{Hayes82}
M. Hayes, "The reconstruction of a multidimensional sequence from the phase or magnitude of its Fourier transform," {IEEE Trans. Acoust. Speech Sign. Proc.} {\bf 30} 140-
154 (1982).

\bibitem{HM82}
M.H. Hayes and J.H. McClellan. ``Reducible Polynomials in More Than One Variable." {Proc. IEEE} {\bf  70}(2):197 Ð 198, (1982).







\commentout{\bibitem{periphery}
A. Jesacher, W. Harm, S. Bernet, and M. Ritsch-Marte,
``Quantitative single-shot imaging of complex objects using phase retrieval with a designed periphery," {\em Opt. Exp.}{\bf 20} (2012) 5470-5480.}

\bibitem{LM79}
P.-L. Lions and B. Mercier,``Splitting algorithms for the sum of two nonlinear operators," {SIAM  J. Num.  Anal.} {\bf 16}, 964-979 (1979).


\bibitem{ptycho2}
A. M. Maiden, M. J. Humphry, F. Zhang and J. M. Rodenburg, ``Superresolution imaging via ptychography,"
{J. Opt. Soc. Am. A} {\bf 28}, 604-612 (2011).

\bibitem{ptycho-rpi}
A.M. Maiden,	 G.R. Morrison,	 B. Kaulich,	 A. Gianoncelli	 \& J.M. Rodenburg,
``Soft X-ray spectromicroscopy using ptychography with randomly phased illumination,"
{\em Nat. Commun.} {\bf 4}, 1669 (2013). 



\bibitem{Miao99} 
J. Miao, P. Charalambous, J. Kirz and D. Sayre, \lq\lq Extending the methodology of X-ray crystallography to allow imaging of micrometre-sized non-crystalline specimens,"  { Nature} {\bf 400},  342--344 (1999).  

\bibitem{Miao00} 
J. Miao, J. Kirz and D. Sayre, \lq\lq The oversampling phasing method,"  { Acta Cryst. D} {\bf 56}, 1312--1315 (2000).  

\bibitem{MSC}
J. Miao, D. Sayre and H.N. Chapman, ``Phase retrieval from the magnitude of the Fourier transforms of nonperiodic objects,"
{\em J. Opt. Soc. Am. A} {\bf 15} 1662-1669 (1998).

\bibitem{Nut03}
K.A. Nugent, A.G. Peele, H.N. Chapman, \&  A.P. Mancuso, ``Unique Phase Recovery for Nonperiodic Objects," 
{\em Phys. Rev. Lett.} {\bf 91}, 203902 (2003).

\bibitem{Nut05}
K. A. Nugent, A. G. Peele, H. M. Quiney and H. N. Chapman, ``Diffraction with wavefront curvature: a path to unique phase recovery,"
{\em  Acta Crystallogr. Sect. A} {\bf 61}, 373-381 (2005).



\bibitem{ptycho08}
P. Thibault, M. Dierolf, A. Menzel, O. Bunk, C. David, F. Pfeiffer, ``High-resolution scanning X-ray diffraction microscopy", { Science} {\bf 321},  379-382 (2008).

\bibitem{Thi09}
P. Thibault, M. Dierolf, O. Bunk, A. Menzel, F. Pfeiffer, 
``Probe retrieval in ptychographic coherent diffractive imaging,"
Ultramicroscopy
{\bf 109},  338Ð343 (2009)
\bibitem{Wil06}
G. J. Williams,
H. M. Quiney,
B. B. Dhal,
C. Q. Tran,
K. A. Nugent,
A. G. Peele,
D. Paterson,
and M. D. de Jonge, ``Fresnel coherent diffractive imaging", {\em Phys. Rev. Lett.}
{\bf 97}, 025506(2006).

\bibitem{Rod10}
F. Zhang and J. M. Rodenburg, ``Phase retrieval based on wave-front relay and modulation,"
{\em Phys. Rev.  B} {\bf  82}, 121104(R) (2010). 

\end{thebibliography}
\end{document}